\documentclass[sigconf,balance=false]{acmart}
\usepackage{popets}

\newif\ifarxiv
\arxivtrue 
\arxivfalse 
\newcommand{\shortlong}[2]{\ifarxiv{#2\xspace}\else{#1\xspace}\fi}

\usepackage{amsthm}
\usepackage[inline]{enumitem}
\theoremstyle{definition}
\newtheorem{theorem}{Theorem}[section]
\newtheorem{definition}[theorem]{Definition}
\newtheorem{lemma}[theorem]{Lemma}
\newtheorem{remark}[theorem]{Remark}

\usepackage[conf={restate,no link to proof}]{proof-at-the-end}
\usepackage{footmisc}

\usepackage[linesnumbered,ruled,vlined]{algorithm2e}

\usepackage{colortbl}
\usepackage{lineno}
\usepackage{multirow}
\usepackage{array}
\usepackage{dsfont}
\allowdisplaybreaks

\usepackage{xspace}
\def\notationcolor{black} 

\newcommand{\notation}[2]{\newcommand{#1}{{\textcolor{\notationcolor}{\ensuremath{#2}}}}}

\newcommand{\term}[2]{\newcommand{#1}{\textcolor{\notationcolor}{#2}\xspace}}

\newcommand{\Var}{\operatorname{Var}} 
\newcommand{\E}{\operatorname{E}} 
\newcommand{\real}{\mathbb{R}}

\notation{\genb}{\beta}
\notation{\alphe}{\genb_{emp}}
\notation{\alphw}{\genb_{wage}}

\notation{\mech}{\mathcal{M}}
\notation{\range}{range}
\notation{\data}{\mathcal{D}}
\notation{\randalg}{\mathcal{A}}
\notation{\sign}{sign}
\notation{\outp}{\omega}
\notation{\typeone}{\alpha}
\notation{\eqdist}{\overset{distr.}{=}} 
\notation{\Fchi}{\operatorname{F}_{\chi^2_1}}

\notation{\estab}{\mathcal{E}} 
\notation{\patt}{A} 
\notation{\catt}{C} 
\notation{\numatt}{k} 
\notation{\nump}{\numatt_{p}} 
\notation{\numc}{\numatt_{c}} 
\notation{\pattv}{a} 
\notation{\cattv}{c} 
\notation{\numest}{n} 
\notation{\record}{r} 
\notation{\sqroot}{\sqrt{\phantom{x}}} 

\notation{\distparam}{\gamma} 
\notation{\noclip}{\widehat{\tau}}
\notation{\neighfun}{\psi} 
\notation{\confparam}{\zeta}

\notation{\lowerbound}{L}
\notation{\upperbound}{U}
\notation{\uncertI}{\mathcal{I}} 
\notation{\identqj}{q_j^{(I)}} 
\notation{\identq}{q^{(I)}} 
\notation{\groupjbyiq}{q_j^{(\patt_i)}} 
\notation{\truncaggjbyiq}{q_j^{(T,\patt_i)}} 
\notation{\truncagggenbyoneq}{q^{(T,\patt_1)}} 
\notation{\truncagggenbyQq}{q^{(T,\patt_{\numquery})}} 

\notation{\pattuniq}{\xi} 
\notation{\aggjbyiv}{v_{\ell}} 
\notation{\groupl}{\mathcal{G}_{\ell}}

\notation{\aggjbyi}{V_{\ell}} 
\notation{\maxjbyi}{b_{\ell}} 
\notation{\numgroup}{k_g} 
\notation{\numell}{k_\ell} 
\notation{\saggv}{v} 
\notation{\sagg}{V} 
\notation{\smax}{m_{*}} 

\notation{\truncval}{\tau} 
\notation{\noisymaxjbyi}{\widetile{M}_{\ell}} 
\notation{\noisymaxjbyiv}{\widetile{m}_{\ell}} 
\notation{\truncdist}{\delta} 
\notation{\numquery}{k_{Q}}
\notation{\addmaxtrunc}{\zeta}

\notation{\sens}{\Delta}
\notation{\senstwo}{\sens_2}
\notation{\sensneighfun}{\sens_{\neighfun}}
\notation{\senssqrt}{\sens_{\sqroot}}
\notation{\gquery}{gsum} 
\notation{\grouper}{\mathcal{G}} 
\notation{\basicmech}{\mech_{ge}}

\newcommand{\queryprivbudget}[1]{\mu_{#1}}
\newcommand{\queryprivbudgetemp}[1]{\mu_{#1,e}}
\newcommand{\queryprivbudgetw}[1]{\mu_{#1,w}}

\term{\neighmech}{\neighfun-mechanism} 
\term{\sqrtmech}{\sqroot-mechanism} 
\term{\pncmech}{pnc-mechanism} 
\term{\shortpnc}{pnc}

\term{\sqn}{sqrt-neighbors}
\term{\gedp}{Gaussian Establishment DP}
\term{\shortdp}{gedp}
\term{\longdp}{Gaussian Establishment Differential Privacy}

\term{\conf}{confidential}
\term{\san}{sanitized}

\newcommand{\pickles}{\ensuremath{\mathbb{G}}}
\newcommand{\bob}{\ensuremath{\mathbb{B}}}
\newcommand{\alice}{\ensuremath{\mathbb{A}}}
\newcommand{\fred}{\ensuremath{\mathbb{F}}}
\newcommand{\wendy}{\ensuremath{\mathbb{C}}}

\setcopyright{popets}
\copyrightyear{YYYY}

\acmYear{YYYY}
\acmVolume{YYYY}
\acmNumber{X}
\acmDOI{XXXXXXX.XXXXXXX}
\acmISBN{}
\acmConference{Proceedings on Privacy Enhancing Technologies}
\begin{document}

\title[Confidentiality Protection for Establishment Data]{Statistics-Friendly Confidentiality Protection for Establishment Data, with Applications to the QCEW}

\author{Kaitlyn Webb, Prottay Protivash, John Durrell, Aleksandra Slavkovi\'c, Daniel Kifer}
\orcid{}
\affiliation{%
\institution{Pennsylvania State University}
\city{}
  \state{}
  \country{}
}

\author{Daniell Toth}
\affiliation{%
\institution{National Institute of Statistical Sciences}
\city{}
  \state{}
  \country{}
}
\renewcommand{\shortauthors}{Webb et al.}

\begin{abstract}
 Confidentiality for business data is an understudied area of disclosure avoidance, where legacy methods struggle to provide acceptable results. Standard formal privacy techniques  for person-level data, like differential privacy, are designed to protect against membership inference and hence do not provide suitable confidentiality/utility trade-offs  due to the highly skewed nature of business data and because extreme outlier records are often important contributors to query answers. Prior proposals, therefore, took a personalized differential privacy approach that allowed privacy parameters to degrade for the outlying records -- larger establishments get weaker membership inference guarantees. However, 
 providing guarantees to some entities that are strictly weaker than guarantees for others is problematic from a policy standpoint. In this paper, we propose a novel confidentiality framework for business data with a focus on interpretability for policy makers. Instead of protecting against membership inference, which is often not a concern in business data, we protect against attribute inferences that are too precise. In our framework, data curators specify a neighbor function that is used to define  uncertainty interval bands around an establishment's attribute values and the privacy parameters govern the strength of indistinguishability between values within the same uncertainty interval.
 We propose two query-answering mechanisms under this framework and evaluate them on: (1) a confidential Quarterly Census of Employment and Wages (QCEW) dataset produced by the U.S. Bureau of Labor Statistics (this was done through a cooperative agreement), and (2) a substitute dataset that we created from public sources (and will publicly release).
\end{abstract}


\maketitle
\section{Introduction}\label{sec:intro}
Business data provide crucial snapshots of various
sectors of a country's economy. The collection and dissemination of business statistics is of great value in both the public and private sectors. As with person-level data, there are strong confidentiality concerns, but the requirements are not the same. Protections for person-level data are designed to mask records that are outliers, as individuals with uncommon characteristics can be vulnerable to disclosure and harassment. However, the ``outliers'' in business data are the large companies whose contributions to employment and wage statistics are important to study. In business data, the analogue to a person is an \emph{establishment}, which is a single physical location where business activity occurs (e.g., a specific restaurant in a restaurant chain).

Disclosure avoidance for person-level data has advanced with the rise of differential privacy (DP) \cite{dwork2006calibrating}, but much less work has focused on protecting establishment data. 
The dominant method in practice is cell suppression \cite{adam1989security}. It is used, for example, to protect the Quarterly Census of Employment and Wages (QCEW), the flagship economics dataset produced by the Bureau of Labor Statistics (BLS) \citep{blshandbook,cohen2006}. Given a set of desired statistics, cell suppression redacts a subset of them so that a weak attacker (who can only use linear algebra and has no access to external data) cannot derive sensitive information about a business establishment. Despite this weak adversarial model, cell suppression suffers from severe loss of utility:  $\approx 60\%$ out of roughly 3.6 million statistics in the QCEW tables are suppressed \citep{cohen2006, yang_et_al_11, toth2014}. Hence there is a need for new technology with rigorous formal privacy guarantees and better utility.


This problem has received little attention from the formal privacy community \cite{haney,splittingmech,finley2024slowly, tran2023differentiallyprivatesyntheticheavytailed}. Viable solutions should (1) provide confidentiality guarantees that are \textbf{interpretable} to mathematical statisticians and policy makers in statistical agencies and (2) should provide \textbf{flexibility}  that allows policy makers to specify how the confidentiality needs for large establishments differ from those of small establishments. Earlier work \cite{haney,tran2023differentiallyprivatesyntheticheavytailed} did not support any customization. Recent work \cite{splittingmech,finley2024slowly} has made progress in this direction, by applying the personalized DP philosophy \cite{jorgensen2015conservative}: every establishment should have its own privacy parameters. Specifically, \cite{splittingmech,finley2024slowly} propose schemes for protecting membership inference (i.e., whether an establishment was included in the data) where larger businesses get worse privacy parameters than smaller businesses, leading to the interpretation that some businesses get strictly weaker protections than others -- a position that can be problematic for policy-makers. In many cases, membership inference, especially for large businesses, is not a concern -- they need to be included in the data for reliable economic estimates, and their existence is public knowledge. For instance, the QCEW is a census and collects data about all businesses.

Instead, we design our approach based on the metric DP philosophy \cite{metricdp}: a entity should have plausible deniability about whether its  true value is $x$ or some ``nearby'' value, where the concept of ``nearby'' is defined using an application-specific metric. We argue that smaller establishments need stronger guarantees on the relative error of an attacker's inference about them, while large establishments need stronger guarantees on the absolute error, where absolute error refers to the standard error of the attacker's estimate and relative error is the absolute error over the confidential value. We derive mathematical conditions on metrics that will guarantee such an interpretation of the confidentiality semantics. The result is that the specification of \emph{what} needs plausible deniability becomes customized to an establishment's size and is separated from the quantification of \emph{how strong} the plausible deniability is (i.e., how hard is it to distinguish $x$ from a nearby value $y$). 

We then propose two query answering mechanisms for this framework. The first mechanism provides query answers that have a data-dependent variance that cannot be released. The second mechanism piggybacks on the first one to answer additional queries with a data-dependent variance that is completely safe to release. This property is very useful to statisticians since they need to assess how much they can trust a noisy query answer. We evaluate our techniques on a real confidential high-stakes dataset, the QCEW, by leveraging a cooperative agreement with the Bureau of Labor Statistics. We also evaluate our techniques on a business establishment dataset that we synthesized from public data sources. We plan to share this dataset after publication in order to help support future research on this topic. 
To summarize, our contributions are:

\begin{itemize}[leftmargin=*]
    \item Inspired by Gaussian Differential  Privacy \cite{gdp} and Metric DP \cite{metricdp}, we propose a privacy definition for business establishment data that we call \gedp (or \shortdp for short). Within the Metric DP framework, we identify a class of metrics suitable for business data and which formalize the plausible deniable needs of establishments of different sizes. We show how the metrics can be interpreted as intended uncertainty intervals around an establishment's true values, emphasizing more relative error uncertainty for smaller establishments and more absolute error uncertainty for larger ones. To measure the strength of the plausible deniability, we use privacy accounting based on Gaussian Differential Privacy as it directly measures resilience to statistical attacks: how much success can the best hypothesis test achieve when trying to distinguish between two plausible values $x$ and $y$ that are in each other's intended uncertainty intervals? 
    \item We advocate for a specific choice of a metric, based on the square root transform, due to its superior privacy and statistical properties compared to the traditional business data approaches that aim to give all establishments equal relative error protections. 
    \item We introduce two mechanisms for \gedp, the \neighmech and \pncmech, for answering group-by queries under this framework. The \neighmech  has data-dependent variance that cannot be revealed exactly. However, the \pncmech piggybacks on top of the \neighmech to provide answers to additional queries, using data-dependent Gaussian noise whose variance is completely safe to release. This is an important property for statisticians. 
    \item We evaluate our mechanisms on real-world data (a historical, confidential QCEW dataset accessed through a cooperative agreement) and on a publicly shareable partially synthetic dataset that we reconstructed from partially redacted statistics published by the U.S. Census Bureau. 
\end{itemize}

Note: our methods apply to skewed data in general. We focus on business data for concreteness to better explain the confidentiality requirements and to spur interest in the topic -- it is an untapped subfield with huge practical and research potential.

The paper is organized as follows. We provide background information on the QCEW in  Section~\ref{sec:qcew} since the design of formal privacy definitions for establishment data require an understanding of the confidentiality concerns and characteristics of such data. In Section~\ref{sec:background}, we present notation and technical background on Gaussian DP \cite{gdp} (which motivates our framework) and discuss related work in Section~\ref{sec:related}. In Section \ref{sec:model}, we propose the privacy definition \gedp. We present two query-answering mechanisms in Section~\ref{sec:mech}. In Section~\ref{sec:microdata}, we explain how to convert noisy query answers into privacy preserving microdata. 
In Section~\ref{sec:comps}, we present experiments on confidential real and publicly shareable partially synthetic data . We conclude and outline future work in Section~\ref{sec:conc}. Proofs can be found in the appendix.

\section{Background on the QCEW}\label{sec:qcew}

The Bureau of Labor Statistics (BLS) Quarterly Census
of Employment and Wages (QCEW) data is an economically and politically important dataset containing wage and employment information about businesses in the United States.
The data provided by QCEW are crucial for measuring labor trends and industry developments, provide a sample-frame for all BLS establishment surveys, and are used to benchmark the Current Employment Statistics survey, which is a primary economic indicator used by the federal reserve \cite{brave2021}. QCEW data has also been used for a wide range of analyses that have helped economists understand the effects of minimum wage, the non-profit sector and the effect of the COVID-19 pandemic on employment levels \citep{meer16, salamon2005, dalton2020}.

The QCEW is collected each quarter under a cooperative program between BLS and the State Employment Security Agencies. 
States collect the data from unemployment insurance (UI) administrative records which firms are legally obligated to submit quarterly. 
The primary unit of analysis is an \emph{establishment}, which is a single physical location where one predominant activity, classified by a single industry code, occurs \citep{sade2016}. Meanwhile, a \emph{firm} is a business made up of one or more establishments. All of the state-collected, administrative data, covering 
 over 10~million establishments, are sent to the BLS \citep{blshandbook,cohen2006}.
Since the UI data are collected for all U.S. workers covered by state unemployment insurance laws or by the Unemployment Compensation for Federal Employees program, they represent a virtual census of employment and wages, providing a complete overview of the national economic picture.

BLS supplements the states' UI data collection by conducting an annual refiling survey to ensure each establishment is labeled under the correct industry code and further cleans the data. The data from each state are then sent back to the state along with BLS produced aggregated estimates protected by a disclosure limitation method. \textbf{Each state retains ownership over its own data under this arrangement.}  Thus, there are  \emph{54 different data owners} (50 states, the District of Columbia, Puerto Rico, Virgin Islands, and BLS), each covered by different laws and policies. In addition to the data published by BLS, each data owner can separately publish statistics about their own parts of the data. Under the current cell suppression methodology, it is feasible that data published by BLS, when combined with a separate release produced by a state, could be used to undo protections for a different state. This is known as a \emph{composition attack} \cite{ganta}. Our motivation for studying a formal privacy alternative to cell suppression is motivated by this problem, since formal privacy methods for person-level data are known to resist composition attacks.

\textbf{Data Schema:} Quarterly data about each establishment include (1) location (e.g., county and state), (2) total quarterly wages,  (3) three monthly employment counts, (4) an ownership type (private, local government, etc.) and (5) a six-digit industry code known as the North American Industry Classification System (NAICS) code. The NAICS code \cite{naics} is hierarchically structured. For example, ``construction'' has the NAICS prefix 23, ``construction of buildings'' has the prefix 236 and the NAICS code for ``new multifamily housing construction'' is 236116. \textbf{The wages and employment data are considered {confidential} while the rest of the attributes (county, NAICS code, ownership type) are considered {public.}}

\textbf{Dissemination:} The QCEW establishment-level data are published in aggregated form, mostly as group-by sum queries, also known as marginals and are identified by an aggregation level code \cite{qcewdata}; e.g., aggregation level 15 (``National, by NAICS 3-digit -- by ownership sector'') provides the national-level total quarterly wages and monthly employment totals for each combination of 3-digit NAICS code and ownership type (private, foreign government, state government, etc.) while aggregation level 55 provides the same information, but broken down by state, and 75 breaks it down by county.
BLS also receives requests for custom tabulations. Thus a highly desirable property for a disclosure avoidance system is the ability to convert noisy query answers into \emph{confidentiality-protected microdata} from which those custom tabulations can be computed. 

\textbf{Skewness:}
A major difference between establishment data and person-level data is the skewness of the data -- there are outliers whose influence on query answers is important to preserve. Establishments can be very small, such as an ice-cream shop with a handful of employees, or very large, like Disneyland Resort, which employs approximately 
36,000 employees in the Orange County, CA \cite{DisneylandResorts_2025}. 

The formal privacy approach for person-level data is to inject enough noise into query answers to mask the effect of any entity. 
Could this idea be directly applied to establishment data by defining an entity to be a person, or by defining it to be an entire establishment? The naive idea of masking the effect of one \emph{person} in establishment data is too weak because it may reveal employment and wage  information about a business that is too precise. For example Disneyland \cite{DisneylandResorts_2025} and Microsoft \cite{msjobs} publicly provide rough estimates of their employment counts (36,000 and 125,000, respectively), that are rounded to the thousands, indicating they may need larger protections.
On the other hand, injecting enough noise to mask the presence an \emph{establishment} like Disneyland Resort  would render regional economic statistics meaningless. 

Proposals for addressing this skewness issue \cite{splittingmech,finley2024slowly} took a personalized DP approach \cite{jorgensen2015conservative} in which larger establishments receive worse privacy parameters than smaller establishments and therefore are given weaker privacy semantics. This can be difficult to justify from a public policy standpoint. Our alternative proposal is given in Section \ref{sec:model}, after discussing background and related work.

\section{Notation and Background}\label{sec:background}


\begin{table}[htbp]
\begin{center}
\caption{Table of Notation}\label{tab:notation}
\begin{tabular}{|cp{0.73\linewidth}|}\hline
$\Phi$: & Gaussian CDF\\
$\mu$: & Gaussian DP and \shortdp Privacy Parameter\\
$\data$: & Data set of records for establishments $\estab_1,\dots,\estab_\numest$\\
$\patt_1,\dots, \patt_\nump$: & Public attributes of an establishment \\
$\catt_1,\dots, \catt_\numc$: & Confidential attributes of an establishment\\
$\record$: & Establishment record. $\record[\catt_i]$ is the value for $\catt_i$\\
$\neighfun$: & Neighbor function (Definition \ref{def:close}).\\
$\distparam$: & Closeness parameter (Definition \ref{def:close}).\\
$\mech$: & Mechanism.\\
$\uncertI_{\neighfun,\distparam}(x)$: & Uncertainty interval around $x$ induced by $\neighfun,\distparam$.\\
\hline
\end{tabular}
\end{center}
\end{table}

The notation used in the paper is summarized in Table \ref{tab:notation}.
 A dataset $\data$ contains records about  establishments $\estab_1,\dots, \estab_{\numest}$. Establishments have \textbf{public attributes} $\patt_1,\dots, \patt_{\nump}$ and \textbf{confidential attributes} $\catt_1,\dots,\catt_\numc$. The confidential attributes are numeric and often nonnegative (e.g., number of employees).  
 To simplify the discussion, we will assume that the domain of the confidential attributes is nonnegative, but in Section \ref{sec:nonon}, we explain how this condition can be completely eliminated.
 
We use $\pattv_i$ (resp., $\cattv_i$) to refer to specific values of attribute $\patt_i$ (resp., $\catt_i$). We can represent an establishment record $r$ as a tuple  $\record=(\pattv_1,\dots, \pattv_\nump, \cattv_1,\dots, \cattv_{\numc})$ and refer to attributes within a record using indexing notation; e.g., the value of attribute $\catt_1$ in $\record$ is $\record[\catt_1]$.

A \emph{mechanism} $\mech$ is a randomized algorithm whose input is confidential data and whose output is intended for public distribution. Our framework for protecting establishment data builds upon ideas from the Metric DP framework \cite{metricdp} and Gaussian DP privacy accounting  \cite{gdp}, which are defined as follows:

\begin{definition}[Metric DP \cite{metricdp}] Given a metric $d_X$ over datasets, a mechanism $\mech$ satisfies (metric) $d_X$-privacy if for all $x\in\range(\mech)$ and all pairs of datasets $\data_1, \data_2$, $$P(\mech(\data_1)=x)\leq e^{d_X(\data_1, \data_2)}P(\mech(\data_2)=x).$$ 
\end{definition}
Note: the Metric DP framework does not specify how to choose a metric $d_X$. Thus, users of this framework must first identify a  metric that is suitable for their application and then design efficient mechanisms.

If $d_H$ is the Hamming distance (number of records on which two datasets differ) and $\epsilon>0$, then using the metric $\epsilon d_H$ recovers the original definition of pure $\epsilon$-DP\cite{dwork2006calibrating}. 
The intuitive meaning behind this definition is that if two datasets $\data_1$ and $\data_2$ are close under a metric $d_X$,
and one of them was the input to $\mech$, then an attacker who has the output of $\mech$ will have difficulty determining
whether the input was $\data_1$ or $\data_2$.

The current state of the art in specifying what it means for an attacker to have difficulty in determining which dataset was the input
is known as Gaussian DP and $f$-DP. It measures difficulty in terms of the best possible hypothesis test that uses the output of $\mech$.
It is defined as follows (here $\Phi$ denote the cumulative distribution function of the standard normal distribution): 

\begin{definition}[$f$-DP and $\mu$-Gaussian DP \cite{gdp}]
Let $f$ be a continuous, concave, non-decreasing function\footnote{Note, this formulation is equivalent to that of Dong et al. \cite{gdp} using trade-off function $1-f$. The formulation presented here is easier to work with in our opinion.} such that $f(x)\geq x$. A mechanism $\mech$ satisfies $f$-DP if for all pairs of datasets $\data_1,\data_2$ that differ on the value of one record and for all $S\subseteq\range(\mech)$,
$$P(\mech(\data_1)\in S) \leq x \Rightarrow P(\mech(\data_2)\in S) \leq f(x).$$
In particular,  when $f(x)\equiv \Phi(\Phi^{-1}(x)+\mu)$ for some  number $\mu\geq 0$, then  $\mech$ satisfies $\mu$-Gaussian DP or $\mu$-GDP for short. Note that the privacy requirement on such an $\mech$ is equivalent to the condition:
$$\Phi^{-1}(P(\mech(\data_2)\in S)) \leq \Phi^{-1}(P(\mech(\data_1)\in S))+\mu.$$
\end{definition}

Intuitively, $f$-DP means that if an attacker is trying to distinguish between $\data_1$ and $\data_2$ based on the output of $\mech$, then no matter what method the attacker uses, if its false positive rate is $x$ then the true positive rate is $\leq f(x)$. Similarly, if the false negative rate is $y$ then the true negative rate is $\leq f(y)$. Thus the function $f$ is a trade-off between true and false positive rates. Note that when $f(x)=x$ then the attacker does no better than random guessing, which represents perfect confidentiality. 
From a statistics perspective, an attacker is using the output of $\mech$ to perform a hypothesis test of whether the input to $\mech$ was $\data_1$ (the \emph{null hypothesis}) or $\data_2$ (the \emph{alternative hypothesis}). GDP guarantees that  if the significance level (probability of incorrectly rejecting the null hypothesis) of the test is at most $\alpha$, then the power of the test (probability of correctly  rejecting the null hypothesis) is at most $f(\alpha)$ \cite{gdp}.

$\mu$-GDP also has the following interpretation: the ability of an attacker to distinguish between $\data_1$ and $\data_2$ is at least as difficult as distinguishing whether a random variable $y$ was sampled from a $N(0,1)$ distribution or a $N(\mu,1)$ distribution \cite{gdp}.

Note: Gaussian DP is not defined in terms of a metric over datasets. However, as we show in Section \ref{sec:model}, we can combine the two ideas to get the best of both worlds: defining plausible deniability requirements using a metric, and specifying how hard it is to distinguish between two plausible values using GDP. The main challenge is in choosing suitable metrics and designing mechanisms.

Gaussian DP for person-level data has several desirable properties that simplify the design of mechanisms. The first is \emph{postprocessing invariance} that says if a mechanism $\mech$ satisfies $\mu$-GDP and $\randalg$ is a (randomized) algorithm whose domain contains the range of $\mech$, then the combined function $\randalg \circ \mech$ (which first runs $\mech$ on the input dataset and then runs $\randalg$ on the output of $\mech$) also satisfies $\mu$-GDP. 
The second important property is \emph{fully adaptive composition} \cite{stfully}. Let $\mech_1,\dots,\mech_k$ be a sequence of mechanisms. They can be adaptively chosen, so that each $\mech_{j}$ and its privacy parameter $\mu_j$ is chosen after seeing the outputs of $\mech_1,\dots,\mech_{j-1}$. If $\sqrt{\sum_{i=1}^k\mu^2_j} \leq \mu$ in all possible situations, then releasing all of their outputs satisfies $\mu$-GDP.
Thus a larger mechanism $\mech$ can be constructed out of smaller mechanisms $\mech_1,\dots,\mech_k$. The parameter $\mu$ is known as the overall \emph{privacy loss budget} and the adaptive composition result allows the privacy budget to be allocated among its components $\mech_1,\dots, \mech_k$. This type of privacy accounting is a desirable feature of formal privacy definitions. 

Another important feature of Gaussian DP is group privacy, which is the guarantee that is achieved for pairs of datasets $\data_1$ and $\data_2$ that differ on the addition/removal of multiple records.
\begin{lemma}[Group Privacy \cite{gdp}]
    Let $\mech$ be a mechanism satisfying $f$-DP. Let $\data_1$ and $\data_2$ be datasets that differ on the values of $k$ people. Let $f^{(k)}$ denote application of $f$ for $k$ times  (e.g., $f^{(3)}(x)=f(f(f(x)))$). Then, for any set $S$, $P(\mech(\data_2)\in S)\leq f^{(k)}(P(\mech(\data_1)\in S))$. In particular, if $\mech$ satisfies $\mu$-GDP, then $\Phi^{-1}(P(\mech(\data_2)\in S)) \leq \Phi^{-1}(P(\mech(\data_1)\in S)) + k\mu$.
\end{lemma}


\section{Related Work}\label{sec:related}

Confidentiality protection for business data has received much less research attention than confidentiality for surveys of individuals. Popular classical techniques are  cell suppression \cite{cox1980,cox1981}, EZS noise infusion \cite{ezs}  and synthetic data \cite{pistner2018synthetic, kinney2011towards}. Cell suppression is the only one with a formal attack model -- an adversary without external information who is restricted to linear operations and should not be able to guess true values to within $p\%$ (where $p$ is a privacy parameter)\cite{cox1980,cox1981}. EZS multiplicative noise \cite{ezs} is another relative error protection scheme. It was evaluated by Yang et al. \cite{yangezs}, who found that it underprotects small cells and adds significant distortions to large aggregates.


One of the earliest formal approaches to business data confidentiality was proposed by Haney et al. \cite{haney}. They modified pure- and approximate-DP to provide relative error protections for confidential aggregates using multiplicative Laplace noise and also additive noise via the smooth sensitivity framework \cite{nissim2007smooth}.  Seeman et al. \cite{splittingmech} generalized this idea with zero-Concentrated DP (zCDP) \cite{zcdp} and introduced the \emph{splitting mechanism} that repeatedly uses any arbitrary underlying zCDP mechanisms to protect data. 

The Personalized DP Framework \cite{jorgensen2015conservative} allows every entity to have its own privacy parameter but does not prescribe how to choose it. Seeman et al. \cite{splittingmech} and Finley et al. \cite{finley2024slowly} follow this framework and propose methods to protect against membership inference by setting the personalized privacy parameters for zero-concentrated DP \cite{zcdp} so that larger businesses get larger (worse) privacy parameters. 
The work of \cite{finley2024slowly} is closest to ours, but with the following distinctions. They are concerned with
membership inference, while our security model is different and seeks to protect attribute inference (absolute error of attribute inference increases with its size, but relative error decreases). They propose a class of mechanisms called \emph{transformation mechanisms}. The $\neighfun$ mechanisms we propose in  Section \ref{sec:mech} are a subset of the transformation mechanisms, but have extra restrictions that prevent paradoxical situation where information about small establishments receives almost no protections, while information about large establishments gets heavily distorted (see Example \ref{ex:morerelative}). Furthermore, we also introduce  the \pncmech, which is novel (Section \ref{sec:mech}).

\section{Confidentiality for Establishments}\label{sec:model}
Recent work on formal privacy for establishments \cite{finley2024slowly,splittingmech} focuses on membership
inference (how well the existence of an establishment is protected). However, we note that in many 
applications,  like the QCEW, the \emph{existence} of establishments is intentionally not protected at all.
Furthermore, establishments  advertise their existence (they need customers to exist) and have public observable physical locations.
In terms of utility, strong membership inference protections are also undesirable -- making the contributions of Disneyland Resort indistinguishable from that of an ice-cream shop would severely distort statistics used to measure regional economies. At the other extreme, masking the effect of just one \textbf{person} does not properly address the confidentiality requirements of establishments (protecting aggregate data like total employment and wages) \cite{haney}.
Hence, we consider a different security model.

\subsection{A security model for establishments}\label{sec:securitymodel}
Instead of membership inference, we focus on protections against attribute inference.
For business data, attackers are often interested in making precise inferences about establishments in order to gain a business advantage. In our security model, an attacker starts with relatively easy-to-collect information about establishments. For example, some attributes are generally publicly known, including the existence of an establishment, its location, and its industry sector. Small establishments can be visited or observed by an attacker to estimate their size (e.g., an attacker may observe that the number of employees is approximately $\leq 10$). The size estimates made by an attacker on a small business are likely to have low absolute error but large relative error. Larger businesses often reveal approximate information about themselves on their own web pages or other public reports. For example, Disneyland Resort shared in 2025 that they employ $\approx 36,000$ employees in the Orange County, CA \cite{DisneylandResorts_2025}. This information for larger establishments tends to have lower relative error but higher absolute error than information observable about smaller establishments. Assuming Disneyland rounded the true employment to the nearest thousand, an attacker may make an estimate in the range [35,500,36,500], which would have at most an absolute error of 1,000 and at most a relative error of 3\%. Other information gathering activities by attackers can include surveys:
\begin{example}\label{example:sampling}
    Suppose an establishment $\estab$ has $x$ employees and a statistician (e.g., an attacker or labor economist) wants to estimate its employment count. 
    Suppose nearly all employees of  $\estab$ live within a region whose total population, $N$, is known with high accuracy (e.g., from Census data). For some $q\in[0, 1]$, the statistician may take a random sample (with replacement) of $n=Nq$ people in the region and record the proportion $p$ that work for $\estab$. The quantity $np$ can be modeled as the sum of $n$ Bernoulli$(x/N)$ random variables. So, an unbiased estimate of the number of employees of $\estab$ is $\hat{x}=np(N/n)$. The variance of this estimate is $\Var(\hat{x})=n(x/N)(1-x/N)N^2/n^2=x\frac{N-x}{n}\leq \frac{1}{q}x$.  If the sample is without replacement, then $np$ has a Hypergeometric$(N, x, n)$ distribution. Using the unbiased estimate $\hat{x}=np(N/n)$, we get  
    $Var(\hat{x})=x\frac{N-x}{Nq}\frac{N-Nq}{N-1}\leq x\frac{N(1-q)}{q(N-1)}$. In both cases, the standard deviation is  $O(\sqrt{x})$ and so confidence intervals would have lengths $O(\sqrt{x})$. Thus, the attacker's estimate has absolute error $O(\sqrt{x})$ and relative error $O(1/\sqrt{x})$. Again we see the trend of \textbf{low absolute error and high relative error for small establishments, and high absolute error but low relative error for large establishments.}
\end{example}
In our security model, an attacker uses this information to create ballpark estimates of an establishment's size. The goal of the data curator is to release approximate query answers that prevent an attacker from significantly sharpening their inference about establishments. That is, in order to get more precise inferences, the attacker must turn to more expensive data gathering activities rather than attacking the published data.

In our framework, a data curator specifies a neighbor function $\neighfun$ which defines what ``ballpark estimate'' means. The neighbor function can be selected by the data curator based on how accurate public information appears to be (in this paper, we argue that $\neighfun=\sqroot$ is a sensible choice). Afterwards, the privacy parameter $\mu$ determines how hard it is to sharpen inference within this ballpark estimate.

%


\subsection{The Neighbor Function}
Metric DP \cite{metricdp} uses a distance metric to help specify which plausible values should be nearly indistinguishable from one another. However, it does not specify how to create application-dependent metrics. Our approach, $\mu$-\gedp, creates the metric indirectly. First, we introduce a \emph{neighbor function} that can be used to specify intended uncertainty intervals whose absolute size increases with an establishment's size, but whose relative size decreases. Then we define the metric in terms of the neighbor function. This process allows us to use techniques from Gaussian DP \cite{gdp} to measure the difficulty of distinguishing between plausible values in the same uncertainty interval instead of the default $\epsilon$-DP \cite{dwork2006calibrating} used by Metric DP or $\rho$-zCDP \cite{zcdp} used in prior work on business data \cite{splittingmech,finley2024slowly}. As we explain in Section \ref{sec:plausiblegdp}, this choice results in more accurate and more interpretable quantification of how strong the plausible deniability guarantees are.


\begin{definition}[Neighbor function, $\neighfun$-close]\label{def:close}
    A real-valued function with nonnegative domain $\neighfun:\real_{\geq 0}\rightarrow \real$ is called a \emph{neighbor function} if it is (1) strictly increasing, (2) continuous, 
    (3) concave, and (4) the function  $g(x)=\neighfun(\exp(x))$ is convex.
    Let $\distparam_1,\dots, \distparam_\numc$ be nonnegative real numbers, which we call \emph{distance parameters}.
    An establishment record $(\pattv_1^{(1)},\dots, \pattv_\nump^{(1)}, \cattv_1^{(1)},\dots, \cattv_{\numc}^{(1)})$ is $\neighfun$-close to an establishment record $(\pattv_1^{(2)},\dots, \pattv_\nump^{(2)}, \cattv_1^{(2)},\dots, \cattv_{\numc}^{(2)})$ with respect to those distance parameters when:
    \begin{itemize}[leftmargin=*]
       \item $\pattv_i^{(1)} = \pattv_i^{(2)}$  for $i=1,\dots, \nump$ (i.e., the public attributes match), and 
      \item  $|\neighfun(\cattv_j^{(1)}) - \neighfun(\cattv_j^{(2)})|\leq \distparam_j$   for $j=1,\dots, \numc$ (i.e., the confidential attributes are close enough in a transformed space defined by $\neighfun$) 
    \end{itemize}
\end{definition}

For example, two potential values $x$ and $y$ for the employment count at establishment $\estab_1$ are considered close  with respect to $\neighfun$ and a distance parameter $\distparam$ if $|\neighfun(x)-\neighfun(y)|\leq \distparam$. This is the same as requiring $\neighfun(y)\in [\max(\neighfun(0), \neighfun(x)-\distparam), ~~\neighfun(x)+\distparam]$. Two candidate records for the same establishment are considered close by Definition \ref{def:close} if all of their corresponding confidential attributes are close. Hence, our metric is the $L_\infty$ metric after the transformation $\neighfun$ has been applied.

\begin{definition}[Uncertainty Interval $\uncertI_{\neighfun,\distparam}$]\label{def:uncert}
    The \emph{uncertainty interval} around $y$ induced by $\neighfun$ and $\distparam$ is defined as $\uncertI_{\neighfun,\distparam}(y)= [\lowerbound, ~\upperbound]$ where $\lowerbound=\neighfun^{-1}\left(\max(\neighfun(0), \neighfun(x)-\distparam)\right)$ and $\upperbound=\neighfun^{-1}\left(\neighfun(x)+\distparam\right)$. Denote the length of the uncertainty interval as $|\uncertI_{\neighfun,\distparam}|=\upperbound-\lowerbound$.
\end{definition}
If $|\neighfun(x)-\neighfun(y)|\leq \distparam$ then $x\in \uncertI_{\neighfun,\distparam}(y)$ and $y\in \uncertI_{\neighfun,\distparam}(x)$ (i.e., $x$ and $y$ are inside each other's intervals). So, if two records $\record_1$ and $\record_2$ are $\neighfun$-close then they are in each other's uncertainty intervals for all confidential attributes. We discuss specific examples of $\neighfun$ in Section \ref{sec:neighfun}.
The conditions on $\neighfun$ in Definition \ref{def:close} ensure that uncertainty intervals behave as intended (increasing absolute error but decreasing relative error as the attribute value increases):

\begin{theoremEnd}[category=model,proof end]{theorem}\label{thm:neighfun}
Let $\neighfun$ be a function satisfying the conditions of Definition \ref{def:close}. Then for any $x,y,t$ such that $0\leq x<y$ and $t\geq 0$
\begin{enumerate}[leftmargin=*]
\item\label{thm:neighfun:item1} $\neighfun^{-1}$ exists and is increasing.
\item\label{thm:neighfun:item2} We have $|\neighfun(y+t)-\neighfun(x+t)|\leq |\neighfun(y)-\neighfun(x)|$.
\item\label{thm:neighfun:item4}  The length of the uncertainty interval around $y$ is at least as big as the interval around $x$: $|\uncertI_{\neighfun,t}(y)|\geq |\uncertI_{\neighfun,t}(x)|$.
\item\label{thm:neighfun:item5} The length of the uncertainty interval around $x$ relative to $x$ is at least as big as the length of the uncertainty interval around $y$ relative to $y$, $\frac{|\uncertI_{\neighfun,t}(y)|}{y} \leq \frac{|\uncertI_{\neighfun,t}(x)|}{x}$
\end{enumerate}
\end{theoremEnd}
\begin{proofEnd}
    To prove Item \ref{thm:neighfun:item1}, we note that since $\neighfun$ is increasing, the inverse must exist and clearly it is increasing as well (since $x<y \Leftrightarrow \neighfun(x) < \neighfun(y)$).

    To prove Item \ref{thm:neighfun:item2}, we take advantage of the concavity of $\neighfun$.  Given numbers $a<b<c$, we note that $b=a\left(\frac{c-b}{c-a}\right)+ c\left(\frac{b-a}{c-a}\right) = a\left(\frac{c-b}{c-a}\right) + c\left(1-\frac{c-b}{c-a}\right)$ 
    \begin{align*}
      \lefteqn{  \frac{\neighfun(c)-\neighfun(b)}{c-b} - \frac{\neighfun(b)-\neighfun(a)}{b-a} }\\
      &= \frac{\neighfun(c)-\neighfun\left(a\left(\frac{c-b}{c-a}\right) + c\left(1-\frac{c-b}{c-a}\right)\right)}{c-b} - \frac{\neighfun\left(a\left(\frac{c-b}{c-a}\right) + c\left(1-\frac{c-b}{c-a}\right)\right)-\neighfun(a)}{b-a} \\
      &\leq 
      \frac{\neighfun(c)-\neighfun(a)\left(\frac{c-b}{c-a}\right) - \neighfun(c)\left(1-\frac{c-b}{c-a}\right)}{c-b} 
      - \frac{\neighfun(a)\left(\frac{c-b}{c-a}\right) + \neighfun(c)\left(1-\frac{c-b}{c-a}\right)-\neighfun(a)}{b-a} \\
 &\text{ by concavity of $\neighfun$ }\\
      &= 
      \frac{(\neighfun(c)-\neighfun(a))\left(\frac{c-b}{c-a}\right) }{c-b} 
      - \frac{(\neighfun(c)-\neighfun(a))\left(1-\frac{c-b}{c-a}\right)}{b-a} \\
      &= 
      (\neighfun(c)-\neighfun(a))\left(\frac{1}{c-a}\right)  
      - \frac{(\neighfun(c)-\neighfun(a))\left(\frac{b-a}{c-a}\right)}{b-a} \\
       &= 
      (\neighfun(c)-\neighfun(a))\left(\frac{1}{c-a}\right)  
      - (\neighfun(c)-\neighfun(a))\left(\frac{1}{c-a}\right)=0 \\
    \end{align*}
    and thus $\frac{\neighfun(c)-\neighfun(b)}{c-b} \leq  \frac{\neighfun(b)-\neighfun(a)}{b-a}$ when $c > b > a$. Noting that item \ref{thm:neighfun:item2} is trivial when $t=0$, we only need to consider $t>0$. The first case we consider is when $y=x+t$, so that the intervals $[x,y]$ and $[x+t, y+t]$ are adjacent. In this case, we directly apply the derived relation to get:
    \begin{align*}
        0\leq \frac{\neighfun(y+t)-\neighfun(x+t)}{y-x}  \leq \frac{\neighfun(x+t)-\neighfun(x)}{x+t-x}=\frac{\neighfun(y)-\neighfun(x)}{y-x}
    \end{align*}
     from which we get $|\neighfun(y+t)-\neighfun(x+t)|\leq |\neighfun(y)-\neighfun(x)|$.
    The second case we consider is when $[x+t, y+t]$ and $[x, y]$ are disjoint (so that $x<y < x+t < y+t$). Applying the derived relation  twice, we get:
    \begin{align*}
        0\leq \frac{\neighfun(y+t)-\neighfun(x+t)}{y-x} \leq \frac{\neighfun(x+t)-\neighfun(y)}{x+t-y} \leq \frac{\neighfun(y)-\neighfun(x)}{y-x}
    \end{align*}
    from which we get $|\neighfun(y+t)-\neighfun(x+t)|\leq |\neighfun(y)-\neighfun(x)|$. The remaining case to consider is when $[x+t, y+t]$ and $[x, y]$ overlap at more than just an endpoint (so that $x< x+t < y < y+t$). In this case,
    \begin{align*}
        \lefteqn{\Big(\neighfun(y+t)-\neighfun(x+t)\Big) - \Big(\neighfun(y)-\neighfun(x)\Big)}\\
        &=\Big(\neighfun(y+t)-\neighfun(y) + \neighfun(y)-\neighfun(x+t)\Big) - \Big(\neighfun(y)-\neighfun(x+t)+\neighfun(x+t)-\neighfun(x)\Big)\\
        &=\Big(\neighfun(y+t)-\neighfun(y) \Big) - \Big(\neighfun(x+t)-\neighfun(x)\Big)
    \end{align*}
    apply the derived relation twice (as we did in the case of the disjoint intervals), we get
    $$0\leq \frac{\neighfun(y+t)-\neighfun(y) }{t} \leq \frac{\neighfun(x+t)-\neighfun(x)}{t}$$
    and this allows us to conclude that $|\neighfun(y+t)-\neighfun(x+t)|\leq |\neighfun(y)-\neighfun(x)|$.

    To prove Item \ref{thm:neighfun:item4}, we see that the case $t=0$ is trivial and so we can assume $t>0$.  Next, we note that since $\neighfun$ is increasing and concave, then $\neighfun^{-1}$ is increasing and convex. To see why, concavity implies $\neighfun(ax+(1-a)y)\geq a\neighfun(x) + (1-a)\neighfun(y)$ and applying $\neighfun^{-1}$ to both sides, we get $ax + (1-a)y \geq \neighfun^{-1}(a\neighfun(x) + (1-a)\neighfun(y))$. Replacing $\neighfun(x)$ with $x^\prime$ and $\neighfun(y)$ with $y^\prime$, we have $a \neighfun^{-1}(x^\prime) + (1-a)\neighfun^{-1}(y^\prime) \geq \neighfun^{-1}(ax^\prime + (1-a)y^\prime)$.

    Next, following the proof of Item \ref{thm:neighfun:item2}, but replacing concavity with convexity, it follows that given  numbers $a,b,c,d$ with $a$ being the smallest and $d$ being the largest, we have $\frac{\neighfun^{-1}(d)-\neighfun^{-1}(c)}{d-c} \geq  \frac{\neighfun^{-1}(b)-\neighfun^{-1}(a)}{b-a}$ (i.e., convexity not only implies that derivatives are non-decreasing, but same with secants). We consider several cases based on the relationship between $\neighfun(x)-t$, $\neighfun(0)$, and $\neighfun(y)-t$. The case is trivial if $\neighfun(x)-t< \neighfun(y)-t\leq \neighfun(0)$. Next we consider the case where $\neighfun(0)\leq \neighfun(x)-t\leq \neighfun(y)-t$. Substituting $d=\neighfun(y)+t$ and $c=\neighfun(y)-t$ and $b=\neighfun(x)+t$ and $a=\neighfun(x)-t$, then noting that $d-c=b-a=2t>0$, we get $\neighfun^{-1}(\neighfun(y)+t)-\neighfun^{-1}(\neighfun(y)-t) \geq \neighfun^{-1}(\neighfun(x)+t)-\neighfun^{-1}(\neighfun(x)-t)\geq 0$. Finally, if $\neighfun(x)-t\leq \neighfun(0)\leq \neighfun(y)-t$, we substitute $a=\neighfun(0)$ and $c=\neighfun(y)-t$ and get $\frac{\neighfun^{-1}(\neighfun(y)+t)-\neighfun^{-1}(\neighfun(y)-t)}{2t} \geq \frac{\neighfun^{-1}(\neighfun(x)+t)}{\neighfun(x)-\neighfun(0)+t}$. Since $\neighfun(x)-\neighfun(0)+t\leq \neighfun(x)-(\neighfun(x)-t)+t=2t$, we get $\neighfun^{-1}(\neighfun(y)+t)-\neighfun^{-1}(\neighfun(y)-t) \geq \neighfun^{-1}(\neighfun(x)+t)-\neighfun^{-1}(\neighfun(x)-t)\geq 0$. 

   To prove Item \ref{thm:neighfun:item5}, we note that if $g(a)=\neighfun(\exp(a))$ is convex then $h(b)=g^{-1}(b)=\log(\neighfun^{-1}(b))$ is concave, and both $g$ and $h$ are increasing functions. Thus we know that for numbers $a,b,c,d$ with $a$ being the smallest and $d$ being the largest, $\frac{h(d)-h(c)}{d-c} \leq  \frac{h(b)-h(a)}{b-a}$. This means that
   \begin{align*}
      \frac{1}{d-c} \log\frac{\neighfun^{-1}(d)}{\neighfun^{-1}(c)} \leq \frac{1}{b-a}\log\frac{\neighfun^{-1}(b)}{\neighfun^{-1}(a)}
   \end{align*}
   Substituting
   $d=\neighfun(y)+t$ and $c=\neighfun(y)$ and $b=\neighfun(x)+t$ and $a=\neighfun(x)$, then noting that $d-c=b-a=t>0$, we get 
   \begin{align}\label{eq:neighfun:item5:step1}
       \frac{\neighfun^{-1}(\neighfun(y)+t)}{y} \leq \frac{\neighfun^{-1}(\neighfun(x)+t)}{x}
   \end{align}
   Thus the result holds when $\neighfun(x)-t<\neighfun(y)-t\leq \neighfun(0)$. When $\neighfun(0)< \neighfun(x)-t<\neighfun(y)-t$, we substitute $d=\neighfun(y)$ and $c=\neighfun(y)-t$ and $b=\neighfun(x)$ and $a=\neighfun(x)-t$, then noting that $d-c=b-a=t>0$, we get $\frac{y}{\neighfun^{-1}(\neighfun(y)-t)} \leq \frac{x}{\neighfun^{-1}(\neighfun(x)-t)}{x}$. Since $\neighfun(x)-t>\neighfun(0)$ implies $\neighfun^{-1}(\neighfun(y)-t)>\neighfun^{-1}(\neighfun(x)-t)>0$, we find
   \begin{align}\label{eq:neighfun:item5:step2}
       \frac{\neighfun^{-1}(\neighfun(y)-t)}{y} \geq \frac{\neighfun^{-1}(\neighfun(x)-t)}{x}
   \end{align}
   Thus, 
   \begin{align*}
       \lefteqn{\frac{\neighfun^{-1}(\neighfun(y)+t)}{y}-\frac{\neighfun^{-1}(\neighfun(y)-t)}{y}}\\
       &\leq \frac{\neighfun^{-1}(\neighfun(x)+t)}{x}-\frac{\neighfun^{-1}(\neighfun(y)-t)}{y}  & &\text{ by \eqref{eq:neighfun:item5:step1}}\\
        &\leq \frac{\neighfun^{-1}(\neighfun(x)+t)}{x}-\frac{\neighfun^{-1}(\neighfun(x)-t)}{x}  & &\text{ by \eqref{eq:neighfun:item5:step2}.}
   \end{align*}
   The result holds when $\neighfun(0)< \neighfun(x)-t<\neighfun(y)-t$. Finally, when $ \neighfun(x)-t\leq \neighfun(0)\leq \neighfun(y)-t$,  we know $\neighfun^{-1}(\neighfun(y)-t)\geq 0$. Thus, 
   \begin{align*}
       \lefteqn{\frac{\neighfun^{-1}(\neighfun(y)+t)}{y}-\frac{\neighfun^{-1}(\neighfun(y)-t)}{y}}\\
       &\leq \frac{\neighfun^{-1}(\neighfun(x)+t)}{x}-\frac{\neighfun^{-1}(\neighfun(y)-t)}{y}  & &\text{ by \eqref{eq:neighfun:item5:step1}}\\
        &\leq \frac{\neighfun^{-1}(\neighfun(x)+t)}{x} 
   \end{align*}
   Therefore $\frac{\neighfun^{-1}(\neighfun(y)+t)}{y}-\frac{\neighfun^{-1}(\max(\neighfun(0),\neighfun(y)-t)}{y}\leq \frac{\neighfun^{-1}(\neighfun(x)+t)}{y}-\frac{\neighfun^{-1}(\max(\neighfun(0),\neighfun(x)-t)}{x}$ for all cases of $0\leq x<y$ and $t\geq 0$.
\end{proofEnd}

\begin{textAtEnd}[category=model]
    It is important to note that a piecewise linear function $\neighfun_{pl}$ with 2 or more pieces \emph{cannot} be a neighbor function, even if it is concave and increasing. The reason is that $\neighfun_{pl}(\exp(x))$ is not convex and hence will not allow relative error for large establishments to decrease. 
\end{textAtEnd}

\begin{theoremEnd}[category=model,all end]{theorem}\label{thm:piecewise}
    Let $\neighfun_{pl}$ be a non-decreasing concave piece-wise linear function with 2 or more pieces. Then $\neighfun_{pl}$ is not a neighbor function.
\end{theoremEnd}
\begin{proofEnd}
    A concave function defined on an interval is continuous except possibly at the endpoints. Thus, it a concave piecewise linear function has at least 2 pieces, there exists a point $x_0$ where two pieces meet. Thus in a small neighborhood to the left of $x_0$, the function $\neighfun_{pl}(x)=a_1x + b_1$ for some $a_1,b_1$ and in a small neighborhood to the right of $x_0$, we have $\neighfun_{pl}(x)=a_2x + b_2$ for some $a_2,b_2$. Importantly, by continuity, $a_1x_0+b_1 = a_2x_0 + b_2$. Since $\neighfun_{pl}(x)=a_1x + b_1$ is concave, $a_1 > a_2$.

    Now, the function $\neighfun_{pl}(\exp(x))$ is equal to $e^{a_1x+b_1}$ in a small neighborhood to the left of $x_0$, and has derivative $a_1e^{a_1x+b_1}$. Meanwhile,  $\neighfun_{pl}(\exp(x))$ is equal to $e^{a_2x+b_2}$ in a small neighborhood to the right and has derivative $a_2e^{a_2x+b_2}$. Thus, left derivative at $x_0$ is $a_1/a_2>1$ times bigger than the right derivative at $x_0$. This means that as we approach and pass $x_0$, the derivative of $\neighfun_{pl}(\exp(x))$ decreases and hence it cannot be convex.

    Thus $\neighfun_{pl}$ is not a neighbor function.
\end{proofEnd}

\begin{remark}\label{remark:multifun}
Definition \ref{def:close} uses one neighbor function $\neighfun$ and then a separate distance parameter $\distparam_i$ for each confidential attribute $\catt_i$. In the most general case, one can have a separate $\neighfun_i$ and $\distparam_i$ for each $\catt_i$. We chose the simpler version to increase readability. However, the more general version is needed for a few results in Section \ref{sec:interaction} about interactions between privacy definitions.
\end{remark}

Using $\neighfun$-closeness, we define neighboring datasets as follows: 
\begin{definition}[$\neighfun$-neighbors]\label{def:neigh}
        Let $\neighfun$ be a function satisfying the requirements of Definition \ref{def:close}. 
    Let $\distparam_1,\dots, \distparam_\numc$ be nonnegative real numbers.
    Two datasets $\data_1$ and $\data_2$ are $\neighfun$-neighbors if $\data_1$ can be obtained from $\data_2$ by replacing one record $\record$ with another record $\record^\prime$ that is $\neighfun$-close to $\record$ (with distance parameters $\distparam_1,\dots, \distparam_\numc$).
\end{definition}

\subsubsection{Extensions when Confidential Attributes can be Negative.}\label{sec:nonon}
The nonnegativity constraint on confidential attributes is used to simplify the discussion in this paper. However, this assumption can be eliminated in several ways, either by modifying the neighbor function or transforming the data:
\begin{itemize}[leftmargin=*]
    \item Given a neighbor function $\neighfun$ defined on nonnegative numbers, one can define an extension $\neighfun^\prime$ over the real numbers as $\neighfun^\prime(x)=\sign(x)\neighfun(|x|)$. This provides the semantics that values near $0$ get uncertainty intervals with high relative errors and values far from $0$ get uncertainty intervals with high absolute errors.
    \item Sometimes an attribute can  be naturally expressed as the difference of two nonnegative attributes; e.g., \emph{profit} can be replaced by  \emph{revenue} and \emph{cost}, since profit is the difference of the two. 
    \item Otherwise, a confidential attribute $\catt$ that is sometimes negative can be replaced with $\catt^\prime=\max(0, \catt)$ and $\catt^{''}=|\min(0,\catt)|$. If $\catt$ is profit, $\catt^\prime$ can be interpreted as the gain and $\catt^{\prime\prime}$ can be interpreted as the loss. Both $\catt'$ and $\catt^{\prime\prime}$ are nonnegative and $\catt=\catt^\prime-\catt^{\prime\prime}$.
\end{itemize}

\subsection{$\mu$-Gaussian Establishment DP}\label{sec:plausiblegdp}

Our approach, $\mu$-\gedp, adapts Gaussian DP \cite{gdp} to the neighbor function $\neighfun$ in order to measure how difficult it should be to distinguish between records that are $\neighfun$-close. 
\begin{definition}[$\mu$-\shortdp]\label{def:gedp}
    Let $\neighfun: \real_{\geq 0}\rightarrow\real$ be a function satisfying the requirements of Definition \ref{def:close} and 
    let $\distparam_1,\dots, \distparam_\numc$ be nonnegative real numbers. A mechanism $\mech$ satisfies $\mu$-\gedp ($\mu$-\shortdp) if for all pairs of $\neighfun$-neighbors  $\data_1,\data_2$ (with distance parameters  $\distparam_1,\dots, \distparam_\numc$) and all sets $S$, the following is satisfied: $\Phi^{-1}(P(\mech(\data_2)\in S)) \leq \Phi^{-1}(P(\mech(\data_1)\in S))+\mu.$
\end{definition}

\begin{theoremEnd}[category=model,proof end]{theorem}[Adaptive Composition]\label{thm:gedpcomp}
Let $\mech_1,\dots, \mech_K$ be a sequence of mechanisms such that $\mech_{i+1}$ can access the output of $\mech_1,\dots, \mech_{i}$. Suppose each $\mech_i$ satisfies $\mu_i$-\shortdp with the same neighbor function $\neighfun$ and distance parameters $\distparam_1,\dots,\distparam_\numc$. Then the mechanism that releases all of their outputs satisfies $\sqrt{\sum_{i=1}^K \mu_i^2}$-\shortdp with the same neighbor function and distance parameters.
\end{theoremEnd}
\begin{proofEnd}
    This is proven in the same way as adaptive composition in Gaussian Differential Privacy \cite{gdp}.
\end{proofEnd}
Thus \shortdp and Gaussian DP share the same accounting method (privacy parameter and the form of the probabilistic inequality). However, they differ in the choice of neighbors $\data_1,\data_2$. Gaussian DP is concerned with equal membership inference guarantees for all entities, which ruins the utility of establishment data (see  experiments in Section \ref{sec:experiment_synth}). Meanwhile, \shortdp protects against attribute inferences being more precise
than the uncertainty interval. 
We discuss specific choices of $\neighfun$ in Section \ref{sec:neighfun}, then analyze the interactions between different neighboring functions  in Section \ref{sec:interaction}.

The semantics of the protections come from the Gaussian DP privacy accounting method and are based on statistical hypothesis testing. Suppose an attacker is trying to
distinguish between two candidate records ($\record_1$ and $\record_2$) for an establishment, and both candidate records are within each other's uncertainty intervals. The null hypothesis is that $\record_1$ is true and the alternative hypothesis is that $\record_2$ is true. Based on the output $\outp$ of a mechanism satisfying $\mu$-\gedp, the attacker guesses either $\record_1$ or $\record_2$. The \emph{significance level} $\alpha$ of the hypothesis test is the probability the attacker incorrectly guesses $\record_2$ when the true record is $\record_1$. The \emph{power} is the probability the attacker correctly guesses $\record_2$ when the input is $\record_2$ and is bounded by  $\Phi(\Phi^{-1}(\alpha)+\mu)$. Given a desired significance level, more privacy means the power should be lower, so lower values of $\mu$ guarantee more privacy.

\begin{figure}[!htp]
    \centering
    \includegraphics[width=0.99\linewidth]{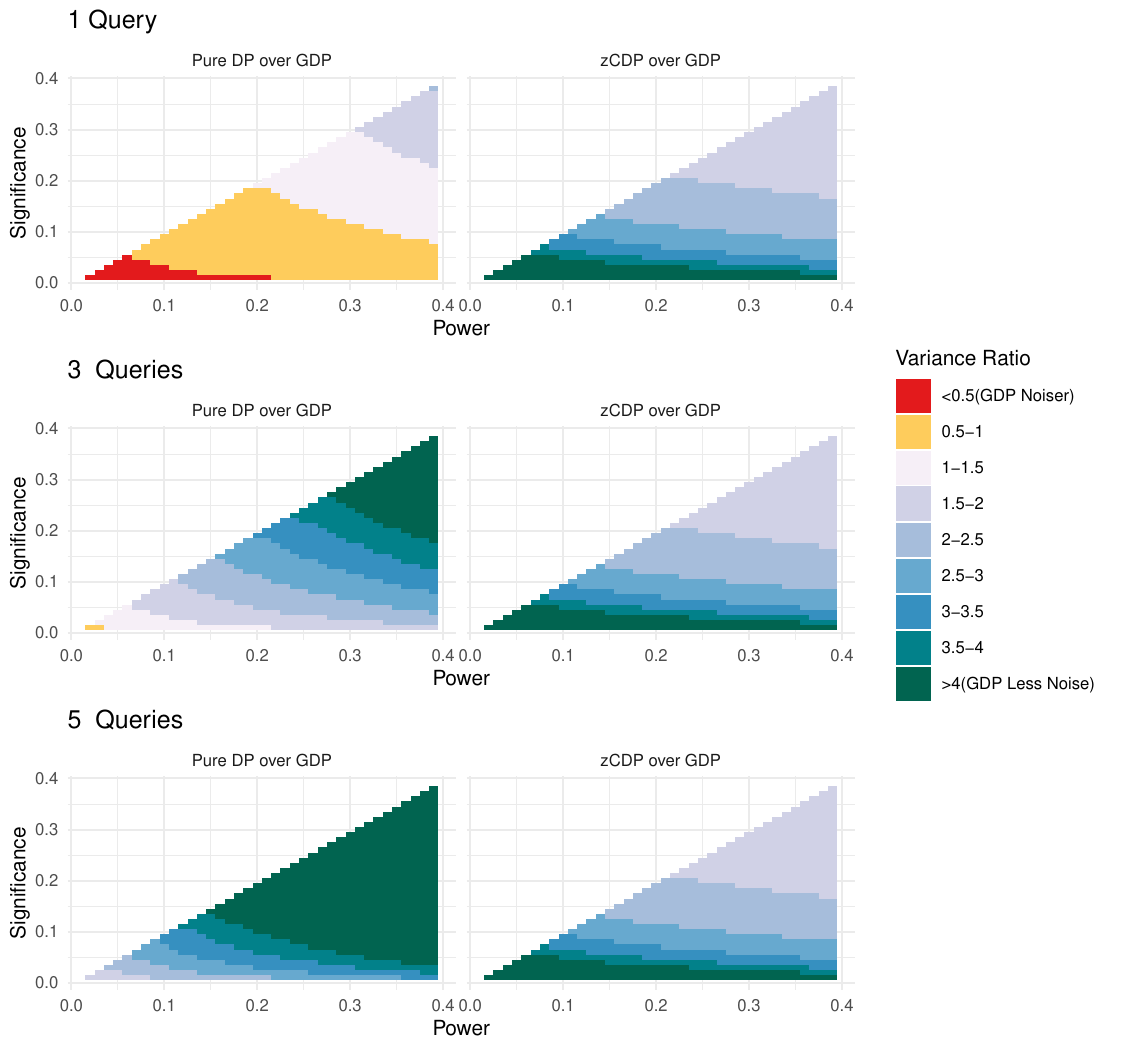}
    \caption{Query variance needed to achieve a given significance level (y-axis) and power (x-axis) when issuing 1 (top), 3 (middle), and 5 (bottom) queries. Left column: $\epsilon$-DP vs. Gaussian DP. Right column: $\rho$-zCDP vs. Gaussian DP. }
    \label{fig:compare_frameworks}
\end{figure}

Note that the Gaussian DP accounting method can be easily swapped out for
  $\epsilon$-DP \cite{dwork2006calibrating}, approximate-DP \cite{dwork2006our}, zCDP \cite{zcdp}, Renyi DP \cite{mironov2017renyi}, and $f$-DP \cite{gdp}. We chose Gaussian  DP because it has tighter privacy accounting than zCDP and Renyi DP for mechanisms that inject Gaussian noise \cite{gdp} and has tighter composition properties than $\epsilon$-DP, approximate-DP, and $f$-DP \cite{gdp,murtagh2015complexity}.
%
%
As an illustration, we pose the following question. Suppose an additive noise mechanism adds noise to $q$ counting queries. Under the 3 different privacy accounting schemes of $\epsilon$-DP (used by metric differential privacy \cite{metricdp}), $\rho$-zCDP (used by prior work on establishment data \cite{finley2024slowly,splittingmech}), and Gaussian DP, what per-query variance is needed to achieve a desired power for a given significance level? For counting queries over person-level data, Figure \ref{fig:compare_frameworks} plots the ratio of per-query variance of $\epsilon$-DP vs. Gaussian DP (left) and $\rho$-zCDP vs. Gaussian DP (right). (Calculation details can be found in Appendix \ref{app:sigcomp}). The Figure shows that Gaussian DP always requires less variance than $\rho$-zCDP to achieve the same semantics. If only 1 query is ever issued, the $\epsilon$-DP is preferable to Gaussian DP. However, if more than 1 query needs to be answered, Gaussian DP requires less per-query variance. 


\subsection{Semantics and Choices of  $\neighfun$}\label{sec:neighfun}
Suppose establishment $\estab$ has $x$  employees. 
Intuitively, $\mu$-\shortdp ensures that $x$ is nearly indistinguishable from any $y$ in the uncertainty interval $\uncertI_{\neighfun,\distparam}(x)$ around $x$ (Definition \ref{def:uncert}). Note that $x$ is also in the uncertainty interval around $y$. That is, $y\in \uncertI_{\neighfun,\distparam}(x) \Leftrightarrow x\in \uncertI_{\neighfun,\distparam}(y)$ which is mathematically equivalent to $|\neighfun(x)-\neighfun(y)|\leq \distparam$.
The level of indistinguishability between such $x$ and $y$ is controlled by the $\mu$ parameter. Suppose an  attacker wishes to determine whether the employment count of $\estab$ is $x$ or $y$. For any attack based on the output of a $\mu$-\shortdp mechanism $\mech$, if the false positive rate (over the randomness in $\mech$) is $\alpha$, then the true positive rate is $\leq \Phi(\mu+\Phi^{-1}(\alpha))$ and if the method's false negative rate is $\alpha$ then the true negative rate is $\leq\Phi(\mu+\Phi^{-1}(\alpha))$. Thus, distinguishing between any $x$ and $y$ in the same uncertainty interval is at least as hard as distinguishing between the Gaussians $N(0,1)$ and $N(\mu, 1)$ based on a data point sampled from one of those two Gaussians.
Note that this is not a separate guarantee for each attribute. It is a simultaneous guarantee; any two records whose attributes belong to the uncertainty intervals of each other have this indistinguishability guarantee. 

\begin{remark}\label{remark:groupp}
    The group privacy properties of Gaussian DP, which are inherited by \shortdp,  also provide indistinguishability guarantees for wider intervals. If $|\neighfun(x)-\neighfun(y)|\leq m\distparam$ then distinguishing between $x$ and $y$ is at least as hard as distinguishing between the Gaussians $N(0,1)$ and $N(m\mu, 1)$ based on a point  sampled from one of the Gaussians. Stated differently, if $\mech$ satisfies $\mu$-\shortdp with respect to $\neighfun$ and distance parameters $\distparam_1,\dots, \distparam_\numc,$ then $\mech$ satisfies $(m\mu)$-\shortdp with respect to $\neighfun$ and distance parameters $m\distparam_1,\dots, m\distparam_\numc$.

    This group privacy result provides a connection to the membership inference: an establishment's value $x$ is indistinguishable from 0 with privacy parameter $\lceil(\neighfun(x)-\neighfun(0))/\distparam\rceil$. It also provides a connection to metric differential privacy, using a discrete metric between records:  $d(\record_1,\record_2)=\max_{i=1,\dots, \numc} \left\lceil\frac{|\neighfun(\record_1[\catt_i])-\neighfun(\record_2[\catt_i])|}{\distparam_i}\right\rceil$ and the privacy parameter decreases with this distance.
\end{remark}
Our algorithms work with any $\neighfun$ compatible with Definition \ref{def:close}. However, we believe  there are two important potential choices for practical applications:\footnote{More generally, $\neighfun(x)=\log(x+a)$ or $\sqrt{x+a}$ for some constant $a\geq 0$.}   $\neighfun=\log$ and $\neighfun=\sqroot$. 
The choice of $\neighfun=\log$ ensures that  the length of the uncertainty interval  $\uncertI_{\neighfun,\distparam}(x)$ is proportional to $x$. This is consistent with historical tradition of attempting to limit attacker inference about establishments up to within a pre-specified relative error (e.g., $1\%$ or $10\%$). Notably, this is the motivation for the $p\%$ rule used in cell suppression \cite{cox1981} and the purpose of multiplicative EZS noise \cite{ezs}. The choice of $\neighfun=\sqroot$ is an alternative we advocate for instead.  It makes the length of an uncertainty interval around  $x$ equal to $O(\sqrt{x})$, which matches the error an attacker would get by alternative means, such as sampling as in Example \ref{example:sampling}.

We next argue that reasonable protections for establishment data should adhere to the following principles (for which $\sqroot$ is more suitable than $\log$). The reason is that they match the type of information an attacker can  obtain from alternate sources, as discussed in Section \ref{sec:securitymodel}.
\begin{description}
\item[\textbf{Principle 1:}] \emph{Relative} error protections should be \emph{decreasing}  with establishment size.
\item[\textbf{Principle 2:}] \emph{Absolute} error protections should be \emph{increasing}  with establishment size.
\end{description}

Why are these principles reasonable? As an example, consider $\neighfun=\log$ with distance parameter $\distparam=0.1$. The resulting uncertainty interval around any confidential $x$ is approximately $[0.9x,~ 1.10x]$ -- meaning all establishments, large and small, are to be protected with the same relative error of $10\%$ . For a small ice-cream shop with 3 employees, the statement that the number of employees is between $2.7$ and $3.3$ offers no protections at all (under-protection). What about a large establishment? In 2025, Disneyland Resort reported they had 36,000 employees \cite{DisneylandResorts_2025}. It is presumably rounded to the nearest thousand and so they may be comfortable with an uncertainty interval somewhere between $\pm 500$ (rounding to the nearest thousand) and $\pm 50$ (rounding to the nearest hundred). This represents a desired relative error/uncertainty somewhere between $0.3\%$ and $2.8\%$. Thus, protecting inference about the true employment for Disneyland to $10\%$ is unnecessary and loses utility. Lowering the distance parameter $\distparam$ will further erode protections for the smaller establishments, while raising it would unnecessarily lose utility for the larger establishments. Thus a fixed relative error seems to be unsuitable --- smaller establishments need higher relative error than larger ones. In contrast to $\neighfun=\log$, the choice of $\neighfun=\sqroot$ with distance parameter, say, $0.5$ naturally increases relative error for smaller establishments while decreasing it for the larger establishments (see Table \ref{tab:neighfun} for comparison of $\neighfun=\sqroot$ vs. $\neighfun=\log$).

\begin{table}[!htp]
    \centering
    \begin{tabular}{c|c|c|}\cline{2-3}
         & \multicolumn{2}{|c|}{Neighbor function}  \\
     \multicolumn{1}{c|}{$x$}  &  \multicolumn{1}{c}{$\sqroot$} & \multicolumn{1}{c|}{$\log$}\\\hline  
     3 &              [1.5 -- 5.0]    &   [2.7 -- 3.3]\\
     36 &                [30.2 -- 42.2]    &   [32.6 -- 39.8]\\
     360 &            [341.3 -- 379.2]   &  [325.7 -- 397.9]\\
     36,000 &         [35,810.5 -- 36,190.0]  & [32,574.1 -- 39.786.2]\\\cline{2-3}
    \end{tabular}
    \caption{Uncertainty interval $\uncertI_{\neighfun,\distparam}(x)$ for $\neighfun(x)=\sqrt{x}$ (with $\distparam=0.5$) and $\neighfun(x)=\log(x)$ (with $\distparam=0.1$).}
    \label{tab:neighfun}
\end{table}

\subsection{Deeper Semantics Analysis}\label{sec:interaction}
We next consider the following more advanced confidentiality semantics questions. 
\begin{enumerate*}[label={$\langle$\arabic*$\rangle$}]
    \item What protections are received by individuals (units smaller than an establishment) and firms (groups of establishments with the same owners)?
    \item Can (person-based) differential privacy protections be expressed in the $\mu$-\shortdp framework?
    \item How can one combine the protections and semantics of two different neighbor functions?
    \item What happens when different data publishers use different neighbor functions when publishing statistics about the same establishments (heterogeneous composition)?
\end{enumerate*}

\subsubsection{Confidentiality for Firms}\label{subsub:firms}
A firm is a business made up of one or more establishments, and its confidentiality is equivalent to the group privacy semantics of \shortdp. Consider a firm that consists of $\ell$ establishments. We can model the ability of an attacker to detect $\ell$ total changes to the data from those establishments as follows. Given $\neighfun$ and distance parameters $\distparam_1,\dots, \distparam_\numc$, suppose $\data_1,\data_2,\dots, \data_\ell$ are a sequence of pairwise-neighboring datasets; i.e., for any $i$, $\data_{i+i}$ is obtainable from $\data_i$ by replacing 1 record with a $\neighfun$-close record (Definition \ref{def:close}). Let $\mech$ be a mechanism that satisfies $\mu$-\shortdp with respect to $\neighfun, \distparam_1,\dots, \distparam_\numc$.  Then for any set $S$, the group privacy properties from Remark \ref{remark:groupp} imply that $\Phi^{-1}(P(\mech(\data_1)\in S)) \leq (\ell-1)\mu+\Phi^{-1}(P(\mech(\data_\ell)\in S))$. Equivalently, distinguishing between $\data_1$ and $\data_\ell$ based on the output of $\mech$ is equivalent to distinguishing between $N(0, 1)$ and $N((\ell-1)\mu, 1)$ based on a sampled point from one of those two distributions.

\subsubsection{Confidentiality for People}\label{subsub:people}
When considering the direct and indirect (group privacy) properties of $\mu$-\shortdp, 
protection for individuals is at least as strong as for establishments having one employee. Employees in larger establishments even benefit from blending in a crowd -- the protections are obtained by taking the uncertainty intervals around the \emph{establishment} attributes, and re-centering them around the \emph{person's} attributes. For example, let $\neighfun=\sqroot$ and $\distparam=100$ for the wages attributes. Suppose person $X$ earns \$20,000 in a quarter and is known to work at establishment $\estab$. 
\begin{itemize}[leftmargin=*]
\item If the total quarterly wages for $\estab$ is $\$20,000$ (i.e., $X$ is the only worker),  the uncertainty interval for wages is $\approx [\$1,715.7 - \$58,284.3]$ and so distinguishing between X's true salary vs. any other salary in that range  is at least as hard as distinguishing between $N(0,1)$ vs. $N(\mu, 1)$.
\item If the total quarterly wages for $\estab$ is $\$1,000,000$, the uncertainty interval for $\estab$'s wages is $\approx [\$810,000 - \$1,210,000]$. Thus any decrease in wages by  $\$190,000$ or increase by up to $\$210,000$ are nearly indistinguishable. So, changes to X's quarterly wages by up to $\$190,000$ are nearly indistinguishable (hence the presence of other workers in $\estab$ help to better hide the wages of X).
\end{itemize}
More generally, and formally, suppose the confidential numerical attributes of individual X are $\catt^{(p)}_1,\dots, \catt^{(p)}_{\numc}$ and for establishment $\estab$, the aggregate values are $\catt^{(e)}_1,\dots, \catt^{(e)}_{\numc}$. Let $\mech$ be a $\mu$-\shortdp mechanism with $\neighfun$ and distance parameters $\distparam_1,\dots, \distparam_\numc$. Define the intervals $I_1,\dots, I_\numc$ as the uncertainty intervals for $\estab$ that are re-centered around the values for $X$; i.e., $I_i = \uncertI_{\neighfun,\distparam_i}(\catt_i^{(e)})-\catt_i^{(e)}+\catt_i^{(p)}$. Then distinguishing between the true values $\catt^{(p)}_1,\dots, \catt^{(p)}_{\numc}$ vs. any other choice $\widehat{\catt}^{(p)}_1,\dots, \widehat{\catt}^{(p)}_{\numc}$ from those intervals is at least as hard as distinguishing between $N(0,1)$ vs. $N(\mu, 1)$.

Group privacy (Remark \ref{remark:groupp}) lets us consider what protections are provided within wider intervals. For instance, given an integer $k\geq 1$, if we define the wider intervals 
$I^{(k)}_1,\dots, I^{(k)}_\numc$ as $I^{(k)}_i = \uncertI_{\neighfun,k\distparam_i}(\catt_i^{(e)})-\catt_i^{(e)}+\catt_i^{(p)}$ then distinguishing between the true values for $X$ and any other choices from those intervals is at least as hard as $N(0,1)$ vs. $N(k\mu, 1)$.

\subsubsection{Encoding (person-based) Gaussian DP in \shortdp}\label{subsub:gdp}
We next consider how to directly encode DP protections in this framework; in Section \ref{subsub:combine}, we consider how to combine protections from different $\neighfun$ choices, such as combining DP protections for people with $\neighfun=\sqroot$ protections for establishments. Suppose the contributions of any person to an establishment's confidential attributes $\catt_i$ is bounded by a constant $d_i$ (e.g., person X can only add at most 1 to an employment count, or salary is clipped by some constant $d$ before being aggregated). Then set neighbor function $\neighfun^{(DP)}(x)=x$ and distance parameters $\distparam^{(DP)}_1,\dots, \distparam^{(DP)}_\numc$ to $d_1,\dots, d_\numc$, respectively. This makes two datasets neighbors if they differ on the contributions of one person to an establishment's values.

\subsubsection{Combining Neighbor Functions}\label{subsub:combine} 
We next consider how to combine the protections provided by some neighbor function $\neighfun^{(A)}$ having distance parameters $\distparam^{(A)}_1, \dots, \distparam^{(A)}_\numc$ with another neighbor function $\neighfun^{(B)}$ having distance parameters $\distparam^{(B)}_1, \dots, \distparam^{(B)}_\numc$. This requires using the more general form of $\mu$-\shortdp, as discussed in Remark \ref{remark:multifun}, in which each confidential attribute $\catt_i$ has its own neighbor function $\neighfun^{*}_i$ (instead of all confidential attributes sharing the same neighbor function). The goal is to obtain a new neighbor function 
$\neighfun^{*}_i$ and distance parameter $\distparam^{*}_i$ for each attribute $\catt_i$ so that for every $x$ and $i$, the resulting uncertainty interval $\uncertI_{\neighfun_i^*,\distparam_i^*}(x)$ contains the uncertainty intervals $\uncertI_{\neighfun^{(A)},\distparam^{(A)}_i}(x)$ and $\uncertI_{\neighfun^{(B)},\distparam^{(B)}_i}(x)$. This non-obvious result is provided by the following theorem.

\begin{theoremEnd}[category=model,proof end]{theorem}\label{thm:neighcombine}
Let $\neighfun^{(A)}$ and $\neighfun^{(B)}$ be neighbor functions satisfying the conditions of Definition \ref{def:close}. 
Let $\distparam^{(A)}$ and $\distparam^{(B)}$ be positive numbers. Then $\neighfun^{(A)}$ and $\neighfun^{(B)}$ are continuously differentiable almost everywhere. 
Set $\distparam^* = 1$ and define $\neighfun^*$ as follows:  
\begin{align*}
    \frac{d\neighfun^*(x)}{dx} &= \min\left(\frac{1}{\distparam^{(A)}}\frac{d\neighfun^{(A)}(x)}{dx},~ \frac{1}{\distparam^{(B)}}\frac{d\neighfun^{(B)}(x)}{dx}\right)\\
    \neighfun^*(x) &=\int_0^{x} \frac{d\neighfun^*(x)}{dx}~dx 
\end{align*}
Then $\neighfun^*$ satisfies the conditions of Definition \ref{def:close} to be a neighbor function. Furthermore, the uncertainty intervals defined by $\neighfun^{(A)}, \distparam^{(A)}$ and  $\neighfun^{(B)}, \distparam^{(B)}$ are contained in the uncertainty interval defined by  $\neighfun^*, \distparam*$. That is for any $x$, we have $\uncertI_{\neighfun^{(A)},\distparam^{(A)}}(x)\subseteq \uncertI_{\neighfun^*,\distparam^*}(x)$ and $\uncertI_{\neighfun^{(B)},\distparam^{(B)}}(x)\subseteq \uncertI_{\neighfun^*,\distparam^*}(x)$.
\end{theoremEnd}
\begin{proofEnd}
Continuous differentiability  almost everywhere is a result of concavity of the functions \cite{rockafellar2015convex} (Theorem 25.5). Thus, the set of points at which $\frac{d\neighfun}{dx}$ can not be defined has measure 0 and the set contains at most countable many points.

To show that $\neighfun^*$ is increasing, we note that $\neighfun^{(A)}$ and $\neighfun^{(B)}$ are increasing, so their derivative is positive almost everywhere, hence the derivative of $\neighfun^*$ is positive almost everywhere and hence $\neighfun^*$ is increasing. 

Next, $\neighfun^*$ is continuous because it is defined as an integral.

Next, we show that $\neighfun^*$ is concave. Since $\neighfun^{(A)}$ and $\neighfun^{(B)}$ are concave, their derivatives are non-increasing. The same is true for $\neighfun^{*}$ since for $x_1<x_2$, we have
\begin{align*}
    \frac{d\neighfun^*(x_1)}{dx}  &= \min\left(\frac{1}{\distparam^{(A)}}\frac{d\neighfun^{(A)}(x_1)}{dx},~ \frac{1}{\distparam^{(B)}}\frac{d\neighfun^{(B)}(x_1)}{dx}\right)\\
    &\geq \min\left(\frac{1}{\distparam^{(A)}}\frac{d\neighfun^{(A)}(x_2)}{dx},~ \frac{1}{\distparam^{(B)}}\frac{d\neighfun^{(B)}(x_2)}{dx}\right)\\
    &=\frac{d\neighfun^*(x_2)}{dx}
\end{align*}
and hence $\neighfun^*$ is also concave.

Next, we show that $\neighfun^*(\exp(x))$ is convex as a function of $x$. We note that
if $y=\exp(x)$ then $\frac{d\neighfun^{(A)}(\exp(x))}{dx} = e^x\frac{d\neighfun^{(A)}(y)}{dy} = y \frac{d\neighfun^{(A)}(y)}{dy}$. Hence the fact that $\neighfun^{(A)}(\exp(x))$ and $\neighfun^{(B)}(\exp(x))$ are convex means that $y\frac{d\neighfun^{(A)}(y)}{dy}$ and $y\frac{d\neighfun^{(B)}(y)}{dy}$ are non-decreasing functions of $y$ when $y>0$. Thus for $0<y_1<y_2$, 
\begin{align*}
    y_1\frac{d\neighfun^*(y_1)}{dy}  &= \min\left(y_1\frac{1}{\distparam^{(A)}}\frac{d\neighfun^{(A)}(y_1)}{dy},~ y_1\frac{1}{\distparam^{(B)}}\frac{d\neighfun^{(B)}(y_1)}{dy}\right)\\
    &\leq \min\left(y_2\frac{1}{\distparam^{(A)}}\frac{d\neighfun^{(A)}(y_2)}{dy},~ y_2\frac{1}{\distparam^{(B)}}\frac{d\neighfun^{(B)}(y_2)}{dy}\right)\\
    &=y_2\frac{d\neighfun^*(y_2)}{dy}
\end{align*}
hence $y\frac{d\neighfun^*(y)}{dy}$ is also non-decreasing when $y>0$ and it follows that $\neighfun^*(\exp(x))$ is convex.

To prove containment of uncertainty intervals, we first show that for $x_1 < x_2$, both $\frac{\neighfun^*(x_2)-\neighfun^*(x_1)}{\distparam^*}\leq \frac{\neighfun^{(A)}(x_2)-\neighfun^{(A)}(x_1)}{\distparam^{(A)}}$ and $\frac{\neighfun^*(x_2)-\neighfun^*(x_1)}{\distparam^{*}}\leq \frac{\neighfun^{(B)}(x_2)-\neighfun^{(B)}(x_1)}{\distparam^{(B)}}$. To see why, note that 
\begin{align*}
    \frac{\neighfun^*(x_2)-\neighfun^*(x_1)}{\distparam^*} &= \neighfun^*(x_2)-\neighfun^*(x_1)\\
    &=\int_{x_1}^{x_2} \frac{d\neighfun^*(x)}{dx} ~dx\\
    &\leq \frac{1}{\distparam^{(A)}}\int_{x_1}^{x_2} \frac{d\neighfun^{(A)}(x)}{dx}\\
    &=\frac{\neighfun^{(A)}(x_2)-\neighfun^{(A)}(x_1)}{\distparam^{(A)}}
\end{align*}
and same with $\neighfun^{(A)}$ replaced by $\neighfun^{(B)}$.

Now pick a point $x$ and let $z^{(+A)}=(\neighfun^{(A)})^{-1}\left(\neighfun^{(A)}(x)+\distparam^{(A)}\right)$ and also let $z^{(+*)}=(\neighfun^{*})^{-1}\left(\neighfun^{*}(x)+\distparam^{*}\right)$. Then:
\begin{align*}
    1 &= \frac{\neighfun^{(A)}(x)+\distparam^{(A)} - \neighfun^{(A)}(x)}{\distparam^{(A)}}\\
    &= \frac{\neighfun^{(A)}(z^{(+A)}) - \neighfun^{(A)}(x)}{\distparam^{(A)}}\\
    &\geq \frac{\neighfun^{*}(z^{(+A)}) - \neighfun^{*}(x)}{\distparam^{*}}\\
    &=\neighfun^{*}(z^{(+A)}) - \neighfun^{*}(x)
\end{align*}
but, $1=\neighfun^{*}(z^{(+*)}) - \neighfun^{*}(x)$ and since $\neighfun^*$ is increasing, this means $z^{(+*)} \geq z^{(+A)}$. Hence the right endpoint of the uncertainty interval determined by the combination $\neighfun^{(A)}, \distparam^{(A)}$ is less than or equal to the right endpoint of the uncertainty interval determined by the combination $\neighfun^{*}, \distparam^*$.

The left endpoint is trickier because it has a more complex expression.
Pick a point $x$. Define $\distparam^{(A)}_L=\min(\distparam^{(A)},\quad\neighfun^{(A)}(x)-\neighfun^{(A)}(0))$. This means the uncertainty interval's left endpoint is 
\begin{align*}
z^{(-A)}&=(\neighfun^{(A)})^{-1}\left(\max(\neighfun^{(A)}(0),\quad\neighfun^{(A)}(x)-\distparam^{(A)})\right)\\
&=(\neighfun^{(A)})^{-1}\left(\neighfun^{(A)}(x)-\distparam_L^{(A)}\right)
\end{align*}
Similarly, define
\begin{align*}
    \distparam^{*}_L &= \min(\distparam^{*},\quad\neighfun^{*}(x)-\neighfun^{*}(0))\\
    z^{(-*)}&=(\neighfun^{*})^{-1}\left(\max(\neighfun^{*}(0),\quad\neighfun^{*}(x)-\distparam^{*})\right)\\
&=(\neighfun^{*})^{-1}\left(\neighfun^{*}(x)-\distparam_L^{*}\right)
\end{align*}
Now, if $z^{(-*)}=0$ then the left endpoint of the uncertainty interval associated with $\neighfun^*$ and $\distparam^*$ is the smallest possible and hence $\leq$ the left endpoint associated with $\neighfun^{(A)}$ and $\distparam^{(A)}$. Hence, \textbf{for the rest of the proof, we just need to consider the case where }$\mathbf{z^{(-*)}=(\neighfun^{*})^{-1}(\neighfun^*(x)-\distparam^*)}$.

Then we have:
\begin{align*}
    1 &= \frac{\neighfun^{(A)}(x) - (\neighfun^{(A)}(x) -\distparam_L^{(A)})}{\distparam_L^{(A)}}\\
    &= \frac{\neighfun^{(A)}(x) - \neighfun^{(A)}(z^{(-A)})}{\distparam_L^{(A)}}\\
    &= \frac{1}{\distparam_L^{(A)}}\int_{r=z^{(-A)}}^{x} \frac{d\neighfun^{(A)}(r)}{dr} ~dr\\
    &\geq \frac{1}{\distparam_L^{(A)}}\int_{r=z^{(-A)}}^{x} \distparam^{(A)}\frac{d\neighfun^{*}(r)}{dr} ~dr \quad\text{(by construction of $\neighfun^*$)}\\
    &= \frac{\distparam^{(A)}}{\distparam_L^{(A)}}\left(\neighfun^{*}(x)-\neighfun^{*}(z^{(-A)})\right)\\
    & \geq \neighfun^{*}(x)-\neighfun^{*}(z^{(-A)}) \quad\text{(since $\distparam^{(A)}\geq \distparam^{(A)}_L$)}
\end{align*}
But, since we are dealing with the case that $z^{(-*)}=(\neighfun^{*})^{-1}(\neighfun^*(x)-\distparam^*)$, we have:
\begin{align*}
    1 &= \neighfun^{*}(x) - (\neighfun^{*}(x) -\distparam^{*})\\
       &= \neighfun^{*}(x) - \neighfun^{*}(z^{(-*)})
\end{align*}
So this means $\neighfun^{*}(x) - \neighfun^{*}(z^{(-*)})\geq \neighfun^{*}(x)-\neighfun^{*}(z^{(-A)})$ and so
$\neighfun^{*}(z^{(-A)})\geq \neighfun^{*}(z^{(-*)})$. Since $\neighfun^{*}$ is increasing, this means $z^{(-A)}\geq z^{(-*)}$ and so the left endpoint associated with $\neighfun^{(A)}$ and $\distparam^{(A)}$ is $\geq$ the left endpoint associated with $\neighfun^*$ and $\distparam^*$.
Hence, we have $\uncertI_{\neighfun^{(A)},\distparam^{(A)}}(x)\subseteq \uncertI_{\neighfun^*,\distparam^*}(x)$.

A similar argument holds for $\neighfun^{(B)}$.

\end{proofEnd}

As an application of Theorem \ref{thm:neighcombine}, we show how to combine the setting $\neighfun=\sqroot$ having distance parameters $\distparam_1,\dots, \distparam_\numc$ with the differential privacy setting (from Section \ref{subsub:gdp}) of neighbor function $\neighfun^{(DP)}(x) = x$ and distance parameters $d_1,\dots, d_\numc$.
\begin{textAtEnd}[category=model]
\begin{example}
    We show how to combine the neighbor function $\neighfun^{(DP)}(x)=x$ with distance parameter $d_1$ with neighbor function $\neighfun^{(E)}(x)=\sqrt{x}$ with distance parameter $\distparam_1$.
First, we set $\distparam^*=1$. Then we note that
\begin{align*}
    \frac{1}{d_1}\frac{d \neighfun^{(DP)}(x)}{dx} &= \frac{1}{d_1}\\
    \frac{1}{\distparam_1}\frac{d\neighfun^{(E)}(x)}{dx} &= \frac{1}{2\distparam_1\sqrt{x}}\\
    \frac{d\neighfun^*(x)}{dx} &= 
    \begin{cases}
        \frac{1}{d_1} & \text{ if }x \leq \frac{d_1^2}{4\distparam_1^2} \\
        \frac{1}{2\distparam_1\sqrt{x}}  & \text{ if }x \geq \frac{d_1^2}{4\distparam_1^2}
    \end{cases}\\
    \neighfun^*(x) &=\int^x_0 \frac{d\neighfun^*(t)}{dt}~dt\\
                   &= \begin{cases}
                       \frac{x}{d_1} & \text{ if }x \leq \frac{d_1^2}{4\distparam_1^2}\\
                       \frac{\sqrt{x}}{\distparam_1}-\frac{d_1}{4\distparam_1^2} &\text{ if }x \geq \frac{d_1^2}{4\distparam_1^2}
                   \end{cases}
\end{align*}
\end{example}
\end{textAtEnd}
Applying the theorem for each attribute, we get  $\distparam^*_i=1$ and 
$$\neighfun^*_i(x) = \begin{cases}
                       \frac{x}{d_i} & \text{ if }x \leq \frac{d_i^2}{4\distparam_i^2}\\
                       \frac{\sqrt{x}}{\distparam_i}-\frac{d_i}{4\distparam_i^2} &\text{ if }x \geq \frac{d_i^2}{4\distparam_i^2}
                   \end{cases}$$ 
for $i=1,\dots,\numc$.
Thus, two records $\record_1$ and $\record_2$ are said to be ``close'' if they have the same value for their public attributes and
\begin{align*}
    \Big|\neighfun^*_i(\record_1[\catt_i])-\neighfun^*_i(\record_2[\catt_i])\Big|\leq \distparam_i^* \text{ for }i=1,\dots,\numc
\end{align*}
and two datasets would be neighbors if one can be obtained from the other by replacing one record with another record that is close. The mechanisms we propose in Section \ref{sec:mech} work directly with this generalization of closeness. However, to keep the amount of notation manageable, the discussion in the rest of the paper uses the same neighbor function for each confidential attribute.

\subsubsection{Heterogeneous Composition}\label{subsub:hetergeneous}
Suppose two different data publishers $A$ and $B$ release statistics about the same underlying data, but with completely different settings. One uses $\mu^{(A)}$-\shortdp with $\neighfun^{(A)}$ and $\distparam^{(A)}_1,\dots, \distparam^{(A)}_\numc$ and the other uses $\mu^{(B)}$-\shortdp with $\neighfun^{(B)}$ and $\distparam^{(B)}_1,\dots, \distparam^{(B)}_\numc$. We call this \emph{heterogeneous} composition. The goal is to find new neighbor function $\neighfun^{*}_i$ and distance parameter $\distparam^*_i$ for each attribute, such that the resulting uncertainty intervals are contained inside  both $\uncertI_{\neighfun^{(A)},\distparam^{(A)}_i}(x)$ and $\uncertI_{\neighfun^{(B)},\distparam^{(B)}_i}(x)$ for
each $i$ and $x$. This means mechanism $\mech^{(A)}$ also satisfies $\mu^{(A)}$-\shortdp  with respect to the (per-attribute) neighbor functions $\neighfun^*_1,\dots, \neighfun^*_\numc$ and distance parameters $\distparam^*_1,\dots,\distparam^*_\numc$ (i.e., the general version of \shortdp discussed in Remark \ref{remark:multifun} and Section \ref{subsub:combine}) and  $\mech^{(B)}$ satisfies  $\mu^{(B)}$-\shortdp for those neighbor functions and distance parameters as well. Then this reduces to the standard composition and the overall confidentiality parameter becomes $\sqrt{(\mu^{(A)})^2 + (\mu^{(B)})^2}$. 
The following theorem shows to create the appropriate $\neighfun^{*}_i$ and  $\distparam^*_i$ for each attribute.

\begin{theoremEnd}[category=model,proof end]{theorem}\label{thm:neighcomposedifferent}
Let $\neighfun^{(A)}$ and $\neighfun^{(B)}$ be neighbor functions satisfying the conditions of Definition \ref{def:close}. 
Let $\distparam^{(A)}$ and $\distparam^{(B)}$ be positive numbers. Then $\neighfun^{(A)}$ and $\neighfun^{(B)}$ are continuously differentiable almost everywhere. 
Set $\distparam^*=1$ and define $\neighfun^*$ as follows:
\begin{align*}
    \frac{d\neighfun^*(x)}{dx} &= \max\left(\frac{1}{\distparam^{(A)}}\frac{d\neighfun^{(A)}(x)}{dx},~ \frac{1}{\distparam^{(B)}}\frac{d\neighfun^{(B)}(x)}{dx}\right)\\
    \neighfun^*(x) &=\int_0^{x} \frac{d\neighfun^*(x)}{dx}~dx 
\end{align*}
Then $\neighfun^*$ satisfies the conditions of Definition \ref{def:close} to be a neighbor function. Furthermore, the uncertainty intervals defined by $\neighfun^{(A)}, \distparam^{(A)}$ and  $\neighfun^{(B)}, \distparam^{(B)}$ contain the uncertainty interval defined by  $\neighfun^*, \distparam^*$. That is for any $x$, we have $\uncertI_{\neighfun^{(A)},\distparam^{(A)}}(x)\supseteq \uncertI_{\neighfun^*,\distparam^*}(x)$ and $\uncertI_{\neighfun^{(B)},\distparam^{(B)}}(x)\supseteq \uncertI_{\neighfun^*,\distparam^*}(x)$.
\end{theoremEnd}
\begin{proofEnd}
Continuous differentiability  almost everywhere is a result of concavity of the functions \cite{rockafellar2015convex} (Theorem 25.5). Thus, the set of points at which $\frac{d\neighfun}{dx}$ can not be defined has measure 0 and the set contains at most countably many points.

To show that $\neighfun^*$ is increasing, we note that $\neighfun^{(A)}$ and $\neighfun^{(B)}$ are increasing, so their derivative is positive almost everywhere, hence the derivative of $\neighfun^*$ is positive almost everywhere and hence $\neighfun^*$ is increasing. 

Next, $\neighfun^*$ is continuous because it is defined as an integral.

Next, we show that $\neighfun^*$ is concave. Since $\neighfun^{(A)}$ and $\neighfun^{(B)}$ are concave, their derivatives are non-increasing. The same is true for $\neighfun^{*}$ since for $x_1<x_2$, we have
\begin{align*}
    \frac{d\neighfun^*(x_1)}{dx}  &= \max\left(\frac{1}{\distparam^{(A)}}\frac{d\neighfun^{(A)}(x_1)}{dx},~ \frac{1}{\distparam^{(B)}}\frac{d\neighfun^{(B)}(x_1)}{dx}\right)\\
    &\geq \max\left(\frac{1}{\distparam^{(A)}}\frac{d\neighfun^{(A)}(x_2)}{dx},~ \frac{1}{\distparam^{(B)}}\frac{d\neighfun^{(B)}(x_2)}{dx}\right)\\
    &=\frac{d\neighfun^*(x_2)}{dx}
\end{align*}
and hence $\neighfun^*$ is also concave.

Next, we show that $\neighfun^*(\exp(x))$ is convex as a function of $x$. We note that
if $y=\exp(x)$ then $\frac{d\neighfun^{(A)}(\exp(x))}{dx} = e^x\frac{d\neighfun^{(A)}(y)}{dy} = y \frac{d\neighfun^{(A)}(y)}{dy}$. Hence the fact that $\neighfun^{(A)}(\exp(x))$ and $\neighfun^{(B)}(\exp(x))$ are convex means that $y\frac{d\neighfun^{(A)}(y)}{dy}$ and $y\frac{d\neighfun^{(B)}(y)}{dy}$ are non-decreasing functions of $y$ when $y>0$. Thus for $0<y_1<y_2$, 
\begin{align*}
    y_1\frac{d\neighfun^*(y_1)}{dy}  &= \max\left(y_1\frac{1}{\distparam^{(A)}}\frac{d\neighfun^{(A)}(y_1)}{dy},~ y_1\frac{1}{\distparam^{(B)}}\frac{d\neighfun^{(B)}(y_1)}{dy}\right)\\
    &\leq \max\left(y_2\frac{1}{\distparam^{(A)}}\frac{d\neighfun^{(A)}(y_2)}{dy},~ y_2\frac{1}{\distparam^{(B)}}\frac{d\neighfun^{(B)}(y_2)}{dy}\right)\\
    &=y_2\frac{d\neighfun^*(y_2)}{dy}
\end{align*}
hence $y\frac{d\neighfun^*(y)}{dy}$ is also non-decreasing when $y>0$ and it follows that $\neighfun^*(\exp(x))$ is convex.

To prove containment of uncertainty intervals, we first show that for $x_1 < x_2$, both $\frac{\neighfun^*(x_2)-\neighfun^*(x_1)}{\distparam^*}\geq \frac{\neighfun^{(A)}(x_2)-\neighfun^{(A)}(x_1)}{\distparam^{(A)}}$ and $\frac{\neighfun^*(x_2)-\neighfun^*(x_1)}{\distparam^{*}}\geq \frac{\neighfun^{(B)}(x_2)-\neighfun^{(B)}(x_1)}{\distparam^{(B)}}$. To see why, note that 
\begin{align*}
    \frac{\neighfun^*(x_2)-\neighfun^*(x_1)}{\distparam^*} &= \neighfun^*(x_2)-\neighfun^*(x_1)\\
    &=\int_{x_1}^{x_2} \frac{d\neighfun^*(x)}{dx} ~dx\\
    &\geq \frac{1}{\distparam^{(A)}}\int_{x_1}^{x_2} \frac{d\neighfun^{(A)}(x)}{dx}\\
    &=\frac{\neighfun^{(A)}(x_2)-\neighfun^{(A)}(x_1)}{\distparam^{(A)}}
\end{align*}
and same with $\neighfun^{(A)}$ replaced by $\neighfun^{(B)}$.

Now pick a point $x$ and let $z^{(+A)}=(\neighfun^{(A)})^{-1}\left(\neighfun^{(A)}(x)+\distparam^{(A)}\right)$ and also let $z^{(+*)}=(\neighfun^{*})^{-1}\left(\neighfun^{*}(x)+\distparam^{*}\right)$. Then:
\begin{align*}
    1 &= \frac{\neighfun^{(A)}(x)+\distparam^{(A)} - \neighfun^{(A)}(x)}{\distparam^{(A)}}\\
    &= \frac{\neighfun^{(A)}(z^{(+A)}) - \neighfun^{(A)}(x)}{\distparam^{(A)}}\\
    &\leq \frac{\neighfun^{*}(z^{(+A)}) - \neighfun^{*}(x)}{\distparam^{*}}\\
    &=\neighfun^{*}(z^{(+A)}) - \neighfun^{*}(x)
\end{align*}
but, $1=\neighfun^{*}(z^{(+*)}) - \neighfun^{*}(x)$ and since $\neighfun^*$ is increasing, this means $z^{(+*)} \leq z^{(+A)}$. Hence the right endpoint of the uncertainty interval determined by the combination $\neighfun^{(A)}, \distparam^{(A)}$ is greater than or equal to the right endpoint of the uncertainty interval determined by the combination $\neighfun^{*}, \distparam^*$.

The left endpoint is trickier because it has a more complex expression.
Pick a point $x$. Define $\distparam^{(A)}_L=\min(\distparam^{(A)},\quad\neighfun^{(A)}(x)-\neighfun^{(A)}(0))$. This means the uncertainty interval's left endpoint is 
\begin{align*}
z^{(-A)}&=(\neighfun^{(A)})^{-1}\left(\max(\neighfun^{(A)}(0),\quad\neighfun^{(A)}(x)-\distparam^{(A)})\right)\\
&=(\neighfun^{(A)})^{-1}\left(\neighfun^{(A)}(x)-\distparam_L^{(A)}\right)
\end{align*}
Similarly, define
\begin{align*}
    \distparam^{*}_L &= \min(\distparam^{*},\quad\neighfun^{*}(x)-\neighfun^{*}(0))\\
    z^{(-*)}&=(\neighfun^{*})^{-1}\left(\max(\neighfun^{*}(0),\quad\neighfun^{*}(x)-\distparam^{*})\right)\\
&=(\neighfun^{*})^{-1}\left(\neighfun^{*}(x)-\distparam_L^{*}\right)
\end{align*}
Now, if $z^{(-A)}=0$ then the left endpoint of the uncertainty interval associated with $\neighfun^{(A)}$ and $\distparam^{(A)}$ is the smallest possible and hence $\leq$ the left endpoint associated with $\neighfun^{*}$ and $\distparam^{*}$. Hence, \textbf{for the rest of the proof, we just need to consider the case where }$\mathbf{z^{(-A)}=(\neighfun^{(A)})^{-1}(\neighfun^{(A)}(x)-\distparam^{(A)})}$.

Then we have:
\begin{align*}
    1 &= \frac{\neighfun^{(A)}(x) - (\neighfun^{(A)}(x) -\distparam^{(A)})}{\distparam^{(A)}}\\
    &= \frac{\neighfun^{(A)}(x) - \neighfun^{(A)}(z^{(-A)})}{\distparam^{(A)}}\\
    &= \frac{1}{\distparam^{(A)}}\int_{r=z^{(-A)}}^{x} \frac{d\neighfun^{(A)}(r)}{dr} ~dr\\
    &\leq \int_{r=z^{(-A)}}^{x} \frac{d\neighfun^{*}(r)}{dr} ~dr \quad\text{(by construction of $\neighfun^*$)}\\
    &= \left(\neighfun^{*}(x)-\neighfun^{*}(z^{(-A)})\right)\\
    & \leq \frac{\distparam^*}{\distparam^*_L}\left(\neighfun^{*}(x)-\neighfun^{*}(z^{(-A)})\right) \quad\text{(since $\distparam^{*}\geq \distparam^{*}_L$)}\\
    &=\frac{1}{\distparam^*_L}\left(\neighfun^{*}(x)-\neighfun^{*}(z^{(-A)})\right) \quad\text{(since $\distparam^{*}=1$)}\\
\end{align*}
But,  we also have 
\begin{align*}
    1 &= \frac{\neighfun^{*}(x) - (\neighfun^{*}(x) -\distparam^{*}_L)}{\distparam^*_L}\\
       &= \frac{\neighfun^{*}(x) - \neighfun^{*}(z^{(-*)})}{\distparam^*_L}
\end{align*}
So this means $\neighfun^{*}(x) - \neighfun^{*}(z^{(-*)})\leq \neighfun^{*}(x)-\neighfun^{*}(z^{(-A)})$ and so
$\neighfun^{*}(z^{(-A)})\leq \neighfun^{*}(z^{(-*)})$. Since $\neighfun^{*}$ is increasing, this means $z^{(-A)}\leq z^{(-*)}$ and so the left endpoint associated with $\neighfun^{(A)}$ and $\distparam^{(A)}$ is $\leq$ the left endpoint associated with $\neighfun^*$ and $\distparam^*$.
Hence, we have $\uncertI_{\neighfun^{(A)},\distparam^{(A)}}(x)\supseteq \uncertI_{\neighfun^*,\distparam^*}(x)$.

A similar argument holds for $\neighfun^{(B)}$.
\end{proofEnd}

\section{Mechanisms for {\lowercase{\shortdp}}}\label{sec:mech}
Mechanisms for \gedp can be designed using the appropriate concept of sensitivity. However, the requirement that absolute error protections increase with establishment size means that the mechanisms have data-dependent variance. 
We first present the sensitivity-based framework for \shortdp. We then propose two mechanisms: (1) the \neighmech, which has data-dependent variance that cannot be released exactly (to avoid compromising the confidentiality guarantees), and (2) the \pncmech, which piggybacks on the \neighmech to answer additional queries with data-dependent variances that are safe to release.

\begin{definition}[$\neighfun$-neighbor sensitivity]\label{def:newsens} Given a vector-valued query function $q$, the $\neighfun$-neighbor sensitivity of $q$ denoted as $\sensneighfun(q)$ is defined as: 
 $\sensneighfun(q)=\sup_{\data_1,\data_2}||q(\data_1)-q(\data_2)||_2$, where the supremum is taken over all pairs of $\neighfun$-neighboring datasets with associated distance parameters $\distparam_1,\dots, \distparam_{\numc}$.
\end{definition}

The resulting Gaussian-style mechanism is similar to the one used by GDP \cite{gdp}, but with $\neighfun$-neighbor sensitivity in place of $L_2$ sensitivity:

\begin{definition}[Estab-Gaussian mechanism]\label{def:attributemech} Let $\mu$ be  a positive real-valued privacy parameter and let $q$ be a vector-valued function. Given a neighbor function $\neighfun$ with associated distance parameters $\distparam_1,\dots, \distparam_{\numc}$ and an input dataset $\data$, the Estab-Gaussian mechanism for \shortdp returns $q(\data)+N\left(0,\frac{\sensneighfun^2(q)}{\mu^2}I\right)$.
\end{definition}

\begin{theoremEnd}[category=mechanism]{theorem}\label{thm:mechprops}
Given a privacy budget $\mu>0$, neighbor function $\neighfun$ and distance parameters $\distparam_1,\dots,\distparam_\numc$, 
the Estab-Gaussian mechanism (Definition \ref{def:attributemech}) satisfies $\mu$-\shortdp. \end{theoremEnd}
\begin{proofEnd}
These results follow directly from the proofs of the Gaussian mechanism of Gaussian Differential Privacy.
\end{proofEnd}

\subsection{The \neighmech.}\label{sec:basicmech}

Group-by sum queries are some of the most important types of queries for establishment data.
We represent them as $\gquery_{\langle\grouper, \catt_i\rangle}$, where $\grouper$ is a  function that creates groups from the public attributes of an establishment and $\catt_i$ is the confidential attribute to sum over. For example, consider the query ``total wages for each combination of county and 2-digit NAICS prefix.'' Here $\grouper$ would return a tuple consisting of the county and 2-digit NAICS prefix of an establishment, and $\catt_i$ would be the wages attribute.

\begin{definition}\label{def:neighmech}
    Let $\gquery_{\langle \grouper, \catt_i\rangle}$ be a groupby-sum query that aggregates over confidential attribute $\catt_i$ having distance parameter $\distparam_i$. Let $\mu>0$ be a privacy parameter. The \neighmech $\mech_\neighfun$ is defined as $\mech_{\neighfun}(\data)=\neighfun(\gquery_{\langle \grouper, \catt_i\rangle}(\data)) + N(\mathbf{0}, \frac{\distparam_i^2}{\mu^2}\mathbf{I})$ -- it applies $\neighfun$ to every entry in the group-by query result, then adds independent Gaussian noise, with standard deviation $\distparam_i/\mu$, to each entry. 
\end{definition}

\begin{theoremEnd}[category=mechanism]{theorem}\label{thm:mechneighfun}
Let $\neighfun$ be a neighbor function satisfying Definition \ref{def:close}. Let $\distparam_1,\dots, \distparam_\numc$ be the associated distance parameters. Then the \neighmech $\mech_\neighfun$ satisfies $\mu$-\shortdp with neighbor function $\neighfun$ and distance parameters $\distparam_1,\dots, \distparam_\numc$.
\end{theoremEnd}
\begin{proofEnd}
    First, note that each group in a group-by sum query $\gquery_{\langle \grouper, \catt_i\rangle}$ is formed from the public attributes. Hence modifying one establishment only changes one of the groups. Let $x$ be the value of $\catt_i$ of  an establishment in such a group before modification, let $y$ be the value of $\catt_i$ for that establishment after modification, and let $t$ be the sum of the $\catt_i$ values in the rest of the establishments in that group. By definition of $\neighfun$-closeness, $|\neighfun(x) - \neighfun(y)|\leq \distparam_i$.
    Theorem \ref{thm:neighfun}, item \ref{thm:neighfun:item2}, guarantees that the $\neighfun$-neighbor sensitivity (Definition \ref{def:newsens}) of $\neighfun\circ\gquery_{\langle \grouper, \catt_i\rangle}$ is $\leq \distparam_i$ since $|\neighfun(x+t) - \neighfun(y+t)|\leq |\neighfun(x) - \neighfun(y)|$. Hence, by Theorem \ref{thm:mechprops}, $\mech_{\neighfun}$ satisfies $\mu$-\shortdp.
\end{proofEnd}
The personalized DP framework for establishment data by Finley et al. \cite{finley2024slowly} also propose mechanisms that first
transform the data and then add Gaussian noise. Every \neighmech from our framework is a valid transformation mechanism in the framework of \cite{finley2024slowly} (i.e., a concave, strictly
increasing function with domain from $[a,\infty)$ with $a\in \mathbb{R}$).
However, only a subset of their transformation mechanisms correspond to a \neighmech in our framework. The reason is that not every transformation in their framework satisfies Definition \ref{def:close}, specifically Condition 4. That condition prevents small establishments from having \emph{lower} relative error protections than larger establishments, as the following example shows:
\begin{example}\label{ex:morerelative}
    Consider the  neighbor function $\neighfun$ and its inverse:
    \begin{align*}
    \neighfun(x)=
    \begin{cases}
        10x &\hspace{-0.8em}\text{if }x\leq 10\\
        \frac{x}{100} + 99.9 &\hspace{-0.8em}\text{if }x\geq 10
    \end{cases}&\quad
    \neighfun^{-1}(y)=
    \begin{cases}
        \frac{y}{10} &\hspace{-0.8em}\text{if }y\leq 100\\
        100(y-99.9) &\hspace{-0.8em}\text{if }y\geq 100
    \end{cases}
    \end{align*}
    This function satisfies the requirements of \cite{finley2024slowly} because it is concave and strictly increasing. It does not satisfy our requirements of a neighbor function because $\neighfun(\exp(x))$ is not convex. We next show that releasing $\neighfun(x)+N(0,1)$ would result in counterintuitive protections where small establishments are unprotected while large establishments get much higher relative error protections.
    
    Let $x_1$ and $x_2$ be the employment counts of two establishments. Suppose $\neighfun(x_1)+N(0,1)=51$. What can we infer about $x_1$? Using the properties of the Gaussian distribution, it means that with probability $>0.999$, $\neighfun(x_1)\in 51\pm 3.3$ and so with overwhelming probability $x_1\in 5.1\pm 0.33$. The relative length of this $0.999$-confidence band is $2\frac{0.33}{5.1}<0.13$. So $x_1$ is almost certainly 5. The corresponding establishment received very little relative and absolute error protections. What if $\neighfun(x_2)+N(0,1)=106.3$? In this case, $\neighfun(x_2)\in 106.3\pm 3.3$ with probability $> 0.999$. Using the inverse $\neighfun^{-1}$, we see that $x_2\in [640\pm 330]$ with overwhelming probability. This confidence band has relative size $2\frac{330}{640}>1.03$ which is over 7 times larger than for the small establishment $x_1$. We can tell that $x_2$ is large, but its relative and absolute errors are much higher. 
\end{example}


Additionally, the above example shows that if we get  a noisy answer $\outp_g$ corresponding to a group $g$ in a group-by query, then if $[-z, z]$ is a $q\%$ confidence interval for the standard Gaussian, then $[\neighfun^{-1}(\outp_g-z\distparam_i/\mu), \quad\neighfun^{-1}(\outp_g+z\distparam_i/\mu)]$ is 
a $q\%$ confidence interval for the true aggregation of $\catt_i$ in that group $g$. The quantity $\neighfun^{-1}(\outp_g)$ is the maximum likelihood estimate of the true value, but is biased. The following theorem provides (1) unbiased estimators\footnote{The results of \cite{finley2024slowly,Washio1956UnbiasedEB} can be used to obtain unbiased estimators and their true variances in a more general setting.} when $\neighfun=\sqroot$ or $\neighfun=\log,$ (2) the distribution
of the unbiased estimators and (3) the true variance. Since the variance is a function of the true value, it cannot be released but may be estimated from the noisy answer. The following theorem shows  that the variance can be accurately estimated when $\neighfun=\sqroot$, but for $\neighfun=\log$, the variance estimates  will never converge.

\begin{theoremEnd}[category=mechanism, proof end]{theorem}\label{thm:unbiased}
Given a group-by sum query $\gquery_{\langle\grouper, \catt_i\rangle}$, let $x$ be the true sum within a given group $g$, and let $\outp_g=\neighfun(x)+N(0, \distparam_i^2/\mu^2)$ be the noisy answer provided by $\mech_{\neighfun}$. Then
\begin{itemize}[leftmargin=*]
    \item If $\neighfun=\sqroot,$ then $\outp_g^2$ has the distribution of a non-central $\chi^2$ random variable with 1 degree of freedom and non-centrality parameter $x\mu^2/\distparam^2$, multiplied by $\distparam^2_i/\mu^2$. Then $\widetilde{x}\equiv\outp_g^2-(\distparam_i^2/\mu^2)$ is an unbiased estimate of $x$. The variance of $\widetilde{x}$ is $v\equiv 2(\distparam_i/\mu)^2\left(2x+(\distparam_i/\mu)^2\right)$. Replacing $x$ with $\widetilde{x}$ in this formula results in a releasable estimate $\widetilde{v}$ of the variance, and $\widetilde{v}/v \rightarrow 1$ in probability as $x\rightarrow\infty$.
    \item If $\neighfun=\log,$ then $e^{\outp_g}$ has the log-normal distribution with location parameter $\log(x)$ and scale $\distparam_i^2/\mu^2$. Then $\widetilde{x}\equiv \exp(\outp_g - \distparam_i^2/(2\mu^2))$ is an unbiased estimate of $x$. The variance of $\widetilde{x}$ is $v\equiv x^2(-1+\exp(\distparam_i^2/\mu^2))$. Replacing $x$ with $\widetilde{x}$ in this formula gives a releasable estimate $\widetilde{v},$ but $\widetilde{v}/v$ does not converge in probability as $x\rightarrow\infty$.
\end{itemize}
\end{theoremEnd}
\begin{proofEnd}
 Let $x$ be the true sum within a given group $g$ and $\widetilde{x}\equiv\outp_g^2-(\distparam_i^2/\mu^2)$ where $\outp_g=\sqrt{x}+N(0, \distparam^2_i/\mu^2)$. 
 
Then $\outp_g^2$ is equal in distribution to $\left(\frac{\distparam_i}{\mu}\right)^2~\chi_{1}^2(\eta)$ where $\eta=\left(\frac{\sqrt{x_i}\mu}{\distparam_i}\right)^2$ and $\chi_{1}^2(\eta)$ is a non-central chi-squared random variable with 1 degree of freedom and non-centrality parameter $\eta$ (since the distribution of $\chi_1^2(\eta)$ is, by definition, the distribution of the square of a $N(\sqrt{\eta}, 1)$ random variable). 
 Since $\E(\chi^2_{1}(\eta))=1+\eta$ and $\Var(\chi_{1}^2(\eta))=2(1+2\eta)$ \cite{ncx2text}, then $\widetilde{x}$ is an unbiased estimator for $x$. 
Additionally, $v=\Var\left(\widetilde{x}\right)=2\left(\frac{\distparam_i}{\mu}\right)^2\left(2x+\left(\frac{\distparam_i}{\mu}\right)^2\right).$  A plug-in estimate for the variance is $\widetilde{v}=2\left(\frac{\distparam_i}{\mu}\right)^2\left(2\widetilde{x}+\left(\frac{\distparam_i}{\mu}\right)^2\right)$. Thus, $$\frac{\widetilde{{v}}}{v}-1=\frac{\widetilde{v}-v}{v}=\frac{2\frac{\distparam_i^2}{\mu^2}\left(\widetilde{x}-x\right)}{\frac{\distparam_i^2}{\mu^2}\left(2x+(\distparam_i/\mu)^2\right)}$$
To show that $\widetilde{v}/v\rightarrow 1$ in probability as $x\rightarrow \infty$, we must show $\lim_{x\to \infty}P\left(\left|\frac{2(\widetilde{x}-x)}{2x+(\distparam_i/\mu)^2}\right|>\epsilon_{lim}\right)=0$ for any $\epsilon_{lim}>0$. 

We can express this probability in terms of $x$ and a standard normal random variable $Z$. For  ease of notation let $a_i^2=\frac{\distparam_i^2}{\mu^2}$.
Since $x>-0$, the resulting probability $$P\left(\left|\frac{2}{2x+a_i^2}\left(a_i^2\left(Z+\sqrt{x}/a_i\right)^2-x-a_i^2\right)\right|>\epsilon_{lim}\right)$$ can be expanded to be $1-p_+(x)+p_-(x)$ 
where 
$$p_+(x)=P\left(\left(Z+\sqrt{x}/a_i\right)^2 \leq \frac{2x(1+\epsilon_{lim})+(2+\epsilon_{lim})a_i^2}{2a_i^2}\right)$$ 
and 
$$p_-(x)=P\left(\left(Z+\sqrt{x}/a_i\right)^2 < \frac{2x(1-\epsilon_{lim})+(2-\epsilon_{lim})a_i^2}{2a_i^2}\right)$$. 
Then $p_+(x)$ can be expressed with the standard normal cumulative distribution function $\Phi$: 
\begin{align*}
p_+(x)&=\Phi\left(\frac{\sqrt{2(1+\epsilon_{lim})x+(2+\epsilon_{lim})a_i^2}-\sqrt{2x}}{a_i\sqrt{2}}\right)\\ 
&-\Phi\left(\frac{-\sqrt{2(1+\epsilon_{lim})x+(2+\epsilon_{lim})a_i^2}-\sqrt{2x}}{a_i\sqrt{2}}\right).
\end{align*} Since $\epsilon_{lim}>0$, we know $\lim_{x\to\infty}p_+(x)=1$. Regarding $p_-(x)$, if $\epsilon_{lim}>1$, then $\lim_{x\to\infty}\frac{(1-\epsilon_{lim})}{a_i^2}x+1-\frac{\epsilon_{lim}}{2}<0$ and $\lim_{x\to \infty}p_-(x)=0$. If $0<\epsilon_{lim}\leq1$, then \begin{align*}p_-(x)&=\Phi\left(\frac{\sqrt{2(1-\epsilon_{lim})x+(2-\epsilon_{lim})a_i^2}-\sqrt{2x}}{\sqrt{2a_i^2}}\right)\\
&-\Phi\left(\frac{-\sqrt{2(1-\epsilon_{lim})x+(2-\epsilon_{lim})a_i^2}-\sqrt{2x}}{\sqrt{2a_i^2}}\right).\end{align*} We know $\pm\sqrt{2(1-\epsilon_{lim})x+(2-\epsilon_{lim})a_i^2}-\sqrt{2x}\rightarrow-\infty$ as $x\rightarrow \infty$ for $0<\epsilon_{lim}\leq 1$. Thus $\lim_{x\to\infty}p_-(x)=0$ for any $\epsilon_{lim}>0$. Therefore, $\widetilde{v}/v$ converges in probability to 1 as $x\rightarrow \infty$.

For $\neighfun=\log$  with $x$ restricted to positive values, $\outp_g=\log(x)+N(0,\distparam_i^2/\mu^2)$. Then $e^{\outp_g}$ is a log-normal random variable with location parameter $\eta=\log(x)$ and scale parameter $\sigma^2=\frac{\distparam_i^2}{\mu^2}$. Using known formulas for expectation and variance of log-normal random variables \cite{CasellaBerger}, we know $\widetilde{x}\equiv \exp\left(\outp_g-\distparam_i^2/(2\mu^2)\right)$ is unbiased for $x$.  The variance of $\widetilde{x}$ is: $v=\Var\left(\exp\left(\outp_g-\distparam_i^2/(2\mu^2)\right)\right)=x^2\left(e^{\distparam_i^2/\mu^2}-1\right)$. The plug-in estimate is $\widetilde{v}=\widetilde{x}^2\left(e^{\distparam_i^2/\mu^2}-1\right)$. We consider $\widetilde{v}/v=\widetilde{x}^2/x^2$. Since $\widetilde{x}\equiv x\exp\left(\frac{\distparam_i}{\mu}Z-\frac{\distparam_i^2}{2\mu^2}\right)$ where $Z\sim N(0,1)$,  $P\left(\left|\frac{\widetilde{v}}{v}-1\right|> \epsilon_{lim}\right)=1-P\left(\exp\left(\frac{2\distparam_i}{\mu}Z-\frac{\distparam_i^2}{\mu^2}\right)\leq 1+\epsilon_{lim}\right)+P\left(\exp\left(\frac{2\distparam_i}{\mu}Z-\frac{\distparam_i^2}{\mu^2}\right)<1-\epsilon_{lim}\right)$. This probability does not depend on the value of $x$ and a selection of any $\epsilon_{lim}>1$ makes $P\left(\left|\frac{\widetilde{v}}{v}-1\right|\geq \epsilon_{lim}\right)>0$. Thus $\widetilde{v}/v$ does not converge in probability.  
    
\end{proofEnd}
Unbiased (noisy) query answers and estimates of their variances are important for the postprocessing step of converting the noisy answers into confidentiality-preserving microdata (see Section \ref{sec:microdata}).  Such microdata are useful to statistical agencies, who are often requested to release additional special tabulations. Having confidentiality-preserving microdata on hand means that the agencies can tabulate query answers from the microdata. It is an advantage over the prevailing cell suppression approach, for which it is difficult to analyze interactions with new data products. 

From Theorem \ref{thm:unbiased}, we see that use of $\neighfun=\log$, which is consistent with historical tradition of constant relative error protections, is poorly suited for the microdata creation process as it does not allow good estimates of noisy query variances -- the variance estimates do not converge no matter how large an establishment is. The requirement to estimate the variance can have serious consequences on the quality of the confidentiality-preserving microdata (as we explain in Section \ref{sec:microdata}). Hence we next propose the \pncmech whose data-dependent variance is safe to release.

\begin{textAtEnd}[category=mechanism]
Figure \ref{fig:error_comparison} compares the two estimators for group-by sum queries given in Theorem \ref{thm:unbiased} for the choices of $\neighfun=\sqroot$ and $\neighfun=\log$ using the same settings examples as in Section \ref{sec:neighfun}. As with the privacy settings (where $\log$ under-protects small establishments and overprotects large ones), the choice of $\neighfun=\log$ has undesirable utility properties: it has larger errors for large aggregates. But the large aggregates (such as total employment in the United States) are so politically sensitive that they require extremely low relative errors in practice. This is yet another reason to prefer $\neighfun=\sqroot$.

\begin{figure}[ht]
     \includegraphics[scale=0.6]{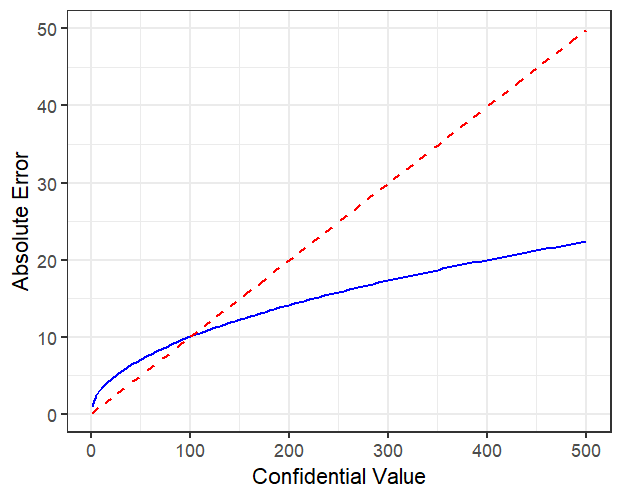}
     \includegraphics[scale=0.6]{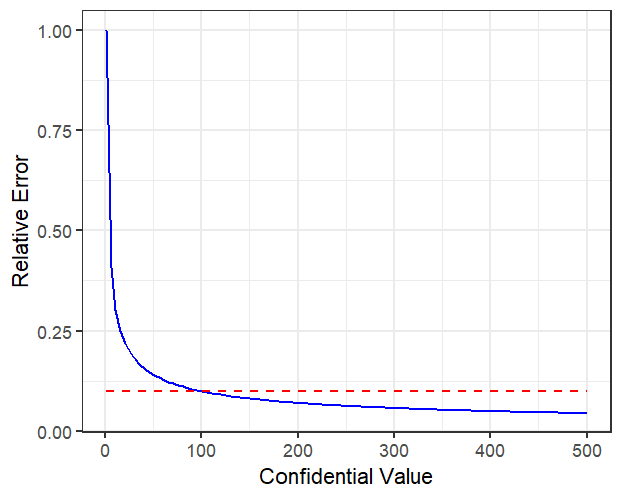}
     \caption{Comparison of the absolute and relative error of unbiased privacy preserving estimators of group-by sum queries, as given by Theorem \ref{thm:unbiased}. The $x$ axis is the true answer. Solid blue line: $\neighfun=\sqroot$ with $\distparam=0.5$ and $\mu=1$. Dashed red line: $\neighfun=\log$ with $\distparam=0.1$.} 
    \label{fig:error_comparison}
\end{figure}
\end{textAtEnd}

\subsection{Probably-No-Clipping (pnc) Mechanism}\label{subsec:pnc}

To answer group-by sum queries $\gquery_{\langle\grouper, \catt_i\rangle}$  with data-dependent Gaussian noise but variances that are safe to release, we propose the probably-no-clipping mechanism (\pncmech). The prerequisite  is  a simultaneous high-probability upper bound $u_{j,i}$ on $\record_j[\catt_i]$ (the value of confidential attribute $\catt_i$ for the record belonging to establishment $\estab_j$) for all $i,j$. \textbf{Privacy is not affected by failure of the upper bound.} Given such an upper bound, we replace the true values $\record_j[\catt_i]$ with  $\min(\record_j[\catt_i], u_{j,i})$ when aggregating records in a group, and add Gaussian noise to satisfy $\mu$-\shortdp.  We obtain the $u_{j,i}$ by running a \neighmech to answer the identity query and applying Theorem \ref{thm:upper}.

\begin{theoremEnd}[category=mechanism]{theorem}\label{thm:upper}
    Given establishment records $\record_1,\dots, \record_\numest$, a neighboring function $\neighfun$ with distance parameters $\distparam_1,\dots, \distparam_{\numc}$, and a privacy loss budget allocations $\mu_1,\dots, \mu_\numc$, define the noisy values $\outp_{j,i}$ as  $\outp_{j,i} = \neighfun(\record_j[\catt_{i}])+N(0, \distparam_i^2/\mu_i^2)$. Let $\confparam\in (0,1)$ be a confidence parameter. First, define the \emph{probably-no-clipping parameter} $\noclip= \Phi^{-1}((1-\confparam)^{1/(\numc \numest)})$. Next, define upper bounds $u_{j,i}$ as $u_{j,i}=\neighfun^{-1}(\outp_{j,i} +  \distparam_i\noclip/\mu_i)$. Then with probability $1-\confparam$ the following inequalities simultaneously hold: $\record_j[\catt_i]\leq u_{j,i}$ for all $j$ and $i$.
\end{theoremEnd}
\begin{proofEnd}
\begin{align*}
    \lefteqn{P\Big(\record_j[\catt_i]\leq u_{j,i} \text{ for }j=1,\dots,\numest \text{ and } i=1,\dots,\numc\Big)}\\
    &= P\Big(\neighfun(\record_j[\catt_i])\leq \outp_{j,i}+\distparam_i\noclip/\mu_i  \text{ for }j=1,\dots,\numest \text{ and } i=1,\dots,\numc\Big)\\
    &= P\Big(-N(0, \sigma_i^2/\mu_i^2)\leq \sigma_i\noclip/\mu_i  \text{ for }j=1,\dots,\numest \text{ and } i=1,\dots,\numc\Big)\\
    &=P(N(0, 1)\leq \noclip \text{ for $\numest \numc$ independent Gaussians})\\
    &=(\Phi(\tau))^{\numest\numc}=(\Phi(\Phi^{-1}((1-\confparam)^{\frac{1}{\numest\numc}}))^{\numest\numc}=1-\confparam
\end{align*}
\end{proofEnd}

The upper bounds $u_{j,i}$  from Theorem \ref{thm:upper} are \emph{public}, since they result from postprocessing the answers to a \neighmech.
We use them to define the \pncmech for group-by sum queries.
\begin{definition}[Probably no clipping mechanism]\label{def:pncmech}
   Let $\gquery_{\langle\grouper, \catt_i\rangle}$ be a group-by sum-query, $g_1,\dots, g_\ell$ be the corresponding disjoint groupings of establishments, and the $u_{j,i}$ be the upper bounds from Theorem \ref{thm:upper}. The \pncmech for $\gquery_{\langle\grouper, \catt_i\rangle}$ with privacy budget $\mu>0$ is: (1) For each group $g_\ell$ compute the maximum establishment contribution: $u^*_\ell = \max_{\estab_j\in g_\ell} u_{j,i}$. (2)
        Compute the per-group sensitivity for each group $g_\ell$: $\sens_\ell=u^*_\ell - \neighfun^{-1}(\max(0, \neighfun(u^*_\ell) - \distparam_i))$.
      (3) Produce the per-group truncated summation for each group $g_\ell$: $T_{g_{\ell}}=\sum_{\estab_j\in g_\ell} \min(\record[\catt_i], u^*_{\ell})$. (4) Release the aggregate values. For each group $g_\ell$ release $T_{g_{\ell}}+N(0, (\sens_\ell/\mu)^2)$ and the variance $(\sens_\ell/\mu)^2$.
\end{definition}

\begin{theoremEnd}[category=mechanism]{theorem}\label{thm:pncmech}
The \pncmech from Definition \ref{def:pncmech} satisfies $\mu$-\shortdp with neighboring function $\neighfun$ and distance parameters $\distparam_1,\dots, \distparam_\numc$.
\end{theoremEnd}
\begin{proofEnd}
    Clearly, the privacy cost of different groups compose in parallel since they affect different establishments, and the choice of which establishment goes into which group is based on the public attributes. Thus we just need to analyze one group, compute the sensitivity of the statistic for that group, and then apply Theorem \ref{thm:mechprops} for the Estab-Gaussian mechanism.

    Given a group $g_\ell$, the statistic being computed is $\sum\limits_{\estab_j\in g_\ell} \min(\record[\catt_i], u^*_{\ell})$. Clearly the sensitivity is the same as the solution to \begin{align*} 
    \max_{x\geq 0,y\geq 0} &|x-y| \\
    &\text{s.t. } |\neighfun(x)-\neighfun(y)|\leq\distparam_i \\
    &\text{ and } x\leq u_{\ell}^*\\
    &\text{ and } y\leq u_{\ell^*}
    \end{align*}
    Without loss of generality, we can set $x>y$. Noting that $\neighfun$ is a neighbor function and hence increasing, we need to make $y$ as small as possible given $x$. Thus we can set $y$ so that $\neighfun(y)=\neighfun(x)-\distparam_i$, meaning that $y=\neighfun^{-1}(\max(0, \neighfun(x)-\distparam_i))$. Thus the sensitivity equals:
    \begin{align*} 
    \max_{x\in [0, u_{\ell}^*]} &|x-\neighfun^{-1}(\max(0, \neighfun(x)-\distparam_i))| 
    \end{align*}
    Next, since $\neighfun$ is concave and increasing, then $\neighfun^{-1}$ is convex and increasing. This means for any $a\geq \distparam_i$, $\neighfun^{-1}(a) - \neighfun^{-1}(a-\distparam_i)$ is an increasing function of $a$. Setting $a=\neighfun(x)$ proves the result.
\end{proofEnd}

The complete workflow is: (1) Run the \neighmech with privacy budgets $\mu^{(a)}_1,\dots, \mu^{(a)}_\numc$ to answer the identity queries for confidential attributes $\catt_1,\dots,\catt_{\numc}$. (2) Compute the $u_{j,i}$ using Theorem \ref{thm:upper}. (3) Use  the \pncmech to answer group-by queries $\gquery_{\langle\grouper_1, \catt_{i_1}\rangle},\dots, \gquery_{\langle\grouper_m, \catt_{i_m}\rangle}$ with budgets $\mu^{(b)}_1,\dots, \mu^{(b)}_m$. The total privacy budget is: 
$$\mu^{(total)}=\sqrt{(\mu^{(a)}_1)^2+\cdots+(\mu^{(a)}_\numc)^2 + (\mu^{(b)}_1)^2+\cdots+(\mu^{(b)}_m)^2}.$$ 

\textbf{Note:} in our experiments, we set the clipping parameter to be $0.01$. This means that with probability $0.99$, no clipping occurs at all when answering a batch of queries.

\section{Postprocessing into Microdata}\label{sec:microdata}
Postprocessing a set of confidentiality-preserving query answers into confidentiality-preserving establishment-level data $\widehat{\data}$ (also known as microdata) is an important task in practice
\cite{li2015matrix,abowd20222020} as it allows agencies to answer additional queries  from $\widehat{\data}$ without spending additional privacy budget. A common approach is inverse-variance-weighted least squares \cite{li2015matrix,abowd20222020,hay2009boosting}. Starting with a collection of queries, $q_1,\dots, q_k$ and their respective noisy answers, $a_1,\dots, a_k$ (e.g., produced via the \neighmech with postprocessing via Theorem \ref{thm:unbiased} or \pncmech),
let $v_1,\dots, v_k$ be the corresponding public estimates of the noisy answer variances.\footnote{The $v_1,\dots, v_k$ depend solely on the mechanisms and privacy parameters, and not on statistical models of the data}
In the case of the \pncmech, the true variances are publicly known; for the \neighmech with $\neighfun=\log$ or $\neighfun=\sqroot$ (followed by postprocessing), the $v_i$ are  approximations (e.g., using Theorem \ref{thm:unbiased}). To create 
$\widehat{\data}$, one solves the following optimization problem:
\begin{align}
    \widehat{\data} \gets \arg\min_{\widehat{\data}} \sum_{i=1}^k\frac{1}{v_i}(q_i(\widehat{\data})-a_i)^2.\label{eqn:microdata}
\end{align}
Non-negativity or other domain-specific constraints can also be added, although their effects on data quality can have unintended consequences \cite{abowd2021uncertainty} that require study for each application.

\begin{textAtEnd}[category=microdata]
It is important to analyze how approximation errors in $v_1,\dots, v_k$ affect the quality of $\widehat{D}$. The \pncmech and \neighmech are too complex to analyze in closed form, so instead we next analyze simple mathematical models that are tractable. They show that if the $v_i$ are approximations to the true variances, the error of queries computed from $\widehat{D}$ can increase. If the approximations are data-dependent, then this introduces bias as well. Numerical simulations in Appendix \ref{app:bias} and Section \ref{sec:comps} support these findings and hence support the use of \pncmech as much as possible when bias is a concern.

\subsubsection*{Case 1: known variances} Suppose there is just one establishment $\estab$ with $X$ employees and we have $n$ independent unbiased noisy estimates $a_1,\dots, a_n$ of the employment count (i.e., $E[a_i]=X$). Each  $a_i$ has known variance $\sigma^2$. Applying Equation \ref{eqn:microdata}, we get the intuitive result that the overall guess of the employment count $X$ should be $(a_1+\cdots+a_n)/n$ and the variance of this guess is $\sigma^2/n$.

\subsubsection*{Case 2: estimated, but data-independent, variances} Next, we examine how error increases when the true variances of the $a_i$ (i.e., $\sigma^2$) are unknown, but noisy variance estimates $v_1,\dots, v_n$ are available. Suppose  the variance estimates $v_i$ have independent inverse Gamma distributions.\footnote{The inverse Gamma distribution IG$(\alpha, \beta)$ is a distribution over nonnegative numbers. It has mean $\beta/(\alpha-1)$ and variance ${\beta^2}/{(\;(\alpha-1)^2(\alpha-2)\;)}$.} If each $v_i$ has distribution IG$(2+\tau,\sigma^2(1+\tau)),$ then $E[v_i]=\sigma^2$, and 
$Var(v_i)=\sigma^4/\tau$ and $\tau$ controls how accurate the $v_i$ are. Solving Equation \ref{eqn:microdata}, the guess for employment count $X$ is $\frac{\sum_{i=1}^n a_i/v_i}{\sum_{j=1}^n 1/v_j}$. It is unbiased (the expected value equals $X$) and the squared error of the guess is:
\begin{align*}
E\left[\left(X - \frac{\sum_i a_i/v_i}{\sum_j 1/v_j}\right)^2\right] 
=\sum_i E\left[\frac{ (X-a_i)^2/v_i^2}{(\sum_j 1/v_j)^2}\right]
=\frac{\sigma^2}{n}\frac{n(3+\tau)}{1+n(2+\tau)},
\end{align*}
where the last equality follows from the following facts: (1) If $v_i$ has the IG$(\alpha,\beta)$ distribution, then $1/v_i$ has the Gamma$(\alpha, \beta)$ distribution. (2) If $1/v_1, \dots, 1/v_n$ are independent samples from a Gamma$(\alpha, \beta)$ distribution then the vector $\frac{(1/v_1, \dots, 1/v_n)}{\sum_j 1/v_j}$ has the Dirichlet$(\alpha, \dots, \alpha)$ distribution and its $i^\text{th}$ component has mean $\alpha/n$, variance $(\alpha/n)(1-\alpha/n)/(1+\alpha/n)$, and second moment $\frac{1}{n}\frac{1+\alpha}{1+n\alpha}$. Comparing to Case 1, we see that \textbf{uncertainty about the variance of the $a_i$ increases the squared error by a factor of} $\frac{n(3+\tau)}{1+n(2+\tau)}$. For example, with $n=2$ and $\tau=1$, this factor is $\approx 1.14$, representing a 14\% increased error simply for not knowing exactly how accurate are our noisy estimates of the employment at establishment $\estab$.

\subsubsection*{Case 3: estimated and data-dependent variances} To make the variances of the $a_i$ both uncertain and data-dependent (like with the \neighmech), suppose $a_1,\dots, a_n$ are samples from an Inverse Gamma IG$(2+X/c, X+X^2/c)$ distribution, where $c>0$ is some constant. Then $E[a_i]=X$ and $Var(a_i)=cX$. Thus the variance estimate is $v_i=a_i/c$. This relation between the mean and estimated variance is similar to when $\neighfun=\sqroot$ (see Theorem \ref{thm:unbiased}).  Solving Equation \ref{eqn:microdata}, the guess for employment count $X$ is $\frac{\sum_{i=1}^n a_i/v_i}{\sum_{j=1}^n 1/v_j}=\frac{n}{\sum_j 1/a_j}$. We use the following facts: (1) if $v_i$ has distribution $IG(\alpha,\beta)$ then $1/v_i$ has the Gamma$(\alpha,\beta)$ distribution, and (2) the sum of $n$ independent Gamma$(\alpha,\beta)$ random variables is a Gamma$(n\alpha,\beta)$ distribution. Then the expected value of the guess is $\frac{n(2+X/c)-n}{n(2+X/c)-1}$, which is slightly \emph{less} than $X$; i.e., the guess has a downward bias.

The above analysis suggests that the \pncmech (whose true variance is publicly releasable) is preferable to the \neighmech (whose variance is \underline{both} data dependent and must be approximated) when bias in the confidentiality-preserving microdata $\widehat{D}$ is a concern. Further numerical results can be found in Appendix \ref{app:bias} and Section \ref{sec:comps}.

\paragraph*{Useful properties about distributions.}
These are the facts needed about the distributions used in the analytical analysis of inverse weighted least squares.
\begin{enumerate}
    \item The Inverse Gamma distribution $IG(\alpha, \beta)$ with shape parameter $\alpha$ and scale parameter $\beta$ has mean $\frac{\beta}{\alpha-1}$ (for $\alpha>1$) and variance $\frac{\beta^2}{(\alpha-1)^2(\alpha-2)}$ (when $\alpha>2$). Note, the variance is the mean squared divided by $(\alpha-2)$.
    \item The Gamma distribution Gamma$(\alpha,\beta)$ with shape parameter $\alpha$ and rate parameter $\beta$ has mean $\frac{\alpha}{\beta}$ and variance $\frac{\alpha}{\beta^2}$.
    \item If $v$ follows the IG$(\alpha,\beta)$ distribution, then $1/v$ follows the Gamma$(\alpha, \beta)$ distribution with shape parameter $\alpha$ and rate parameter $\beta$.
    \item If $1/v_1,\dots, 1/v_n$ are independent and each $1/v_i$ follows the Gamma$(\alpha_i, \beta)$ distribution (with the same $\beta$ but possibly different $\alpha_i$) then $\sum_{i=1}^n 1/v_i$ follows the Gamma$(\sum_{i=1}^n \alpha_i, \beta)$ distribution.
    \item If $1/v_1,\dots, 1/v_n$ are independent and each $1/v_i$ follows the Gamma$(\alpha_i, \beta)$ distribution then the joint distribution of the $n$ quantities $\frac{1/v_1}{\sum_{j=1}^n 1/v_j},\dots, \frac{1/v_n}{\sum_{j=1}^n 1/v_j}$ is the Dirichlet$(\alpha_1,\dots, \alpha_n)$ distribution.
    \item If the joint distribution of $w_1,\dots, w_n$ is Dirichlet$(\alpha_1,\dots, \alpha_n)$, then $E[w_i]=\frac{\alpha_i}{\sum_{j=1}^n \alpha_j}$ and $Var(w_i)=\frac{\frac{\alpha_i}{\sum_{j=1}^n \alpha_j}\left(1-\frac{\alpha_i}{\sum_{j=1}^n \alpha_j}\right)}{1+\sum_{i=j}^n\alpha_j}$ and 
    \begin{align*}
    E[w_i^2] &=Var(w_i) + E[w_i]^2\\
    &=\frac{\alpha_i^2}{(\sum_{j=1}^n \alpha_j)^2} + \frac{\frac{\alpha_i}{\sum_{j=1}^n \alpha_j}\left(1-\frac{\alpha_i}{\sum_{j=1}^n \alpha_j}\right)}{1+\sum_{i=j}^n\alpha_j}\\
    &=\frac{\alpha_i^2 + \alpha_i^2\sum_{j=1}^n \alpha_j + \alpha_i(-\alpha_i + \sum_{j=1}^n \alpha_j)}{(\sum_{j=1}^n \alpha_j)^2(1+\sum_{i=j}^n\alpha_j)}\\
    &= \frac{\alpha_i^2\sum_{j=1}^n \alpha_j + \alpha_i \sum_{j=1}^n \alpha_j}{(\sum_{j=1}^n \alpha_j)^2(1+\sum_{i=j}^n\alpha_j)}\\
    &=\frac{\alpha_i^2 + \alpha_i }{(\sum_{j=1}^n \alpha_j)(1+\sum_{i=j}^n\alpha_j)}\\
    &=\frac{\alpha_i(1 + \alpha_i) }{(\sum_{j=1}^n \alpha_j)(1+\sum_{i=j}^n\alpha_j)}
    \end{align*}
\end{enumerate}

\paragraph*{Analysis of Case 2:}
    Recall $a_i,\dots, a_n$ are random variables where $E[a_i]=X$ and $Var(a_i)=\sigma^2$. Recall that each $v_i$ has distribution IG$(2+\tau,\sigma^2(1+\tau))$. Let $\widehat{X}=\frac{\sum_{i=1}^n a_i/v_i}{\sum_{j=1}^n 1/v_j}$. Then
    \begin{align*}
        E[\widehat{X}] &= E\left[\frac{\sum_{i=1}^n a_i/v_i}{\sum_{j=1}^n 1/v_j}\right]
        = \sum_{i=1}^n E\left[\frac{ a_i/v_i}{\sum_{j=1}^n 1/v_j}\right]\\
        &=\sum_{i=1}^n E\left[\frac{ X/v_i}{\sum_{j=1}^n 1/v_j}\right] =  X E\left[\frac{\sum_{i=1}^n 1/v_i}{\sum_{j=1}^n 1/v_j}\right]=X\\
        E\left[\left(X - \frac{\sum_i a_i/v_i}{\sum_j 1/v_j}\right)^2\right] 
        &=E\left[\left(\frac{\sum_i (X-a_i)/v_i}{\sum_j 1/v_j}\right)^2\right]\\ 
        &=\sum_i E\left[\frac{ (X-a_i)^2/v_i^2}{(\sum_j 1/v_j)^2}\right]\\ 
        &\text{(because the cross terms have expected value}\\
        &\text{0 when expanding the square)}\\
       &= n\sigma^2 E\left[\frac{1/v_1^2}{(\sum_j 1/v_j)^2}\right]\\
       &\text{(because the $v_i$ have the same distribution)}\\
       &\text{($v_1$ has IG$(2+\tau, \sigma^2(1+\tau))$ distribution)}\\
       &\text{($1/v_1$ has Gamma$(2+\tau, \sigma^2(1+\tau))$ distribution)}\\
       &\hspace{-2.5cm}{\text{($\frac{v_1}{\sum_j 1/v_j},\dots, \frac{v_n}{\sum_j 1/v_j}$ has Dirichlet$(2+\tau,\dots,2+\tau)$ distribution)}}\\
       &= n\sigma^2\frac{2+\tau}{n(2+\tau)}\frac{1+2+\tau}{1+n(2+\tau)}\\
       &= \sigma^2\frac{1+2+\tau}{1+n(2+\tau)}\\
       &= \frac{\sigma^2}{n}\frac{n+n(2+\tau)}{1+n(2+\tau)}\\
       %
    \end{align*}

\subsection*{Analysis of Case 3: }  
Let $a_1,\dots, a_n$ be samples from an Inverse Gamma IG$(2+X/c, X+X^2/c)$ distribution, where $c>0$ is some constant.
Then
\begin{align*}
    E\left[\frac{n}{\sum_j 1/a_j}\right] &= n E[1/\text{Gamma}(n(2+X/c), X+X^2/c)]\\
    &=n E\left[\text{IG}(n(2+X/c), X+X^2/c)\right]\\
    &=n\frac{X+X^2/c}{n(2+X/c)-1}\\
    &=X\frac{n(2+X/c)-n}{n(2+X/c)-1}\equiv \theta\\ 
    E\left[\left(X-\frac{n}{\sum_j 1/a_j}\right)^2\right] &= E\left[\left(X-\theta + \theta -\frac{n}{\sum_j 1/a_j}\right)^2\right] \\
    &=(X-\theta)^2 + E\left[\left(\theta - \frac{n}{\sum_j 1/a_j}\right)^2\right]\\
    &\text{(cross terms have 0 expected value)}\\
    &=(X-\theta)^2 + Var(\text{IG}(n(2+X/c), X+X^2/c))\\
\end{align*}
\end{textAtEnd}

\section{Experiments}\label{sec:comps}

We evaluate our proposed methods on real and synthetic data. The real dataset is a  historical QCEW dataset which we used via a cooperative agreement with the BLS. The state, year, and quarter cannot be disclosed for legal and policy reasons. 
We generated synthetic  data  by combining publicly available data sources \cite{rawcbp,rawqwi,imputecbpdata} with statistical techniques like imputation; the details of this process can be found in Appendix \ref{app:synthetic}. The synthetic data, the code to generate it, and the code to evaluate it can be found online \citep{githubRep}. 
In both datasets, the public attributes of an establishment are its NAICS code and county, while the 4 private attributes are the employment counts for the first, second and third months in the quarter, and the total wages paid during the quarter.  The neighbor function for \gedp is set to $\neighfun=\sqroot$. We compare the accuracy of (1) privacy-preserving microdata generated from queries entirely answered by the \sqrtmech (i.e., \neighmech with $\neighfun=\sqroot$) combined with Theorem \ref{thm:unbiased}, followed by conversion into microdata to (2) privacy-preserving microdata generated from queries answered by the \pncmech (except for the identity query, which must be answered by the \sqrtmech, as explained in Section \ref{subsec:pnc}).

\subsection{Evaluations on Confidential QCEW Data}\label{sec:experiments_realQCEW}
\begin{table}[!htp]
\addtolength{\tabcolsep}{-0.3em}
\small
\begin{tabular}{|r|rrrr|rrrr|}
\hline
& \multicolumn{4}{c|}{\bf{\sqrtmech}}  & \multicolumn{4}{c|}{\bf{\pncmech}} \\
      Level   & 1st & Median & Mean & 3rd & 1st & Median & Mean & 3rd \\
  \hline
  \multicolumn{9}{|l|}{\bf{State Level Employment}}\\ \hline
Total  & -6,198 & -6,198 & -6,198 & -6,198 & -118 
& -118 & -118 & -118 \\

Supersector& -1,083 & -293 & -563 & -233 & -100 & -48 & -11 & 24 \\

NAICS-5& -32 & -9 & -9 & 8 & -14 & -1 & -0 & 12 \\ \hline
\multicolumn{9}{|l|}{\bf{County Level Employment}}\\  \hline
Total&  -420 & -252 & -4,604 & -143 & -64 & -11 & -4,331 & 27 \\

Supersector& -69 & -18 & -419 & 16 & -43 & -4 & -394 & 30 \\

NAICS-5& -7 & -1 & -10 & 4 & -6 & -1 & -9 & 5  \\
\hline
\end{tabular} \caption{Quartiles of \emph{signed} differences between group-by query answers computed from confidentiality-preserving microdata vs. ground truth. Note super-sector is essentially 2-digit NAICS prefix.}
 \label{tab:qcew_comp_diff}
 \addtolength{\tabcolsep}{0.3em}
\end{table} 

 \begin{figure}[h] 
\includegraphics[width=0.98\linewidth,clip=true,trim=0.5cm 0.5cm 1.2cm 0.8cm]{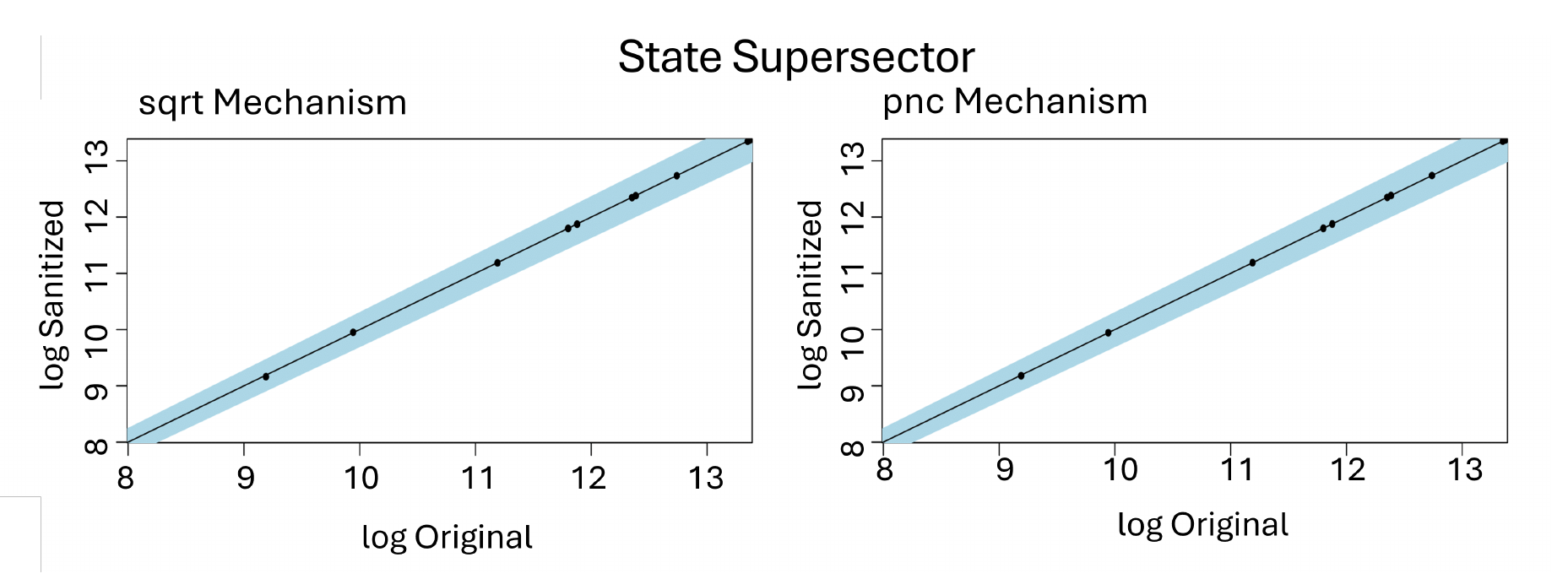}

\includegraphics[width=0.98\linewidth,clip=true,trim=0.5cm 0.5cm 1.0cm 0.8cm]{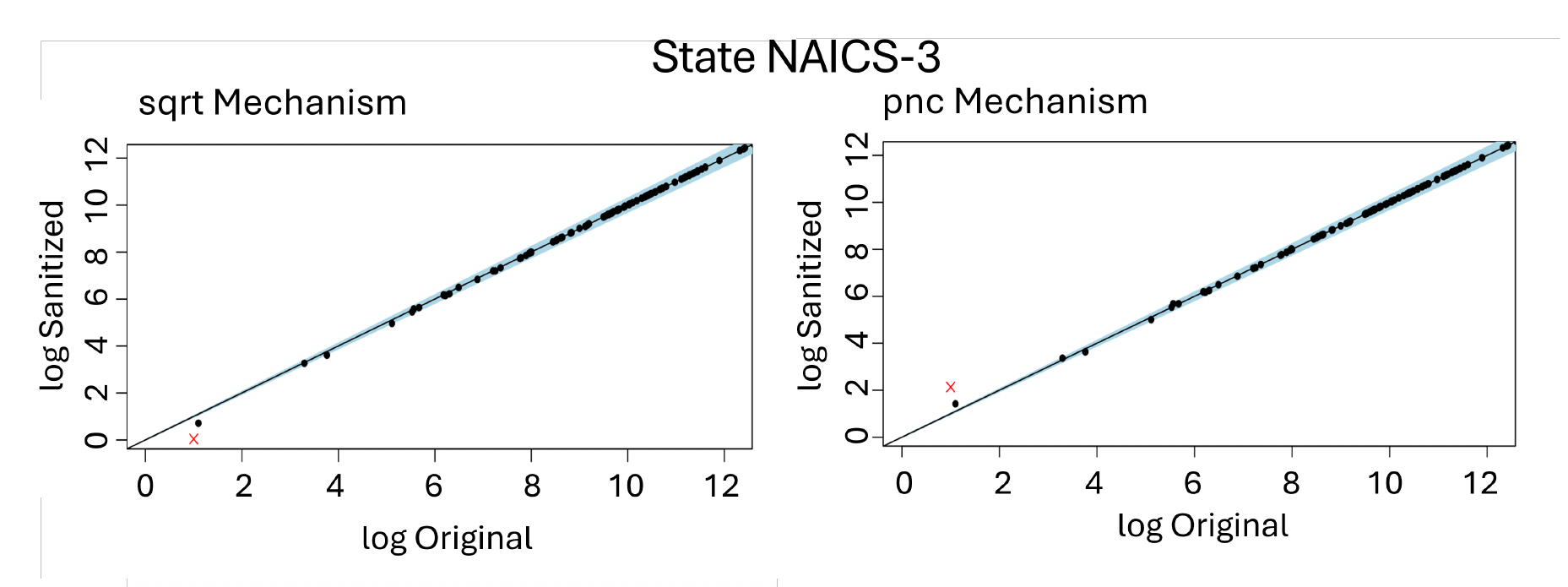}

\includegraphics[width=0.98\linewidth,clip=true,trim=0.5cm 0.5cm 1.0cm 0.8cm]{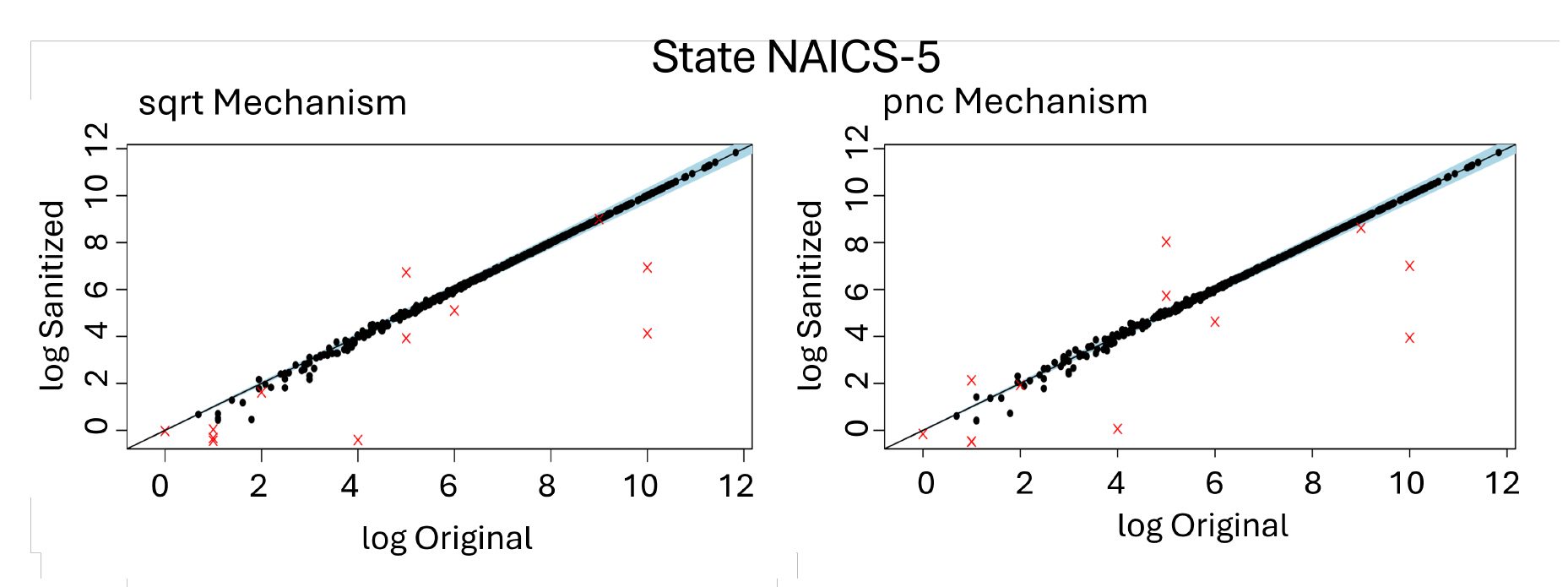}
\caption{(Employment) Scatter plot of confidentiality-preserving microdata vs. ground truth for the groups in group-by queries for the workflow based on the \sqrtmech (first column) and \pncmech (second column).  The blue band shows $\pm3\%$ of the original value. Points outside this band are colored red. 
Only the smallest groups (where error must be large to protect confidentiality) have values that are further than $3\%$ from the ground truth.}
\label{fig:agg55}
\end{figure}

We next describe the settings for the experiment on QCEW data for which we have permission to share results, which only cover employment data (not wages).
The \textbf{measurement queries} are the group-by queries for which the mechanisms produced noisy answers for the 3 monthly employment counts using the following groupings: (Q1) Identity -- no grouping, (Q2) State total -- grouping all establishments in the (redacted) state covered by the data, (Q3) 5-digit NAICS prefix, (Q4) County, (Q5) County by 5-digit NAICS prefix. For each of the 3 months, the respective query budgets were: (Q1) $0.7$, (Q2) $0.2$, (Q3) $0.6$, (Q4) $0.6$, (Q5) $0.7$. This results in an overall privacy budget (square root of the sum of squares) $\queryprivbudget{}=\sqrt{0.7^2*3+0.2^2*3+0.6^2*3+0.6^2*3+0.7^2*3}\approx 2.28$. The distance parameter for employment was $\distparam_{emp}=0.5$. The \pncmech confidence parameter was $\confparam=0.01$ (i.e., with probability $0.99$, no group-by query was changed by clipping). The noisy measurement query answers are postprocessed into confidentiality-preserving microdata. The \textbf{evaluation queries} are the group-by queries for which we evaluated accuracy -- they can be different from the measurement queries.

Table~\ref{tab:qcew_comp_diff} shows the \emph{signed} difference between employment group-by-queries  computed from the confidentiality-preserving microdata and the ground truth (i.e., we don't take the absolute value). Negative numbers indicate a negative difference (i.e., downward bias). In addition to state total, the groupings used for evaluation are supersector (2-digit NAICS prefix) and 5-digit NAICS prefix. The table summarizes the differences by computing the quartiles across groups in the group-by-queries (for state total, all establishments are grouped together into 1 group, so all quartiles are the same). We see that using the \pncmech produces greater accuracy and less bias. This serves as additional empirical validation for Section \ref{sec:microdata}, as it shows the \pncmech reduces bias in the microdata. Some bias is still present because usage of the \pncmech requires that the identity query be answered by the \neighmech (with all other group-by queries answered by the \pncmech).

Figure~\ref{fig:agg55} shows a scatter-plot of true vs. noisy answers for individual groups within employment group-by-queries (supersector, 3-digit NAICS prefix, 5-digit NAICS prefix), with a 3\% band around the true value. Points outside the band are colored red. The results show both methods are accurate. All but the smallest groups have answers within $3\%$ of the true value. The groups with small employment counts, especially if they contain few establishments, would raise the most confidentiality concerns, and therefore should  exhibit large errors. This is reflected by more red points in fine-grained groupings (e.g., group by  5-digit NAICS prefix) compared to coarser groupings (e.g., group by 3-digit NAICS prefix).

\subsection{Evaluations on Synthetic Data}\label{sec:experiment_synth}

\begin{figure*}[!htp]

\includegraphics[width=0.3\linewidth,clip=true,trim=0.5cm 0.5cm 0cm 0cm]{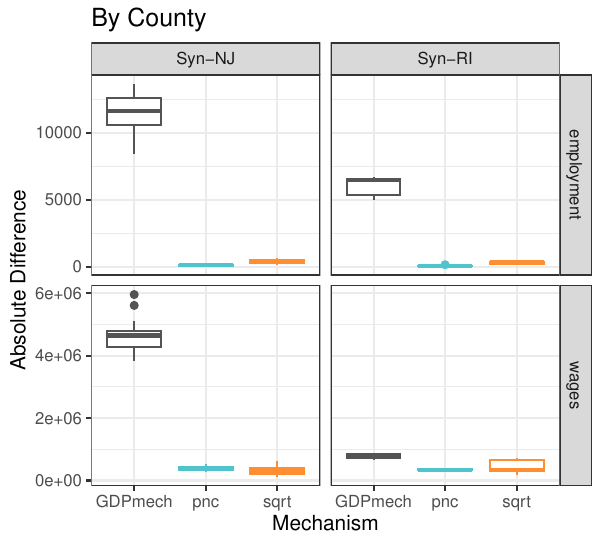}
\includegraphics[width=0.3\linewidth,clip=true,trim=0.5cm 0.5cm 0cm 0cm]{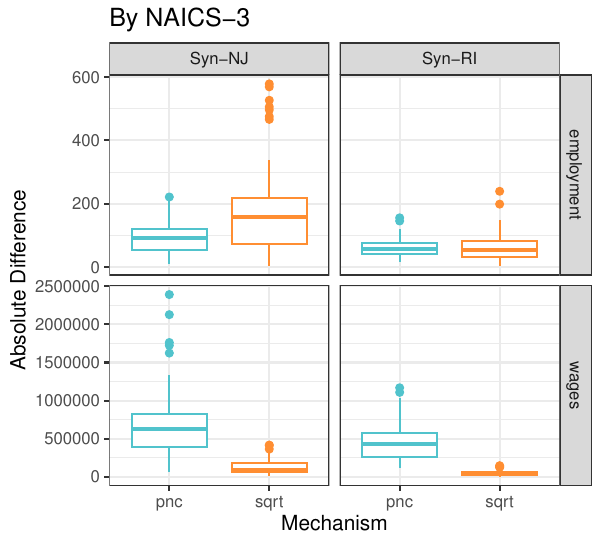}
\includegraphics[width=0.3\linewidth,clip=true,trim=0.5cm 0.5cm 0cm 0cm]{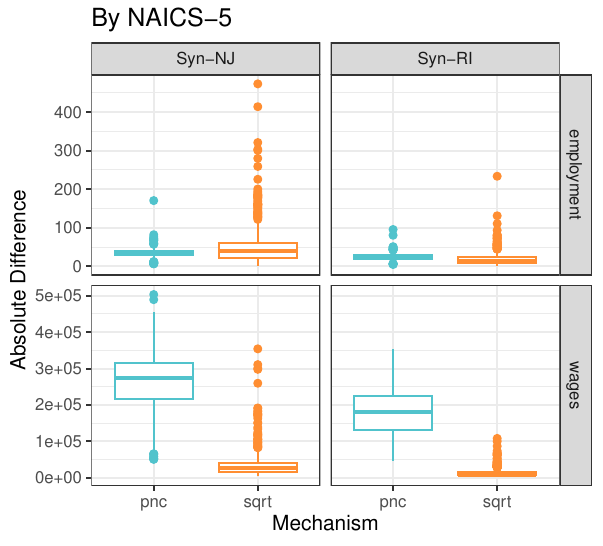}
\caption{Absolute error averaged across 34 replications for employment group-by queries (top) and wages (bottom) for Synthetic New Jersey and Synthetic Rhode Island data.}\label{fig:eqempwage}
\end{figure*}

Since experiments on confidential QCEW data are limited by policy, we synthesized a dataset with the same schema using public information \cite{rawcbp,rawqwi,imputecbpdata} and statistical imputation (see the Appendix \ref{app:synthetic} for details). 
We use synthetic New Jersey (Syn-NJ) and Synthetic Rhode Island (Syn-RI) data, which serve as examples of fairly large and fairly small states, for further experimentation. This allows us to add results for wages as well. \shortlong{Due to space constraints, only a subset of our results is presented.}{}

\begin{figure}[!htp]
\includegraphics[width=0.63\linewidth,clip=true,trim=0.5cm 0.53cm 0cm 0cm]{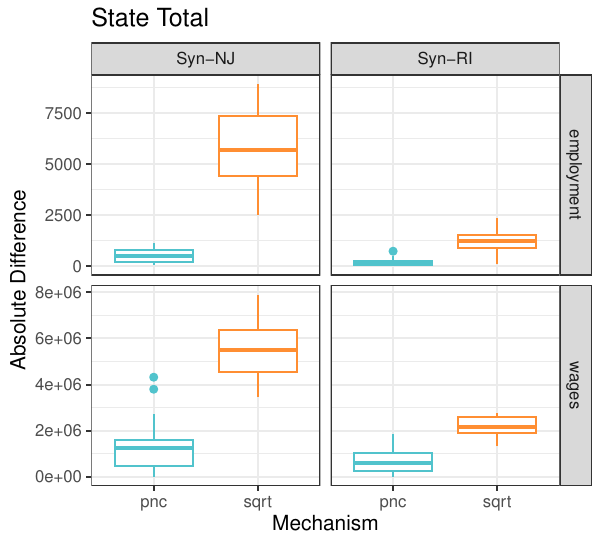}
\caption{Absolute error of trivial grouping at state-level for employment (top) and wages (bottom) for Synthetic New Jersey and Synthetic Rhode Island data across 34 replicates.}\label{fig:eqempwage2}
\end{figure}

Synthetic New Jersey (Syn-NJ) includes 208,868 establishments and 21 counties. 
The number of establishments in a county ranges from 999 to 28,689 with mean 9,894, median 9,385 and standard deviation 6,968. 
The number of establishments for different NAICS codes present in Syn-NJ ranges from 1 to 8,172 with mean 364 and median 79 -- so this distribution is highly skewed. The largest establishment's employment is 7,629 and the largest wages is 3,136,080. 
%
Synthetic Rhode Island has only 5 counties and 25,420 establishments. The number of establishments in a county ranges from 1,102 to 14,414 with mean 5,084, median 3,354, and standard deviation 5,340. The number of establishments for different NAICS codes present in Syn-RI ranges from 1 to 610, with mean 13, median 4 and standard deviation 34.  The largest Syn-RI establishment's employment is 4,050 and the largest wages is 510,039.
As a baseline, we can compare against the Gaussian mechanism of GDP at the establishment level (i.e., providing equal defense against membership inference for all establishments). We use the values of the largest establishment in each state as sensitivity bounds used by the GDP Gaussian mechanism (giving it a slight non-private advantage).

For the privacy settings, we set the privacy budget allocation for wages to be the same  as  the QCEW allocations for employment  in Section \ref{sec:experiments_realQCEW} and then rescaled everything so that the total privacy budget was nearly the same as in Section \ref{sec:experiments_realQCEW}. Specifically, each month of employment, and also the quarterly wages had the following privacy budgets for the group-by query groupings: (Q1) Identity -- no grouping: 0.611, (Q2) State total -- grouping all establishments in the same: 0.179, (Q3) 5-digit NAICS prefix: 0.525, (Q4) County: 0.525, (Q5) County by 5-digit NAICS prefix: 0.611. We used the distance parameter $\distparam_w=50$. The overall privacy budget was $\queryprivbudget{}=\sqrt{0.611^2*4 + 0.179^2*4 + 0.525^2*4 + 0.525^2*4 + 0.611^2*4}\approx 2.31$.

For each workflow (\pncmech vs. \sqrtmech   vs. GDP mechanism), we compute the absolute error of query answers and then repeat the experiment for 34 repetitions in order to assess variability of the errors. 

Figures \ref{fig:eqempwage} and \ref{fig:eqempwage2} compare the absolute errors for employment and wages on a variety of group-by queries on Synthetic New Jersey and Rhode Island. The figures summarize the errors using a box plot, which shows the median, upper and lower quartiles (the top and bottom of the box) and the spread of points outside the quartiles (the whiskers and dots outside the boxes). Figure \ref{fig:eqempwage} summarizes the absolute error for each group within a query is averaged across the replicates. The group-by queries cover (1) grouping by county to get county-level employment and wage totals, (2) grouping by 3-digit NAICS prefix -- this is an important query that is not a
measurement query, and (3) grouping by 5-digit NAICS prefix -- this is an important query and is expected to be sparse (i.e., many of the groups have very few establishments). Figure \ref{fig:eqempwage2} summarizes the absolute error from each replicate of the trivial grouping in which all establishments are combined together to get state-level employment and wages totals. 

 Both of our proposed mechanisms perform better than the GDP mechanism (left plot of Figure \ref{fig:eqempwage}). The scale of the noise from the GDP mechanism renders the data fairly useless and thus is omitted from the rest of our discussion. For employment counts (the top rows of the Figures \ref{fig:eqempwage} and \ref{fig:eqempwage2} ), the \pncmech performs substantially better than \sqrtmech. This is because the amount of noise injected into a group by \sqrtmech depends on the total within that group. However, the amount of noise added to a group by \pncmech depends on the value of the largest establishment in the group. Specifically, $\pncmech$ uses the noisy identity query to estimate simultaneous high-probability upper bounds on each establishment and then clips each establishment's wages and employment to ensure that the upper bounds are not exceeded. It adds up the clipped values in a group,\footnote{Since the upper bound is high probability, it means that all of the clipped values are probably identical to the true values} and then adds noise based on the largest establishment upper bound in the group. If no establishment dominates the group (i.e., is much larger than the rest), then the largest of the establishment upper bounds estimated by \pncmech is significantly smaller than the total within a group, leading to less noise compared to \sqrtmech. 

Thus, for high-level aggregations like state total (Figure \ref{fig:eqempwage2}), there are so many establishments in a single group that no establishment can dominate the employment or wages. Hence \pncmech significantly outperforms \sqrtmech on state totals. However, as the number of establishments in a group decreases, a single establishment  is more likely to dominate the total within its group, giving the \pncmech an advantage over the \sqrtmech.

For medium-level aggregations, like  County  (first plot of Figure \ref{fig:eqempwage}), there is an increasing chance that some groups in the group-by queries contain a single establishment that can dominate the rest. Hence, the advantage of \pncmech over \sqrtmech starts to decrease. It still performs better on employment and also Syn-RI wages, but the \sqrtmech is slightly better for Syn-NJ wages.

The comparison for the NAICS-3 (second plot of Figure \ref{fig:eqempwage}) and NAICS-5 (last plot of Figure \ref{fig:eqempwage}) is particularly interesting because NAICS-5 is a sparse group-by query (this should benefit the \sqrtmech) containing many groups, and NAICS-3 is a non-measure-ment query (all the information about it contained in the confidentiality-preserving microdata mostly comes from the noisy answers to the group-by 5-digit NAICS prefix query). Surprisingly, we continue to see that \pncmech outperforms \sqrtmech on employment counts for NAICS-5 and NAICS-3, although the difference between the performance of the two mechanisms has become very small. However, for total wages grouped by NAICS-3 or NAICS-5, we see that \sqrtmech clearly outperforms \pncmech. This suggests that wages data are more highly skewed than employment count and an establishment is more likely to dominate the total wages in its industry than the total employment.

The exact relationship between a group (in a group-by query) and which mechanism works best for it depends on the skewness in a group, the distance parameters,  and privacy parameters. An interesting direction for future work is to use the confidentiality-preserving answers to the identity query to help estimate skewness in all groups of all group-by queries and (along with the public privacy parameters) to determine whether the \pncmech or \sqrtmech should be used for a particular query.

\section{Conclusions and Future Work}\label{sec:conc}

In this paper, we introduced \gedp, a privacy definition for establishment data, designed to be interpretable to mathematical statisticians in statistical agencies.  We proposed two mechanisms, the \neighmech with non-releasable data-dependent variance, and the \pncmech with \emph{releasable} data-dependent variance. 

Future work includes mechanisms that answer queries over multiple confidential attributes.
Another interesting direction is how to optimize a workload -- adaptively selecting measurement queries and privacy budgets in order to maximize accuracy. It is likely to be an iterative process, where prior query answers are used to estimate the expected gain in accuracy that can be obtained from candidate future measurement queries.

\begin{acks}
This work was supported by a NSF BAA grant (NCSES-BAA 49100421C0022 to The Pennsylvania State University). We thank David Talan for help with data access and Spencer Jobe for help with running our experiments on internal BLS data.
\end{acks}

\bibliographystyle{ACM-Reference-Format}
\bibliography{main}

\appendix

\clearpage
\onecolumn
\appendix
\section{Analyzing attacker error due to sampling.}\label{app:trade}
The following example demonstrates why $\psi =\sqrt{}$ makes statistical sense and how it relates to estimating establishment count. 
\begin{example}
    Suppose an establishment $\estab$ has $x$ employees and a statistician (e.g., an attacker or labor economist) wants to estimate its employment count. 
    Suppose nearly all employees of  $\estab$ live within a region whose total population, $N$, is known with high accuracy (e.g., from Census data). For some $q\in[0, 1]$, the statistician may take a random sample of $n=Nq$ people in the region and record the proportion $p$ that work for $\estab$. The quantity $np$ can be modeled as the sum of $n$ Bernoulli$(x/N)$ random variables. So, an unbiased estimate of the number of employees of $\estab$ is $\hat{x}=np(N/n)$. The variance of this estimate is $\Var(\hat{x})=n(x/N)(1-x/N)N^2/n^2=x\frac{N-x}{n}\leq \frac{1}{q}x$. The standard deviation is therefore $O(\sqrt{x})$ and confidence intervals would therefore have lengths $O(\sqrt{x})$.
\end{example}

\section{Discussion of Cell Suppression Weaknesses}\label{app:cell}
The BLS is obligated to use statistical disclosure limitation methods to protect sensitive values (employment and wages) of individual establishments. Internal policy and federal law, specifically the Confidential Information Protection and Statistical Efficiency Act (CIPSEA) of 2002, require confidentiality protections to all data acquired and maintained by the BLS under a pledge of confidentiality. 

The current disclosure avoidance method for the QCEW is cell suppression.
According to the cell suppression methodology \cite{adam1989security}, sensitive cells are initially identified by either the $p\%$ rule, the $qp$-rule or the $nk$-rule \cite{schm2020}. The $p\%$ rule, for example, considers the question of whether the second largest establishment in a cell can use the cell total to learn the largest establishment's confidential value up to \textbf{$p\%$ relative error} (a cell is flagged as sensitive if the cell total minus the contributions of the two largest establishments is less than $p\%$ of the largest establishment). These cells are suppressed in a step called \emph{primary suppression}. The \emph{secondary suppression} phase then identifies combinations of cells from which the suppressed cells can be deduced via linear algebra. This phase  removes additional cells so that such combinations no longer exist. 

Primary and secondary suppression procedures result in the BLS suppressing the majority of the cells before they are released \citep{blshandbook}. Over 60\% of more than 3.6 million cells in the QCEW tables are suppressed \citep{cohen2006, yang_et_al_11, toth2014}.  Apart from  concerns about loss of data, secondary suppression can be complex to implement and often requires solving computationally expensive linear programs \cite{cox1980}.

Cell suppression also has two major confidentiality vulnerabilities. First, it is brittle against external knowledge --
it is possible for, say Pennsylvania, to release additional data about its establishments that, when combined with cell-suppressed QCEW data, allows attackers to reverse engineer other cells (e.g., compromising confidentiality for businesses in New York). Second, cell suppression is only designed to protect against \emph{linear attackers} -- those who only use addition, subtraction, and scalar multiplication to attack data. It is not designed to protect against more sophisticated statistical attacks \cite{holantoth10}, which is why the $p$ parameter from the $p\%$ rule is also kept confidential. It is therefore not possible to realistically determine what confidential values might become vulnerable to disclosure when new types of data tabulations are published.  This prevents the BLS from publishing new statistics and analyses from the QCEW microdata because of the unknown ways that new data products could impact the disclosure protections of existing products.

\begin{table}[ht]
\caption{Establishment microdata and three marginals computed from it. Primary suppression must remove the cells for County X and Sector 52 (red), as they are directly revealing of the wages \bob. Secondary suppression is needed to prevent $\bob$ from being reconstructed by the remaining cells.}\label{tab:suppression}
\begin{tabular}{|lccc|}\hline
\textbf{Establishment} &\textbf{County} &\textbf{Sector} & \textbf{Wages}\\\hline
Bob's Building &X &52 &\bob\\
Carlos's Construction &Y&23 &\wendy\\
Green Hills Farming & Y & 11 & \pickles\\
Isaac's Foundations & Z & 23 & \fred\\
Alice's Apples & Z & 11 & \alice\\\hline
\end{tabular}

\vspace{0.5em}
\begin{tabular}{|c|}
\multicolumn{1}{c}{\textbf{Total Wages}}\\\hline
$\bob\! +\! \wendy\! +\! \pickles\! +\! \fred\! + \!\alice$
\\\hline
\multicolumn{1}{c}{\phantom{A}}\\
\end{tabular}$~$
\begin{tabular}{|c|c|c|}
\multicolumn{3}{c}{\textbf{By County}}\\\hline
\textbf{X} & \textbf{Y} & \textbf{Z}\\\hline\cline{1-1}
\textcolor{red}{\bob}  & $\wendy\!+\! \pickles$ & $\fred\! +\!\alice$\\\hline\cline{1-1}
\end{tabular}$~$
\begin{tabular}{|c|c|c|}
\multicolumn{3}{c}{\textbf{By Sector}}\\\hline
\textbf{52} & \textbf{23} & \textbf{11}\\\hline
\textcolor{red}{\bob}  & $\wendy\! +\!\fred$ &  $\alice\!+\!\pickles$\\\hline
\end{tabular}
\end{table}

 Table \ref{tab:suppression} shows a simplified microdata set and 3 marginals computed from it: total wages, wages by county, and wages by sector (2-digit NAICS code) prefix. The values in cells for County X and also for Sector 52 (displayed in red) are directly disclosive because they only contain one establishment, and hence would undergo primary suppression.
 
 The secondary suppression phase then identifies which non-suppressed cells can be used to deduce the cells suppressed in the first phase, or can isolate additional establishments. For example, the total wages minus the combined wages of counties Y and Z, or total wages minus the combined wages of sectors 23 and 11 recover the wages \bob. Meanwhile, the wages in counties Y and Z along with the wages in sectors 23 and 11 form a linear system of 4 equations and 4 unknowns and hence is uniquely solvable. A subset of these problematic cells needs to be suppressed with the goal that the remaining cells cannot be used to reconstruct sensitive values. Thus, for example, one could additionally suppress the total wages table, along with total wages of County Y. All that remains to be published are the wages of County Z, Sector 23, and Sector 11. In our example 50\% of the cells end up being suppressed. This is not an anomaly. Primary and secondary suppression procedures results in the BLS suppressing the majority of the cells before they are released \citep{blshandbook}. 

\section{Deriving Variance from Power and Significance}\label{app:sigcomp}

We can frame differential privacy definitions through power and significance level similar to the GDP framework in Section \ref{sec:background}. Let $y$ be the output of a privacy-preserving mechanism $\mech$. Consider the hypothesis test:
\begin{align}
    H_0:&y \text{ from }\mech(\data_1)& H_1:y\text{ from }\mech(\data_2) \label{eq:hyp_test}
\end{align}
for any pair of neighboring $\data_1$ and $\data_2$.

Fixing a significance and power level, we calculate the smallest variance for the privacy noise added by a standard mechanism to satisfy pure DP, Gaussian DP and zCDP.

\subsection{Pure DP}

If a mechanism,$\mech$, satisfies $\epsilon$-DP, also known as pure differential privacy, by definition 
$$P(\mech(\data_1)\in S)\leq e^{\epsilon}P(\mech(\data_2)\in S)$$
for any pair of neighboring datasets, $\data_1,\data_2$ and $S\in \operatorname{Range}(\mech)$ \cite{dwork2006calibrating}.

Since this result must hold for any pair of neighboring datasets and any $S$ in the range of $\mech$. Let $y$ be the output of $\mech(D_1)$ and let $S$ be the rejection region. By definition of significance, $P(\mech(\data_1)\in S)=\alpha$, where $\alpha$ is the significance level (or Type I error). Thus $\alpha\leq e^\epsilon P(\mech(\data_2)\in S)$. However, $P(\mech(\data_2)\in S)\leq 1-\beta$, by definition of power. Since the definition must hold for any $S$ in the range of $\mech$, a similar result holds for the complement of $S$\footnote{This result is also derived by Wasserman \& Zhou \cite{wasserman10}.}.

\begin{align}
    1-\alpha&\leq e^\epsilon \beta\label{eq:ineq1}\\
    \alpha &\leq e^{\epsilon}(1-\beta)\label{eq:ineq2}
\end{align}

If we fix significance level $\alpha$ and power ($1-\beta$), we find an lower bound for $\epsilon$. 

From \eqref{eq:ineq1} we have
\begin{align*}
    \frac{1-\alpha}{\beta}\leq e^\epsilon \implies \log\left(\frac{1-\alpha}{\beta}\right)\leq \epsilon
\end{align*}

From \eqref{eq:ineq2}, we have
\begin{align}
    \alpha\geq e^{-\epsilon}(1-\beta) \implies \epsilon\geq \log\left(\frac{1-\beta}{\alpha}\right)
\end{align}
Thus $\epsilon\geq \max\left(\log\left(\frac{1-\alpha}{\beta}\right),\log\left(\frac{1-\beta}{\alpha}\right)\right)$

Let $m$ be the number of queries. By composition properties of pure DP, each query should have a $\epsilon_i= \frac{\max\left(\log\left(\frac{1-\alpha}{\beta}\right),\log\left(\frac{1-\beta}{\alpha}\right)\right)}{m}$ to obtain the fixed significance and power levels.

If $\mech$ is the commonly used Laplace mechanism \cite{dwork2006calibrating}, then the noise added to each query result has variance $2\left(\frac{m}{\epsilon_i}\right)^2$. 

\subsection{GDP}

Once again, let $S$ be a rejection region and $y\gets \mech(\data_1)$. A mechanism $\mech$ satisfies $\mu$-GDP if
$$\Phi^{-1}(P(\mech(\data_1) \in S)) \leq \mu + \Phi^{-1}( P(\mech(\data_2) \in S) )$$
Thus
$$\Phi^{-1}(\alpha) \leq \mu + \Phi^{-1}( 1-\beta )$$
This means $\mu=\Phi^{-1}(\alpha)-\Phi^{-1}(1-\beta)$ is the smallest privacy parameter that satisfies the given significance and power levels. For $m$ queries, by composition properties each query must satisfy $\mu_i=\mu/\sqrt{m}$. The Gaussian mechanism satisfies $\mu_i$-GDP if the variance of the infused noise is $1/(\mu_i^2)$.

\subsection{zCDP}

We omit the original definition of $\rho$-zCDP, but include the hypothesis testing interpretation of it \cite{zCDP_hypTest}. Once again, let $S$ be a rejection region and $y\gets \mech(\data_1)$. A mechanism $\mech$ satisfies $\rho$-zCDP if
$$ P(\mech(\data_1)\in S)^\eta P(\mech(\data_2)\in S)^{1-\eta} + P(\mech(\data_1)\notin S)^{\eta} + P(\mech(\data_2)\notin S)^{1-\eta} \leq e^{\rho\eta(\eta-1)}
$$  
for any $\eta>1$ \cite{zCDP_hypTest}. This is equivalent to
\begin{equation}
    \alpha^\eta(1-\beta)^{1-\eta}+(1-\alpha)^\eta\beta^{1-\eta}\leq e^{\rho \eta(\eta-1)} \label{eq:zcdp}
\end{equation}
Thus 
$$\rho\geq \max_{\eta>1}\left(\frac{\log\left(\alpha^\eta(1-\beta)^{1-\eta}+(1-\alpha)^\eta\beta^{1-\eta}\right)}{\eta(\eta-1)}\right)$$

This is solved numerically to find the smallest $\rho$ for a fixed power and significance level. Denote this solution as $\rho^*$. By composition properties, for $m$ queries, each query should satisfy $\rho_i=\rho^*/m$. 

A typical mechanism that satisfies $\rho_i$-zCDP is the Gaussian mechanism with variance $1/2\rho_i$ \cite{zcdp}.


\section{Proofs from Section \ref{sec:model}}\label{app:model}
\printProofs[model]

\section{Proofs from Section \ref{sec:mech}}\label{app:mech}
Several results in Section \ref{sec:mech} rely on existing properties of Gaussian Differential Privacy \cite{gdp} and adapt definitions associated with the GDP framework. The \neighfun-neighbor sensitivity (Definition \ref{def:newsens}) adapts $L_2$-sensitivity (Definition \ref{def:l2sens}) to our proposed framework.

\begin{definition}[$L_2$ sensitivity]\label{def:l2sens}
The $L_2$ sensitivity of a vector-valued function $q$ is defined as $\senstwo(q)\equiv\sup_{\data_1\sim\data_2} ||q(\data_1)-q(\data_2)||_2$, where the supremum is taken over all pairs of datasets that differ on the value of one record.
\end{definition}

 The $\neighfun$-neighbor sensitivity restricts the pairs of dataset to any which are $\neighfun$-neighbors (Definition \ref{def:neigh}) where as $L_2$-sensitivity considers all pairs of datasets differing on one record. Then the  Estab-Gaussian mechanism in Definition \ref{def:attributemech} adapts the Gaussian Mechanism (Definition \ref{def:gaussianmechanism}) using the $\neighfun$-neighbor sensitivity (Definition \ref{def:newsens}).

\begin{definition}[Gaussian Mechanism \cite{gdp}]\label{def:gaussianmechanism}
Let $q$ be a query represented by a vector-valued function. Given a privacy parameter $\mu$ and input dataset $\data$, the Gaussian Mechanism returns $q(\data) + N\left(0,\frac{\senstwo^2(q)}{\mu^2}\mathbf{I}\right)$ and satisfies $\mu$-GDP.
\end{definition}

\printProofs[mechanism]

\section{Approximation Errors in Estimating Query Answer Variances can Impact on Postprocessed Data Quality}\label{app:microdata}
\printProofs[microdata]

\section{An Empirical Exploration of Bias Caused by Estimated Variances}\label{app:bias}
For a simple, but illuminating scenario, consider a dataset with 100 counties and  2 establishments per county. To simplify interpretation of the outcome, set the ground truth  to be 10 employees in each establishment. The queries of interest are the \textbf{ID}: the identity query (how many employees are in each establishment), \textbf{County}: the number of employees in each county, and \textbf{Total:} the total employment. In the first setting, we add noise to each query using \neighmech with $\neighfun=\sqroot$ ($\distparam=0.5$ and $\mu=1$); in the second setting, $\neighfun=\log$ ($\distparam=0.1$ and $\mu=1$). In both cases, Theorem \ref{thm:unbiased} is used to recover unbiased query answers. We consider 3 postprocessing strategies:
\begin{itemize}[leftmargin=*]
    \item \textbf{Est}: Postprocessing with Equation \ref{eqn:microdata} using estimated variances.
    \item \textbf{Act}: (Non-privacy-preserving) ablation experiment using the true variances in Equation \ref{eqn:microdata}.
    \item \textbf{Hybrid}: (Non-privacy-preserving) ablation experiment that uses Equation\ref{eqn:microdata} with estimated variances for the \textbf{ID} query the true variances for the rest. By comparing to \textbf{Act}, we will see the effect of having exact variances for non-identity queries. The full recommended setup, using \neighmech (with estimated variances) for the identity query and \pncmech (whose exact variances are safe to release) will be studied in Section \ref{sec:comps}.
\end{itemize}
Table \ref{tab:toysqrootlog} shows that the impact of uncertain variances is significant -- the  total employment query, which often is the most important, has the most significant accuracy degradation (\textbf{Est}  vs. \textbf{Act} columns). The effects are ameliorated when the true variance is unknown only for the identity query (\textbf{Hybrid} column). Hence, one should avoid mechanisms with non-releasable variance whenever possible.

\begin{table}[th!]
{
\begin{tabular}{|c||ccc||ccc|}\hline
               & \multicolumn{6}{|c|}{\textbf{Ablation Setting, Reporting Mean Squared Errors}}\\
               & \multicolumn{3}{|c||}{$\mathbf{\neighfun=\sqroot}$} 
               & \multicolumn{3}{c|}{$\mathbf{\neighfun=\log}$}\\
\textbf{Query} & \textit{Est} & \textit{Act} & \textit{Hybrid} & \textit{Est} & \textit{Act} & \textit{Hybrid}\\\cline{1-7}
\textbf{ID}    &     $7.5$   &   $7.6$      &  $7.5$ &     
                    $18.0$   &   $23.7$     &  $19.2$ \\
\textbf{County}&    $10.5$   &   $10.0$     &  $10.1$ &
                    $40.0$   &   $37.9$     &  $35.0$\\
\textbf{Total} &   $624.4$   & $335.2$      &  $407.4$ &
                $22,215.3$   & $1,882.8$    & $4,842.6$\\\hline
\end{tabular}
}
    \caption{Postprocessing ablation study for evaluating the impact of using estimated variances with Equation \ref{eqn:microdata}. When $\neighfun=\sqroot$, then $\distparam=0.5, \mu=1$. When $\neighfun=\log$, then $\distparam=0.1, \mu=1$. Metric:  mean-squared errors averaged over $1,000,000$ trials.}
    \label{tab:toysqrootlog}
\end{table}

\section{Generation of Synthetic Evaluation data}\label{app:synthetic}
We aim to create a synthetic establishment-level dataset that mirrors the structure of the QCEW data from publicly accessible data sources in order to run experiments to access our proposed privacy mechanisms and postprocessing procedures. The QCEW structure includes 18 variables, as described in Table \ref{tab:synthdata_vars}.

We use three main sources of publicly accessible data for this task: the Quarterly Workforce Indicators (QWI) \cite{rawqwi}, the County Business Patterns (CBP) dataset \cite{rawcbp}, and an employment count imputed CBP created by \citeauthor{imputecbpdata}\cite{imputecbpdata,imputecbppaper}. The variables of these datasets are summarized in Table \ref{tab:publicdata}.  We focus on privately-owned establishment data from the 50 states for quarter 1 of 2016 since this matches the most recent imputed CBP dataset \cite{imputecbpdata}. 

There are four main steps to create the synthetic data. First, we combine the data sources by county and industry code. Next, we extract and impute monthly employment counts and quarterly wages for each county and NAICS-4 combination. Then, we get month 1 and 3 employment counts, quarterly wages, and number of establishments for each NAICS-6 by county combination. Finally, we generate establishment-level data that matches the QCEW structure.

\begin{table}[!htb]
    \centering
    \small
    \begin{tabular}{|r l|}
    \toprule
    Variable & Description \\ \midrule
        \texttt{year} & year of data; all entries $=2016$ \\
         \texttt{qtr} & quarter of data; all entries $=1$\\
         \texttt{state} & SOC numeric state/territory value from 1 to 56 \\
         \texttt{cnty} & numeric county value, numbered within each state\\
         \texttt{naics} & 6-digit industry code using North American Industry Classification 
	System (NAICS)\\
    \texttt{naics5, naics4, naics3} & 5, 4, and 3 digit industry code respectively\\
    \texttt{sector} & 2-digit NAICS sector code\\
    \texttt{supersector} & BLS sector code groups NAICS sectors\\
    \texttt{own} & ownership code (our data is all code-$5 =$ ``privately owned") \\
    \texttt{m1emp, m2emp, m3emp} & monthly employment counts\\
    \texttt{wage} & total quarterly wages \\
    \texttt{primary\_key} & unique identifier for each establishment\\
    \texttt{can\_agg} & (metadata variable) all entries ="Y"\\
    \texttt{rectype} & (metadata variable) all entries ="C"\\
    \bottomrule
    \end{tabular}
    \caption{Variables for the establishment-level synthetic microdata.}
    \label{tab:synthdata_vars}
\end{table}

\subsection{Public Access Data}

The U.S. Census offers many publicly accessible datasets that contain information similar to the QCEW data, namely the Quarterly Workforce Indicators (QWI) \cite{rawqwi} and the County Business Patterns (CBP) \cite{rawcbp}. The CBP dataset has values aggregated at the County by NAICS-6 level up to the County by Sector level, while county by NAICS-4  is the most precise aggregation level of the QWI dataset. Additionally, both datasets have many missing and suppressed values (Table \ref{tab:suppressioncounts}).
The CBP dataset includes values for number of establishments (\texttt{estnum}), a mid-March (month 3 of quarter 1) employment count (\texttt{emp}) and quarterly total payroll values (\texttt{qp1})  (Table \ref{tab:publicdata}). The CBP dataset is not ideal for creating the monthly employment counts for the synthetic QCEW data, since it only has one of the three monthly employment counts that the QCEW data has.  The QWI dataset has several employment variables discussed in Table \ref{tab:publicdata}: the employment at the beginning and end of the quarter (\texttt{Emp} and \texttt{EmpEnd} respectively) and the number of workers stably employed through the whole quarter (\texttt{EmpS}).  The CBP data can help fill suppressed QWI month 3 employment values, especially since \citeauthor{imputecbppaper} have imputed the 2016 suppressed CBP employment values up to the county by NAICS-6 level and released the imputed dataset \cite{imputecbpdata}. The quarterly payroll value of the CBP data is a good base for the total quarterly data. However, 50\% of \texttt{qp1} are suppressed at the county by NAICS-4 level and the imputed CBP dataset does not include wages (Table \ref{tab:suppression}. The QWI dataset has the average monthly earnings of workers employed at the beginning of the quarter (\texttt{EarnBeg}) and only 2\% of the county by NAICS-4 cells are suppressed. Using \texttt{EarnBeg} and the monthly employment counts we can fill in suppressed values at the county by NAICS-4 level. For denoting a variable from here on, we refer to it by its data source, public attribute grouping, and then variable name. For example, \texttt{QWI.cnty.EmpEnd} is the employment at the end of the quarter at the county level from the QWI dataset. 

\begin{table}[!htb]
    \centering
    \small
    \begin{tabular}{|>{\raggedleft\arraybackslash}p{0.18\linewidth} >{\raggedright\arraybackslash}p{0.63\linewidth}| }
    \toprule
    \multicolumn{2}{|l|}{\cellcolor{gray!25}All Datasets (never suppressed)}\\ \midrule
    Public Attributes & \texttt{state},  \texttt{county}, \texttt{sector}, \texttt{NAICS3}, \texttt{NAICS4} \\
    \midrule
    \multicolumn{2}{|l|}{\cellcolor{gray!25}County Business Patterns (CBP) \citep{rawcbp}}\\ \midrule
       Public Attribute Levels & 
         County by Sector (\texttt{cntySect}), County by NAICS-3 (\texttt{cntyN3}) to County by NAICS-6 (\texttt{cntyN6}) \\
       Variables  &  \texttt{estnum}: the number of establishments \\
    & \texttt{qp1}: quarterly total payroll values  \\ \midrule
    \multicolumn{2}{|l|}{\cellcolor{gray!25}Imputed CBP (imputeCBP) \citep{imputecbpdata}}\\ \midrule
     Public Attribute Levels & County by Sector (\texttt{cntySect}), County by NAICS-3 (\texttt{cntyN3}) to County by NAICS-6 (\texttt{cntyN6})\\
            Variable &\texttt{emp}: the imputed month 3 employment count 
             \\ \midrule
            \multicolumn{2}{l}{\cellcolor{gray!25}Quarterly Workforce Indicators (QWI) \citep{rawqwi}}\\ \midrule
        Public Attribute Levels &  County (\texttt{cnty}), County by NAICS4 (\texttt{cntyN4}) \\
           Variables  &  \texttt{Emp}: Beginning of quarter employment \\
        & \texttt{EmpEnd}: End of quarter employment \\
        & \texttt{EmpS}:  number of employees who work the whole quarter \\
        & \texttt{EarnBeg}: average month 1 earnings of workers \\ \bottomrule
    \end{tabular}
    \caption{The three public access datasets we use have a variety of variables describing the number of establishments, the number of employees, and the quarterly wages aggregated at various county and industry levels. Within the text, we refer to a variable by its data source, public attribute grouping, and then variable name. For example, \texttt{CBP.cntyN4.qp1} is the quarterly total payroll aggregated by county and NAICS4 from the CBP dataset.}
    \label{tab:publicdata}
\end{table}

\begin{table}[!htp]
    \centering
    \small
    \begin{tabular}{ | r|r r r r r |}
    \toprule
       &   \multicolumn{5}{c|}{Geographic Level}  \\ 
       & \multicolumn{1}{c}{\texttt{cntySect}} & \multicolumn{1}{c}{\texttt{cntyN3}} & \multicolumn{1}{c}{\texttt{cntyN4}} & \multicolumn{1}{c}{\texttt{cntyN5}} & \multicolumn{1}{c|}{\texttt{cntyN6}} \\ \midrule
      \texttt{CBP.estnum}   & 0 (0\%)& 0 (0\%)& 0 (0\%)& 0 (0\%)& 0 (0\%) \\
     \texttt{CBP.qp1} & 10,813 (18\%) & 67,190 (36\%) & 210,047 (50\%) & 393,634 (58\%) & 480,987 (61\%) \\
    \texttt{imputeCBP.emp} & 0 (0\%) & 0 (0\%)& 0 (0\%)& 0 (0\%)& 0 (0\%) \\\bottomrule
    \end{tabular}
    \hspace{0.5em}
    \begin{tabular}{ | r|r r|}
    \toprule
       &   \multicolumn{2}{c|}{Geographic Level}  \\
       & \multicolumn{1}{c}{\texttt{cnty}} & \multicolumn{1}{c|}{\texttt{cntyN4}}\\ \midrule
    \texttt{QWI.Emp}   & 1 (0\%) & 187,814 (43\%)\\
      \texttt{QWI.EmpEnd} & 1 (0\%) & 189,021 (43\%) \\
      \texttt{QWI.EmpS} & 1 (0\%) &  184,082 (42\%)\\
     \texttt{QWI.EarnBeg}  & 0 (0\%) & 8,100 (2\%)  \\ \bottomrule
    \end{tabular}
    \caption{The number suppress values at each public attribute group level for each variable. The percent of rows at a given level that are suppressed in shown in parentheses.}
    \label{tab:suppressioncounts}
\end{table}

\subsection{Synthetic County by NAICS-4 Data}
Since the the NAICS-4 by county level is the most specific level of public attribute groups in the QWI data, we start by creating a dataset of monthly employment counts and quarterly wages at this level. We describe the process in detail and in Algorithms \ref{algo:N4empalgo} to \ref{algo:adjustqp1up}. Algorithm \ref{algo:N4empalgo} gets the monthly employment values and uses Algorithm \ref{algo:m2empalgo} to get the month 2 employment count. Algorithm \ref{algo:m2empalgo} is used in the final stage of the data generation as well to get the establishment level month 2 employment counts. Algorithm \ref{algo:N4wagealgo} is for the wage values and uses Algorithm \ref{algo:adjustqp1up} to bound the wages above using hierarchical industry values.

We use a combination of \texttt{QWI.cntyN4.Emp}, \texttt{QWI.cntyN4.EmpEnd}, and \texttt{imputeCBP.cntyN4.emp} to generate the monthly employment counts for the county by NAICS-4 code cells (Algorithm \ref{algo:N4empalgo}). We denote these monthly employment counts as \texttt{cntyN4.m1emp}, \texttt{cntyN4.m2emp}, and \texttt{cntyN4.m3emp} to avoid confusion with the monthly employment counts at the establishment level or at the county by NAICS-6 level.

When \texttt{QWI.cntyn4.Emp} values are not suppressed, we set \texttt{cntyn4.m1emp} to be equal to them. However, when they are suppressed we use a regression model to impute the values. 

First, we fit the model on the rows where \texttt{QWI.cntyN4.Emp} and \texttt{QWI.cntyN4.EmpEnd} are available. The following linear model is used:
\begin{equation}\label{eq:modelEmpN4}
    \texttt{QWI.cntyN4.Emp}_{c,k}=\beta_0+\beta_1\texttt{QWI.cntyN4.EmpEnd}_{c,k}+\beta_2\texttt{CBP.cntyN4.estnum}_{c,k}+\beta_3\texttt{cntyN4.state}_{c,k}+\beta_4\texttt{cntyN4.sector}_{c,k}+\epsilon_{c,k}
\end{equation}
where $\epsilon_{c,k}\overset{iid}{\sim}N(0,\sigma^2)$ for all $(c,k)$ such that $\texttt{QWI.cntyN4.Emp}_{c,k}$ and $\texttt{QWI.cntyN4.EmpEnd}_{c,k}$ are not missing for the combination of county code $c$ and NAICS-4 code $k$. This model was selected to include both geographic and industry code features as well as the number of establishments in cell. For the geographic predictor variable, we cannot use county as a predictor since there are counties that suppress all \texttt{QWI.cntyN4.Emp} values. 
State is used as the geographic predictor instead. Sector is used for similar reasons. Once the model is fitted, we check that the residuals for normality and equal variance, and identify one outlier. The model is refitted without the outlier. If \texttt{QWI.cntyN4.EmpEnd} is suppressed, we use the \texttt{impute.cntyN4.emp} as the predictor. Then we simulate the missing \texttt{cntyN4.m1emp} values using the refitted model. The \texttt{QWI.cntyN4.EmpEnd} filled with the \texttt{imputeCBP.cntyN4.emp} values are used as month 3 employment counts. 

Then we generate month 2 employment counts using Algorithm \ref{algo:m2empalgo} and noise parameter $\eta=1.69$. This algorithm sets \texttt{cntyN4.m2emp} to zero if month 1 and 3 employment counts are zero. Otherwise the change from the midpoint of month 1 and 3 is randomly generated from a normal distribution with mean $0$ and variance proportional to the ratio of the absolute difference and the average of the monthly employment values. More specifically, the variance is $\sigma^2=\eta\frac{|\texttt{cntyN4.m3emp}-\texttt{cntyN4.m1emp}|}{\texttt{cntyN4.m1emp}+\texttt{cntyN4.m3emp}}$. Thus there is more variation from the midpoint when the monthly employment counts are vastly different. When the change between month 1 and month 3 is large relative to the monthly employment counts, the variation from the midpoint will be larger than if the same change is small relative to the monthly employment counts. If industries with a large number of employees are likely to have more steady changes in employment than small but fast-growing industries, then this algorithm mimics that behavior. The Month 2 Employment Algorithm is also used later when generating the establishment level month 2 employment counts.

 \SetKwComment{Comment}{/* }{ */}
\begin{algorithm}
    \caption{\textit{(Employment for County by NAICS-4) } We use \texttt{QWI.cntyN4.Emp} and \texttt{QWI.cntyN4.EmpEnd} as \texttt{m1emp} and \texttt{m3emp} when available. Otherwise, we use \texttt{impute.CBP.cntyN4.emp} as \texttt{m3emp} and predict \texttt{m1emp} from a model fitted on a combination of \texttt{QWI.cntyN4} and \texttt{CBP.cntyN4} variables. Then \texttt{m2emp} is derived from Algorithm \ref{algo:m2empalgo}. Adjustments to the employment counts are made based on \texttt{QWI.cnty.Emp} and \texttt{QWI.cntyN4.EmpS} when available. For an set of indices such as $\mathcal{I}_{*}$ let the complement be denoted $\mathcal{I}^C_{*}$. Let the sum of values over an empty set be equal to $0$.}\label{algo:N4empalgo}
\KwIn{\texttt{QWI.cntyN4.Emp}, \texttt{QWI.cntyN4.EmpEnd}, \texttt{QWI.cntyN4.EmpS}\Comment*[r]{county by NAICS-4 employment values from QWI}\\ \texttt{QWI.cnty.Emp} \Comment*[r]{county-level employment values from QWI}\\
\texttt{imputeCBP.cntyN4.emp}, \texttt{CBP.cntyN4.estnum} \Comment*[r]{county by NAICS-4 values from CBP and imputed CBP}\\
$\mathcal{I}=\{(c,k): (c,k)\in \texttt{QWI.cntyN4} \cap \texttt{imputeCBP.cntyN4}\}$, \Comment*[r]{county by NAICS-4 indices in QWI and imputed CBP}\\
\texttt{cntyN4.state}, \texttt{cntyN4.sector}, \Comment*[r]{state and sector corresponding to county by NAICS-4 cells}\\
$\eta>0$ \Comment*[r]{noise parameter for \texttt{m2emp}}}

\Comment{Defining sets of indices based on missing values}
Let $\mathcal{I}_{/\texttt{Emp}}=\{(c,k)\in \mathcal{I}: \texttt{QWI.cntyN4.Emp}_{c,k} \text{ is missing}\}$\;
Let $\mathcal{I}_{/\texttt{EmpEnd}}=\{(c,k)\in \mathcal{I}: \texttt{QWI.cntyN4.EmpEnd}_{c,k} \text{ is missing}\}$\;
Let $\mathcal{I}_{*}=\{(c,k)\in \mathcal{I}: \texttt{QWI.cntyN4.Emp}_{c,k} \text{ and }\texttt{QWI.cntyN4.EmpEnd}_{c,k} \text{ are not missing}\}$\;

\Comment{Fit a model to predict \texttt{QWI.cntyN4.Emp} from \texttt{QWI.cntyN4.EmpEnd} and other predictors.}
Fit model \eqref{eq:modelEmpN4} with $(c,k)\in \mathcal{I}_{*}$\;
If necessary, identify outliers based on residual normality plot and fitted  verses residual plot. Remove them and refit model \eqref{eq:modelEmpN4}\;
Let $\hat{f}_{\texttt{cntyN4.Emp}}(\cdot)$ be the function that takes in the predictor variable values and returns the fitted value of the model\;
Let $se_{pred}(\hat{y})$ be the function that takes in a fitted value, $\hat{y}$, from the model and returns the standard error of the prediction\;
\ForEach{$(c,k)\in \mathcal{I}_{*}$}{ 
$\texttt{m3emp}_{c,k}\gets \texttt{QWI.cntyN4.EmpEnd}_{c,k}$\Comment*[r]{use \texttt{QWI.cntyN4.Emp} and \texttt{QWI.cntyN4.EmpEnd} is available}
$\texttt{m1emp}_{c,k}\gets \texttt{QWI.cntyN4.Emp}_{c,k}$
}
\ForEach{$(c,k)\in \mathcal{I}_{/\texttt{EmpEnd}}$}{ 
$\texttt{m3emp}_{c,k}\gets \texttt{imputeCBP.cntyN4.emp}_{c,k}$ \Comment*[r]{use \texttt{imputeCBP.cntyN4.emp} if \texttt{QWI.cntyN4.EmpEnd} is missing}
}
\ForEach{$(c,k)\in \mathcal{I}_{/\texttt{Emp}}$}{ 
$\hat{y}_{c,k}\gets \hat{f}_{\texttt{cntyN4.Emp}}(\texttt{m3emp}_{c,k},\texttt{CBP.cntyN4.estnum},\texttt{QWI.cntyN4.state},\texttt{QWI.cntyN4.sector})$\; 
$\texttt{m1emp}_{c,k}\gets \hat{y}_{c,k}+Z_{c,k}$ where $Z_{c,k}\overset{indep.}{\sim}N(0,se_{pred}^2(\hat{y}_{c,k}))$\Comment*[r]{use refitted model if \texttt{QWI.cntyN4.Emp} is missing}
}
\ForEach{$(c,k)\in \mathcal{I}$}{
$\texttt{m2emp}_{c,k}\gets \text{Month 2 Employment}(\eta, \texttt{m1emp}_{c,k},\texttt{m3emp}_{c,k})$ where Month 2 Employment is Algorithm \ref{algo:m2empalgo}

\If{$\texttt{QWI.cntyN4.EmpS}_{c,k}$ is not missing}{  
$\texttt{m}M\texttt{emp}_{c,k}=\max(\texttt{m}M\texttt{emp}_{c,k},\texttt{QWI.cntyN4.EmpS}_{c,k})$ for $M=1,2,3$ \Comment*[r]{ \texttt{QWI.cntyN4.EmpS} lower bound if available}
}
\Else{
$\texttt{m}M\texttt{emp}_{c,k}=\max(\texttt{m}M\texttt{emp}_{c,k},0)$ for $M=1,2,3$ \Comment*[r]{lower bound by 0 if \texttt{QWI.cntyN4.EmpS} is missing} 
}
}

\Comment{Adjust \texttt{m1emp} values from model using the county total value \texttt{QWI.cnty.Emp} if available.}
\ForEach{$c'$ such that $(c',k)\in \mathcal{I}_{/\texttt{Emp}}$}{
\If{$\texttt{QWI.cnty.Emp}_{c'}$ is not missing}{ 
Let $\mathcal{I}_{c',model}=\{k:(c',k)\in \mathcal{I}_{/\texttt{Emp}}\}$\; 
$\Sigma_{c',model}\texttt{m1emp}\gets \sum_{k\in \mathcal{I}_{c',model}}\texttt{m1emp}_{c',k}$\;
$\Sigma_{c',QWI}\texttt{QWI.cntyN4.Emp}\gets \sum_{k\in \mathcal{I}_{c',QWI}}\texttt{QWI.cntyN4.Emp}_{c',k}$ where $\mathcal{I}_{c',QWI}=\{k:(c',k)\in \mathcal{I}^C_{/\texttt{Emp}}\}$\;
\If{$\Sigma_{c',model}\texttt{m1emp}\neq 0$}{
$\texttt{m1emp}_{c',k}\gets \texttt{m1emp}_{c',k}+(\texttt{QWI.cnty.Emp}_{c'}-\Sigma_{c',QWI}\texttt{QWI.cntyN4.Emp})\frac{\texttt{m1emp}_{c',k}}{\Sigma_{c',model}\texttt{m1emp}}$ for each $k\in \mathcal{I}_{c',model}$
}
\Else{
$\texttt{m1emp}_{c',k}\gets \texttt{m1emp}_{c',k}+\frac{1}{|\mathcal{I}_{c',model}\texttt{m1emp}|}(\texttt{QWI.cnty.Emp}_{c'}-\Sigma_{c',QWI}\texttt{QWI.cntyN4.Emp})$ for each $k\in \mathcal{I}_{c',model}$
}
}
}
Let $\texttt{cntyN4.m1emp}$ be a vector of $\max(0,\texttt{m1emp}_{c,k})$ for $(c,k)\in \mathcal{I}$\;
Let $\texttt{cntyN4.m2emp}$ be a vector of $\max(0,\texttt{m2emp}_{c,k})$ for $(c,k)\in \mathcal{I}$\;
Let $\texttt{cntyN4.m3emp}$ be a vector of $\max(0,\texttt{m3emp}_{c,k})$ for $(c,k)\in \mathcal{I}$\;
\KwResult{$\texttt{cntyN4.m1emp},\texttt{cntyN4.m2emp},\texttt{cntyN4.m3emp}$}
\end{algorithm}

Now that all three monthly employment counts have been determined, we check and adjust these values. First, if a stable employment count \texttt{QWI.cntyN4.EmpS} is available, we force this value to be a lower bound on each monthly employment count. We then use in county total \texttt{QWI.cnty.Emp} values to adjust any model-simulated \texttt{m1emp} values. Denote the value for county $c$ as $\texttt{QWI.cnty.Emp}_c$.  We are unable to adjust the values in Kalawao County, Hawaii (FIPS code 15005), since the county total is suppressed. For each other county, we sum the \texttt{QWI.cntyN4.Emp} values in the county and sum the \texttt{m1emp} values that came from the fitted model for that county. For county $c$, denote the sum of the \texttt{QWI.cntyN4.Emp} data as $\Sigma_{c,QWI}\texttt{Emp}$ and the sum from the fitted model as $\Sigma_{c,model}\texttt{m1emp}$. If $\Sigma_{c,model}\texttt{m1emp}\neq 0$, then each model-fitted $\texttt{m1emp}$ is adjusted proportional to its share of the model-fitted sum: $$\texttt{cntyN4.m1emp}_{c,k}=\texttt{cntyN4.m1emp}_{c,k}+\left(\texttt{QWI.cnty.Emp}_c-\Sigma_{c, QWI}\texttt{QWI.cntyN4.Emp}\right)\frac{\texttt{cntyN4.m1emp}}{\Sigma_{c,model}\texttt{m1emp}}.$$ Otherwise \texttt{m1emp} is adjusted by $(\texttt{QWI.cnty.Emp}_c-\Sigma_{c, QWI}\texttt{QWI.cntyN4.Emp})$ over the number of cells that were simulated from the model.

\begin{algorithm}
    \caption{\textit{(Month 2 Employment)} Using a noise parameter $\eta$, we make a \texttt{m2emp} value from the \texttt{m1emp} and \texttt{m3emp} values. This algorithm is used within Algorithms \ref{algo:N4empalgo} and \ref{algo:estempalgo}}\label{algo:m2empalgo}
\KwIn{$\eta>0$, \texttt{m1emp}$\geq 0$, and \texttt{m3emp}$\geq 0$}
\If {$\texttt{m1emp}=\texttt{m3emp}=0$}{
$\texttt{m2emp}=0$
}
\Else{
$\sigma^2=2\eta\frac{|\texttt{m3emp}-\texttt{m1emp}|}{\texttt{m3emp}+\texttt{m1emp}}$\;
Generate $Z\sim \operatorname{Normal}(0,\sigma^2)$\;
$\texttt{m2emp}=\texttt{m1emp}+\frac{\texttt{m3emp}-\texttt{m1emp}}{2}+Z$\;
\If{$\texttt{m2emp}\leq 0$}{
\If{$\texttt{m3emp}=0|\texttt{m1emp}=0$}{
$\texttt{m2emp}=0$
}
\Else{
$\texttt{m2emp}=1$
}
}
}
\KwResult{$\texttt{m2emp}$}
\end{algorithm}


Finally, we can consider the wage values using Algorithm \ref{algo:N4wagealgo}. If the \texttt{CBP.cntyN4.qp1} value is not suppressed, it is used as the \texttt{wage} value. Otherwise, if the average monthly employee earnings at the beginning of the quarter \texttt{QWI.cntyN4.EarnBeg} is available,then we set $\texttt{wage}=\texttt{QWI.cntyN4.EarnBeg}(\texttt{cntyN4.m1emp}+\texttt{cntyN4.m2emp}+\texttt{cntyN4.m3emp})$ where the monthly employment counts are from Algorithm \ref{algo:N4empalgo}. For the remaining missing NAICS-4 by county cells, we use a regression model. Instead of predicting the wage value outright, we predict the difference between the county by NAICS-4 wage value and the sum of available county by NAICS-6 wage value within the NAICS-4 code.

First, we get the sum of the available NAICS-6 \texttt{CBP.cntyN6.qp1} values and the count of suppressed NAICS-6 cells corresponding to the NAICS-4 by county cell. For NAICS-4 code $k$ and county $c$, denote the sum of available NAICS-6 values as $\Sigma^{(6,4)}_{c,k}$, the count of suppressed cells as $M_{c,k}$. If there are no available county by NAICS-6 values for county $c$ and NAICS-4 code $k$, then $\Sigma^{(6,4)}_{c,k}=0$. For cells that have a \texttt{CBP.cntyN4.qp1} value available, we fit a model:
\begin{align}\label{eq:modelqp1N4}
    \sqrt{\texttt{CBP.cntyN4.qp1}_{c,k}-\Sigma^{(6,4)}_{c,k}}&=\beta_0+\beta_1\texttt{cntyN4.state}_{c,k}+\beta_2\texttt{cntyN4.sector}_{c,k}+\beta_3 M_{c,k}+\beta_4 \texttt{cntyN4.m3emp}_{c,k} \nonumber\\
    &\phantom{=}+\beta_5 \texttt{cntyN4.m3emp}_{c,k}M_{c,k}+\beta_6\texttt{cntyN4.m3emp}_{c,k}^2+\beta_7 \texttt{cntyN4.m3emp}_{c,k}^3+\epsilon_{c,k}
\end{align}
where $\epsilon_{c,k}\overset{iid}{\sim}\operatorname{Normal}(0,\sigma^2)$. The predictors \texttt{cntyN4.state} and \texttt{cntyN4.sector} were chosen for similar reasons as the model in \eqref{eq:modelEmpN4}. The number of suppressed cells is important when predicting the amount missing from a cell higher in the hierarchy of industry codes. The \texttt{cntyN4.m3emp} is chosen over the other monthly employment counts since it was not imputed by our model \eqref{eq:modelEmpN4}. The square root transformation of the response variable is select trial and error until the residual plots suggested the assumptions on the residuals seemed reasonable. The structure of the equation for the predictors was chosen based on trial and error and comparing the AIC and residual plots across considered models.
Once the model is fitted, we find five outliers (FIPS\_NAICS-4: 21033\_4452\/\/, 21107\_5619\/\/, 36007\_2361\/\/, 6007\_3399\/\/, 6109\_5231\/\/). These are removed and the model is refitted. For the \texttt{wage} cells at the NAICS-4 by county level that are not generate from \texttt{CBP.cntyN4.qp1} or \texttt{QWI.cntyN4.EarnBeg}, we simulate $\sqrt{\texttt{qp1}_{c,k}-\Sigma_{c,k}^{(6,4)}}$ using the refitted model. Since $\Sigma^{(6,4)}_{c,k}$ is known for all combinations of counties and NAICS-4 codes, we transform the simulated value from the model to generate $\texttt{wage}_{c,k}$ which becomes the \texttt{wage} value for county $c$ and NAICS-4 code $k$. Then we bound all $\texttt{wage}_{c,k}$ values below by $\Sigma_{c,k}^{(6,4)}$. In Algorithm \ref{algo:adjustqp1up}, we find an upper bound for $\texttt{wage}_{c,k}$ using information from the county by NAICS-3, and county by NAICS sector and county levels. Denote the sum of available $d$-digit NAICS  \texttt{qp1} values in the code $k$ of NAICS $r$ level for county $c$ as $\Sigma_{k,c}^{(d,r)}$. If the NAICS-3 by county $\texttt{CBP.cntyN3.qp1}_{c,j}$ value is not suppressed where $j$ is the NAICS-3 code corresponding to NAICS-4 code $k$, then the maximum $\texttt{wage}_{c,k}$ is $\texttt{CBP.cntyN3.qp1}_{c,j}-\Sigma_{c,k}^{(4,3)}$. Otherwise, if the NAICS sector (i.e. 2 digit NAICS) by county $\texttt{CBP.cntySect.qp1}_{c,i}$ value is not suppressed where $i$ is the NAICS Sector corresponding to NAICS-4 code $k$, then the maximum is $\texttt{CBP.cntySect.qp1}_{c,i}-\Sigma_{c,k}^{(3,2)}$. If both the NAICS 3-digit and NAICS sector are suppressed, we use the county total wages minus the sum of available NAICS sector wages as the maximum.

 \SetKwComment{Comment}{/* }{ */}
\begin{algorithm}
    \caption{\textit{(Wage for County by NAICS-4)} For an set of indices such as $\mathcal{I}_{/\texttt{cntyN4.qp1}}$ let the complement be denoted $\mathcal{I}^C_{/\texttt{cntyN4.qp1}}$.Let any sum of values over an empty set be equal to $0$.}\label{algo:N4wagealgo}
\KwIn{\texttt{QWI.cntyN4.EarnBeg}, \Comment*[r]{county by NAICS-4 average earnings values from QWI}\\ \texttt{cntyN4.m1emp},\texttt{cntyN4.m2emp}, \texttt{cntyN4.m3emp} \Comment*[r]{county by NAICS-4 employment values from Algorithm \ref{algo:N4empalgo}}\\
\texttt{CBP.cntyN6.qp1},\texttt{CBP.cntyN4.qp1}, \texttt{CBP.cntyN3.qp1},\texttt{CBP.cntySect.qp1}, \texttt{CBP.cnty.qp1} \Comment*[r]{quarterly payroll values}\\
$\mathcal{I}=\{(c,k): (c,k)\in \texttt{QWI.cntyN4} \cap \texttt{CBP.cntyN4}\}$, \Comment*[r]{county by NAICS-4 indices in QWI and CBP}\\
\texttt{cntyN4.state}, \texttt{cntyN4.sector}, \Comment*[r]{state and sector corresponding to county by NAICS-4 cells}}
\ForEach{$(c,k)\in \mathcal{I}$}{
Let $\mathcal{I}_{c,k}^{(6,4)}=\{\ell\in \text{NAICS-6 codes within }k: \texttt{CBP.cntyN6.qp1}_{c,\ell} \text{ is not missing}\}$\;
$\Sigma_{c,k}^{(6,4)}\gets \sum_{\ell \in \mathcal{I}_{c,k}^{(6,4)}}\texttt{CBP.cntyN6.qp1}_{c,\ell}$
$M_{c,k}\gets |\{\ell\in \text{NAICS-6 codes within }k: \texttt{CBP.cntyN6.qp1}_{c,\ell} \text{ is suppressed}\}|$\;
}
Let $\mathcal{I}_{/\texttt{cntyN4.qp1}}=\{(c,k)\in \mathcal{I}: \texttt{CBP.cntyN4.qp1}\text{ is missing}\}$ \;
Let $\mathcal{I}_{/\texttt{EarnBeg}}=\{(c,k)\in \mathcal{I}: \texttt{QWI.cntyN4.EarnBeg}\text{ is  missing}\}$\;
\ForEach{$(c,k)\in \mathcal{I}^C_{/\texttt{cntyN4.qp1}}$}{
$\texttt{wage}_{c,k}\gets \texttt{CBP.cntyN4.qp1}_{c,k}$ \Comment*[r]{use \texttt{CBP.cntyN4.qp1} if available}
}
Fit model \eqref{eq:modelqp1N4} with $(c,k)\in \mathcal{I}^C_{/\texttt{cntyN4.qp1}}$\;
If necessary, identify outliers based on residual normality plot and fitted  verses residual plot. Remove them and refit model \eqref{eq:modelqp1N4}\;
Let $\hat{f}_{\texttt{cntyN4.qp1}}(\cdot)$ be the function that takes in the predictor variable values and returns the fitted value of the model\;
Let $se_{pred}(\hat{y})$ be the function that takes in a fitted value, $\hat{y}$, from the model and returns the standard error of the prediction\;
\ForEach{$(c,k)\in \mathcal{I}_{/\texttt{cntyN4.qp1}}\cap \mathcal{I}_{/\texttt{EarnBeg}}$}{
$\hat{y}_{c,k}\gets \hat{f}_{\texttt{cntyN4.qp1}}(\texttt{cntyN4.state},\texttt{cntyN4.sector},M_{c,k}, \texttt{cntyN4.m3emp})$\;
$\texttt{wage}_{c,k}\gets (\hat{y}_{c,k}+Z_{c,k})^2+\Sigma_{c,k}^{(6,4)}$ where $Z_{c,k}\overset{indep.}{\sim}N(0,se_{pred}^2(\hat{y}_{c,k}))$ \Comment*[r]{predict wage value with model}
}
\ForEach{$(c,k)\in \mathcal{I}_{/\texttt{cntyN4.qp1}}\cap \mathcal{I}^C_{/\texttt{EarnBeg}}$}{
$w\gets \texttt{QWI.cntyN4.EarnBeg}_{c,k}(\texttt{cntyN4.m1emp}_{c,k}+\texttt{cntyN4.m2emp}_{c,k}+\texttt{cntyN4.m3emp}_{c,k})$\;
$\texttt{wage}_{c,k}\gets \max(w,\Sigma_{k,c}^{(6,4)})$ \Comment*[r]{lower bound is $\Sigma_{k,c}^{(6,4)}$}
}
\Comment{Adjust based on an upper bound for the wage value}
\ForEach{$(c,k)\in \mathcal{I}_{/\texttt{cntyN4.qp1}}$}{
$u_{c,k}\gets \text{Wage Upper Bound}(c,k,\texttt{CBP.cntyN3.qp1},\texttt{CBP.cntySct.qp1},\texttt{CBP.cnty.qp1})$ where Wage Upper Bound is Algorithm \ref{algo:adjustqp1up}\;
$\texttt{wage}_{c,k}=\min(u_{c,k},\texttt{wage}_{c,k})$
}
\texttt{cntyN4.wage} is a vector of $\texttt{wage}_{c,k}$ for $(c,k)\in \mathcal{I}$\;
\KwResult{\texttt{cntyN4.wage}}
\end{algorithm}

\begin{algorithm}
    \caption{\textit{(Wage Upper Bound)} Using the \texttt{CBP} data, we get an upper bound on the quarterly wages for county $c$ and NAICS-4 code $k$. Let the sum of $\texttt{qp1}$ values over an empty set be set equal to $0$.}\label{algo:adjustqp1up}
    \KwIn{$(c,k)$\Comment*[r]{county and NAICS-4 code}\\
\texttt{CBP.cntyN3.qp1},\texttt{CBP.cntySect.qp1}, \texttt{CBP.cnty.qp1}\Comment*[r]{quarterly payroll values from CBP data}}
$i$ and $j$ are the Sector and the NAICS-3 codes corresponding to NAICS-4 code $k$, respectively\; 
\If{$\texttt{CBP.cntySect.qp1}_{c,i}$ is not missing}{
Let $\mathcal{I}_{c,k}^{(3,2)}=\{j'\in \text{NAICS-3 codes within Sector code }i: \texttt{CBP.cntyN3.qp1}_{c,j'} \text{ is not missing}\}$\;
$u_{c,k}\gets \texttt{CBP.cntySect.qp1}_{c,i}- \sum_{j' \in \mathcal{I}_{c,k}^{(3,2)}}\texttt{CBP.cntyN3.qp1}_{c,j'}$
}
\If{$\texttt{CBP.cntyN3.qp1}_{c,j}$ is not missing}{
Let $\mathcal{I}_{c,k}^{(4,3)}=\{k'\in \text{NAICS-4 codes within NAICS-3 code }j: \texttt{CBP.cntyN4.qp1}_{c,k'} \text{ is not missing}\}$\;
$u_{c,k}\gets \texttt{CBP.cntyN3.qp1}_{c,j}-\sum_{k' \in \mathcal{I}_{c,k}^{(4,3)}}\texttt{CBP.cntyN4.qp1}_{c,k'}$
}
\If{$\texttt{CBP.cntyN3.qp1}_{c,j}$ and $\texttt{CBP.cntySect.qp1}_{c,i}$ are missing}{
Let $\mathcal{I}_{c,k}^{(2)}=\{i'\in \text{NAICS Sectors codes that appear in county }c: \texttt{CBP.cntySect.qp1}_{c,i'} \text{ is not missing}\}$\;
$u_{c,k}\gets \texttt{CBP.cnty.qp1}_c-\sum_{i' \in \mathcal{I}_{c,k}^{(2)}}\texttt{CBP.cntySect.qp1}_{c,j'}$
}
\KwResult{$u_{k,c}$}
\end{algorithm}

\subsection{Synthetic County by NAICS-6 Data}

We use the NAICS-4 by county data to create a NAICS-6 by county dataset (Algorithm \ref{algo:cntyN6algo}). First we identify any NAICS-4 and county combinations which only have one NAICS-6 code within them according the CBP data. For these cells we set \texttt{cntyN6.m1emp}, \texttt{cntyN6.m3emp}, and \texttt{cntyN6.wage} to be the values from the corresponding NAICS-4 level and add a column for the number of establishments \texttt{CBP.cntyN6.estnum}. Then we focus on the NAICS-6 by county cells that have not been filled by the NAICS-4 by county values.

Since we are going to use the CBP imputed \texttt{imputeCBP.cntyN6.emp} value for the \texttt{cntyN6.m3emp} value at the NAICS-6 by county level, we need to adjust our \texttt{imputeCBP.cntyN6.emp} values to match the scale of the \texttt{QWI.cntyN4.EmpEnd} values from the QWI that inform some of the NAICS-4 \texttt{cntyN4.m3emp} values. We create a column in the NAICS-4 by county data, $$\texttt{EmpScale}=\begin{cases}\frac{\texttt{imputeCBP.cntyN4.emp}}{\texttt{cntyN4.m3emp}} & \text{ if }\texttt{cntyN4.m3emp}>0\\
1 & \text{otherwise}\end{cases}.$$ For the NAICS-6 codes that did not automatically inherit the NAICS-4 value, we set \texttt{cntyN6.m3emp} to be the NAICS-6 \texttt{imputeCBP.cntyN6.emp} from the imputed CBP over the \texttt{EmpScale} for the corresponding county by NAICS-4 code. For example, in some county, consider the NAICS-4 code 2361, if the CBP \texttt{imputeCBP.cntyN4.emp} for NAICS-4 2361 is $1000$ and the QWI $\texttt{QWI.cntyN4.EmpEnd}=1250$, let our $\texttt{cntyN4.m3emp}=1250$ for that county and NAICS-4 combination. Then the $\texttt{EmpScale}=0.8$. If the county has CBP imputed $\texttt{imputeCBP.cntyN6.emp}=700$ for NAICS-6 code 236115 and $\texttt{imputeCBP.cntyN6.emp}=300$ for NAICS-6 code 236116, then our adjusted \texttt{cntyN6.m3emp} for that county are $875$ for NAICS 236115 and $375$ for NAICS 236116.

\begin{algorithm}
    \caption{\textit{(County by NAICS-6)}}\label{algo:cntyN6algo}
    \KwIn{
    \texttt{cntyN4.m1emp}, \texttt{cntyN4.m2emp}, \texttt{cntyN4.m3emp}, \texttt{cntyN4.wage}, \Comment*[r]{from Algorithms \ref{algo:N4empalgo} and \ref{algo:N4wagealgo}}\\
    \texttt{CBP.cntyN6.qp1},\texttt{CBP.cntyN6.estnum},\texttt{imputeCBP.cntyN6.emp} \Comment*[r]{county by NAICS-6 values}
    $\mathcal{I}_{\texttt{cntyN6}}=\{(c,k,\ell): \text{county, NAICS-4, and NAICS-6 code combination corresponding to }\texttt{imputeCBP.cntyN6}\text{ rows}\}$,\\
    $\mathcal{I}_{\texttt{cntyN4}}=\{(c,k): \text{county, NAICS-4 code combination corresponding to }\texttt{cntyN4.m3emp}\text{ rows}\}$
    }
    \ForEach{$(c,k)\in \mathcal{I}_{\texttt{cntyN4}}$}{
    $\mathcal{I}_{c,k}=\{\ell: (c,k,\ell)\in \mathcal{I}_{\texttt{cntyN6}}\}$\;
    \If{$|\mathcal{I}_{c,k}|=1$}{$\texttt{m1emp}_{c,\ell}=\texttt{cntyN4.m1emp}_{c,k}$ for $\ell\in \mathcal{I}_{c,k}$\;
    $\texttt{m3emp}_{c,\ell}=\texttt{cntyN4.m3emp}_{c,k}$ for $\ell\in \mathcal{I}_{c,k}$\;
    $\texttt{wage}_{c,\ell}=\texttt{cntyN4.wage}_{c,k}$ for $\ell\in \mathcal{I}_{c,k}$
    }
    \Else{
    \If{$\texttt{cntyN4.m3emp}_{c,k}>0$ and $\texttt{imputeCBP.cntyN4.emp}_{c,k}>0$}{
    $\texttt{EmpScale}_{c,k}=\frac{\texttt{imputeCBP.cntyN4.emp}_{c,k}}{\texttt{cntyN4.m3emp}_{c,k}}$\;
    $\texttt{m3emp}_{c,\ell}=\texttt{imputeCBP.cntyN6.emp}_{c,\ell}/\texttt{EmpScale}_{c,k}$ for all $\ell\in \mathcal{I}_{c,k}$
    }\Else{
    $\texttt{m3emp}_{c,k}=\texttt{imputeCBP.cntyN6.emp}_{c,\ell}$ for all $\ell\in \mathcal{I}_{c,k}$
    }
    $(\texttt{m1emp}_{c,\ell}: \ell \in \mathcal{I}_{c,k})^T\gets \text{Dirichlet Divider}( (\texttt{m3emp}_{c,\ell}: \ell \in \mathcal{I}_{c,k})^T, \texttt{cntyN4.m1emp})$ where Dirichlet Divider is Algorithm \ref{algo:dirichletdivider}\;
    $\mathcal{I}_{c,k}^{(6,4)}=\{\ell\in \mathcal{I}_{c,k}: \texttt{CBP.cntyN6.qp1}_{c,\ell}\text{ is not missing}\}$\;
    $\Sigma_{c,k}^{(6,4)}=\sum_{\ell \in \mathcal{I}_{c,k}^{(6,4)}}\texttt{CBP.cntyN6.qp1}_{c,\ell}$\;
    $\texttt{wage}_{c,\ell}=\texttt{CBP.cntyN6.qp1}_{c,\ell}$ for $\ell\in \mathcal{I}_{c,k}^{(6,4)}$ \Comment*[r]{use \texttt{CBP.cntyN6.qp1} if available}
    $A_{c,k}=\texttt{cntyN4.wage}_{c,k}-\Sigma_{c,k}^{(6,4)}$\;
    $(\texttt{wage}_{c,\ell}:\ell\in \mathcal{I}_{c,k}\cap \ell \notin\mathcal{I}_{c,k})^T\gets \text{Dirichlet Divider}((\texttt{m3emp}_{c,\ell}:\ell\in \mathcal{I}_{c,k}\cap \ell \notin\mathcal{I}_{c,k})^T, A_{c,k})$
    }
    }
    \texttt{cntyN6.wage} is a vector of $\texttt{wage}_{c,\ell}$ for $(c,\ell)\in \mathcal{I}_{\texttt{cntyN6}}$\;
    \texttt{cntyN6.m1emp} is a vector of $\texttt{m1emp}_{c,\ell}$ for $(c,\ell)\in \mathcal{I}_{\texttt{cntyN6}}$\;
    \texttt{cntyN6.m3emp} is a vector of $\texttt{m3emp}_{c,\ell}$ for $(c,\ell)\in \mathcal{I}_{\texttt{cntyN6}}$\;
    $\texttt{cntyN6.estnum}\gets \texttt{CBP.cntyN6.estnum}$\;
    \KwResult{\texttt{cntyN6.m1emp}, \texttt{cntyN6.m3emp}, \texttt{cntyN6.wage}, \texttt{cntyN6.estnum}}
    
\end{algorithm}

To get a \texttt{cntyN6.m1emp} for each NAICS-6 by county we use an algorithm based on the Dirichlet distribution which we call the Dirichlet Divider (Algorithm \ref{algo:dirichletdivider}). The algorithm uses a concentration vector parameter of size $d$ where $d$ is the desired output size. The algorithm randomly divides an inputted total value into $d$ subtotals proportional to the concentration parameter. In this case, our concentration parameter is \texttt{cntyN6.m3emp} at the NAICS-6 level corresponding to a given NAICS-4 code. If all concentration vector elements are $0$, the algorithm sets them all to be $1$. The algorithm generates an observation from a Dirichlet distribution that is a random vector of $d$ proportions that sum to 1.  A inputted total value is multiplied by the random vector to get $d$ non-negative subtotals which sum to to the total. In this case we use the \texttt{cntyN4.m1emp} value from the corresponding NAICS-4 code as the total and the resulting subtotals become the \texttt{cntyN6.m1emp} values for the NAICS-6 by county codes. We will not get \texttt{m2emp} values again until we are at the establishment level.

\begin{algorithm}
    \caption{\textit{(Dirichlet Divider)} Using a concentration input vector $\mathbf{b}\in \mathbb{R}^d_{\geq 0}$, the Dirichlet Divider  randomly splits a total value $A$ into $d$ values, $a_1,a_2,\dots,a_d$ such that $A=\sum_{i=1}^da_i$ and $a_i\geq 0$ for $i=1,\dots, d$.}\label{algo:dirichletdivider}
\KwIn{$\mathbf{b}=(b_1~ b_2~\dots ~b_d)^T\in \mathbb{R}^d$ such that $b_i\geq 0$, total value $A\geq0$}
\If{$b_1=b_2=\dots b_d=0$}{
$\mathbf{b}\gets \mathbf{1}\in \mathbb{R}^d$
}
Generate $\mathbf{p}\sim \operatorname{Dirichlet}(\mathbf{b})$ \;
$(a_1~a_2~\dots~a_d)^T\gets \mathbf{p}A$\;
\KwResult{$\mathbf{a}=(a_1~a_2~\dots~a_d)^T$}
\end{algorithm}

Finally, we just need to derive wage values at the NAICS-6 by county level. We get the sum of NAICS-6-level \texttt{CBP.cntyN6.qp1} for a specific NAICS-4 code, $\Sigma_{c,k}^{(6,4)}$. Then we get the difference between that sum and the \texttt{cntyN4.wage} value at the NAICS-4 level. For NAICS-6 codes within the specified NAICS-4 code that do not have \texttt{CBP.cntyN6.qp1} values we use the Dirichlet Divider with the \texttt{cntyN6.m3emp} values as the concentration vector elements and $\texttt{CBP.cntyN4.qp1}_{c,k}-\Sigma_{c,k}^{(6,4)}$ as the total.

\subsection{Establishment-level Synthetic Data}
\begin{figure}
    \centering
    \includegraphics[width=0.6\linewidth]{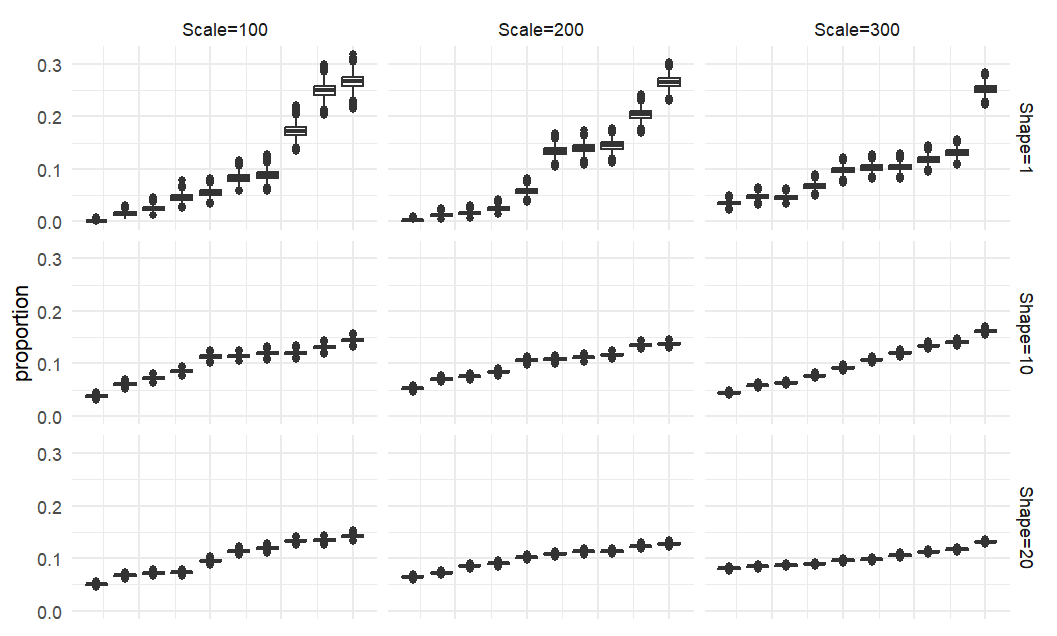}
    \caption{For each of nine different combinations of shape and scale parameters of a Gamma distribution a set of concentration parameters is generated for ten establishments. With each set of concentration parameters, the corresponding proportions are generated over 10,000 replications.}
    \label{fig:appendix-gamma-param-justification}
\end{figure}

The last stage uses the County by NAICS-6 cell values and the Dirichlet Divider Algorithm \ref{algo:dirichletdivider} from previous steps (Algorithm \ref{algo:estempalgo}). We randomly generate the concentration parameters for the Dirichlet Divider as independent and identically distributed random variables from a Gamma distribution with a shape of 10 and a scale parameter of 200 for each establishment in a NAICS-6 by county cell. We use these randomly generated concentration parameters across three applications of the Dirichlet Divider, one for wages, month 1 employment, and month 3 employment. The Gamma distribution produces positive-valued random variables with probability 1 which matches the parameter space of the Dirichlet distribution. Establishment data for wages and employment is often right-skewed \citep{sadeghi2021business,toth2014} and the gamma distribution has a right skew can be easily adjusted using the shape and scale parameters, making it an appropriate choice. Using the same shape parameters for the three applications of the Dirichlet Divider means an establishment should get roughly similar proportions of the wages and employment totals while still allowing the proportion to vary across the wage and employment values. This preserves a relationship between the number of employees and the quarterly wages, but allows the wage per employee to vary across establishments. The shape and scale parameters of the gamma distribution were chosen through trial and error to find values that would result in generated sets of proportions did not vary too much across replications and resulted in values that where not so close to zero that they would be often be rounded down to zero when applied to the total value. A low shape value such as 1 results in low and near zero proportions being generated and varies heavily across applications of the Dirichlet divider. In Figure \ref{fig:appendix-gamma-param-justification}, the proportions up to $0.1$ across repetitions. High shape and scale parameters do not produce much variation across the establishments. 

Using month 1 and 3 employment values from the Dirichlet Divider, month 2 employment is generated with Algorithm \ref{algo:m2empalgo} and noise parameter $0.5$. The state and higher levels of industry codes can be extracted from the county and NAICS-6 codes. In Table \ref{tab:synthdata_vars}, five variables take constant values in our synthetic QCEW data. We complete the generation of the synthetic establishment dataset by adding these constant-valued variables to the dataset.

\begin{algorithm}
    \caption{\textit{(Employment and Wages Microdata) This algorithm uses the Dirichlet Divider (Algorithm \ref{algo:dirichletdivider}} and the Month 2 Employment (Algorithm \ref{algo:m2empalgo} to generate synthetic establishment-level monthly employment and quarterly wage values based on the output from County by NAICS-6 (Algorithm \ref{algo:cntyN6algo}). Then from the NAICS-6 code and county code the remaining variable described in Table \ref{tab:synthdata_vars} are generated. }\label{algo:estempalgo}
    \KwIn{\texttt{cntyN6.m1emp}, \texttt{cntyN6.m3emp}, \texttt{cntyN6.wage}, \texttt{cntyN6.estnum} \Comment*[r]{From Algorithm \ref{algo:cntyN6algo} output}\\
    $\eta>0$ \Comment*[r]{noise parameter for \texttt{m2emp}}\\
    $\alpha_{prior},\theta_{prior}$\Comment*[r]{shape and scale parameters for the Dirichlet prior respectively}}
    $n=0$ \Comment*[r]{initialize index for establishments}
    \ForEach{$(c,\ell)\in \texttt{cntyN6}$}{
    $d\gets \texttt{cntyN6.estnum}_{c,\ell}$\;
    $\mathbf{a}=(a_1,\dots,a_d)$ where $a_j\overset{i.i.d.}{\sim}\operatorname{Gamma}(\alpha_{prior},\theta_{prior})$\;
    $\texttt{m1emp}_{n+1},\dots \texttt{m1emp}_{n+d}\gets \text{Dirichlet Divider}(\mathbf{a},\texttt{cntyN6.m1emp})$ where Dirichlet Divider is Algorithm \ref{algo:dirichletdivider}\;
    $\texttt{m3emp}_{n+1},\dots \texttt{m3emp}_{n+d}\gets \text{Dirichlet Divider}(\mathbf{a},\texttt{cntyN6.m3emp})$ \;
    $\texttt{wage}_{n+1},\dots \texttt{wage}_{n+d}\gets \text{Dirichlet Divider}(\mathbf{a},\texttt{cntyN6.wage})$ \;
    $\texttt{cnty}_{n+j}=c$ for $j=1,\dots, d$\;
    $\texttt{naics}_{n+j}=\ell$ for $j=1,\dots, d$\;
    $\texttt{primary\_key}_{n+j}=n+j$ for $j=1,\dots,d$\;
    $\texttt{m2emp}_{n+j}\gets \text{Month 2 Employment}(\eta, \texttt{m1emp}_{n+j}, \texttt{m3emp}_{n+j})$ for $j=1,\dots,d$ where Month 2 Employment is Algorithm \ref{algo:m2empalgo}\;
   $n=n+d$\;
    }
    \Comment{fill in other variables from \texttt{naics} and \texttt{cnty} values}
    \ForEach{$\texttt{id} \in \texttt{primary\_key}$}{
    Extract $\texttt{supersector}_{\texttt{id}}, \texttt{sector}_{\texttt{id}},\texttt{naics3}_{\texttt{id}}, \texttt{naics4}_{\texttt{id}},$ and $\texttt{naics5}_{\texttt{id}}$ from $\texttt{naics}_{\texttt{id}}$\;
    Extract $\texttt{state}_{\texttt{id}}$ from $\texttt{cnty}_{\texttt{id}}$\;
    Set constant-valued variables $\texttt{year}_{\texttt{id}}=2016,~\texttt{qtr}_{\texttt{id}}=1,~\texttt{own}_{\texttt{id}}=5, \texttt{can\_agg}_{\texttt{id}}="Y",$ and $\texttt{rectype}_{\texttt{id}}="C"$
    }
    
    \KwResult{dataset with a row for each \texttt{primary\_key} and columns for each variable from Table \ref{tab:synthdata_vars}}       
\end{algorithm}

Potential improvements to this method would be to utilize the size class information from the CBP data to inform the generation of monthly employment values at the establishment level. Additionally, a more sophisticated imputation technique could be employed rather than simulating from  regression models. The regression models could also be improved by incorporating more publicly-accessible county-level data or by using more subject-area informed variable selection.

\section{An additional result for QCEW}\label{qpp:qcew}
\begin{figure}[!htp] 
\includegraphics[width=0.24\linewidth]{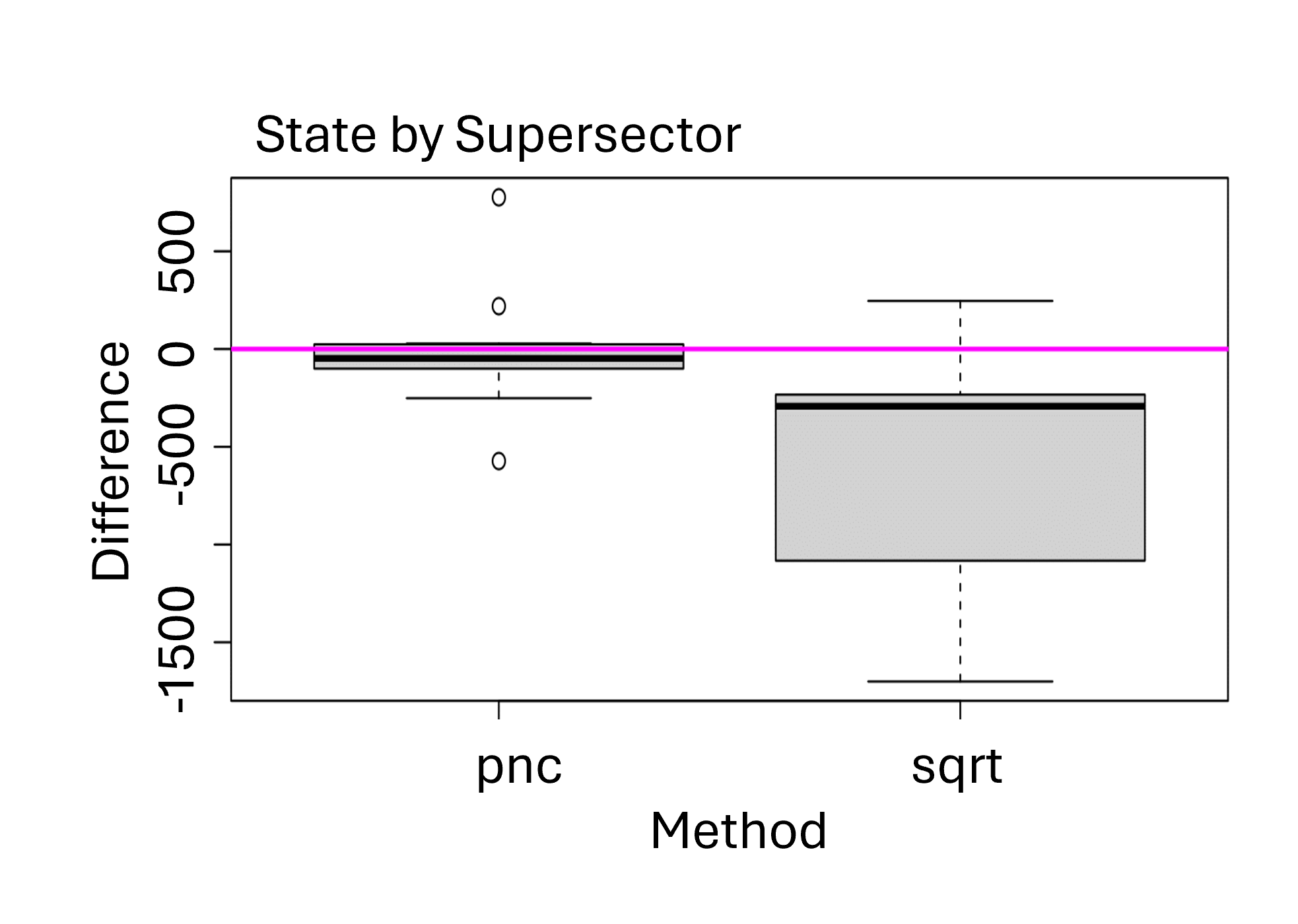}
\includegraphics[width=0.24\linewidth]{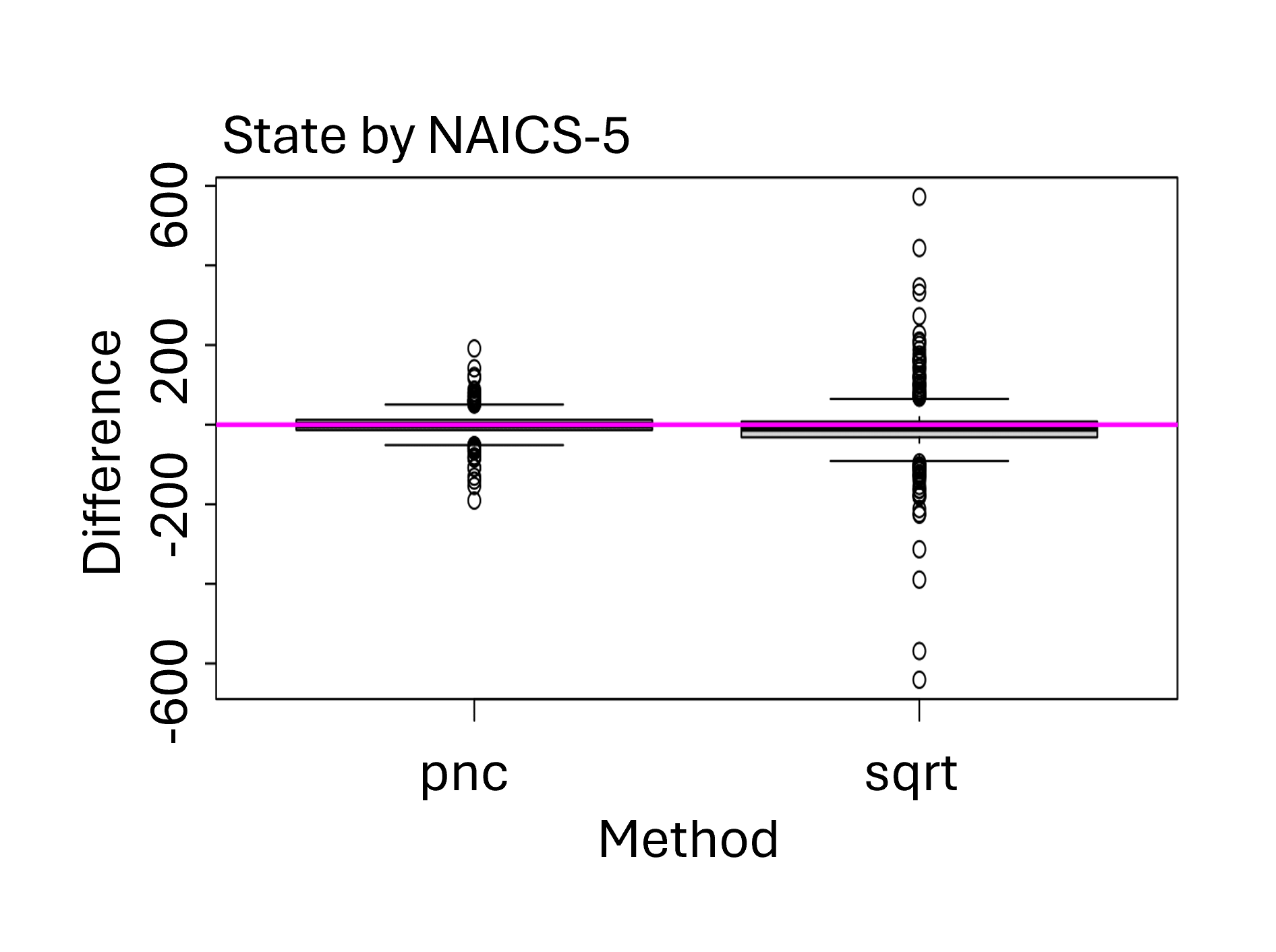}
\includegraphics[width=0.24\linewidth]{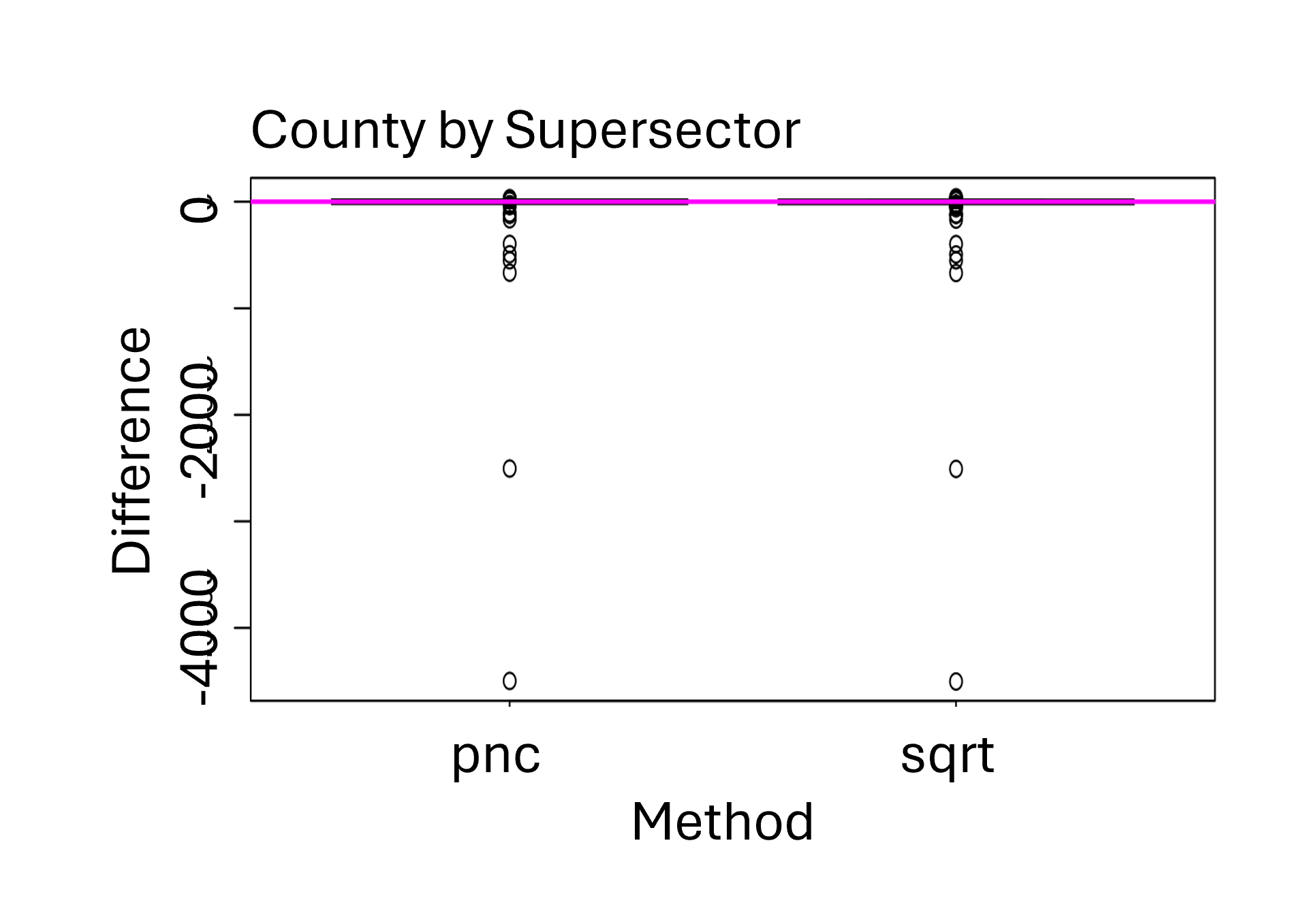}
\includegraphics[width=0.24\linewidth]{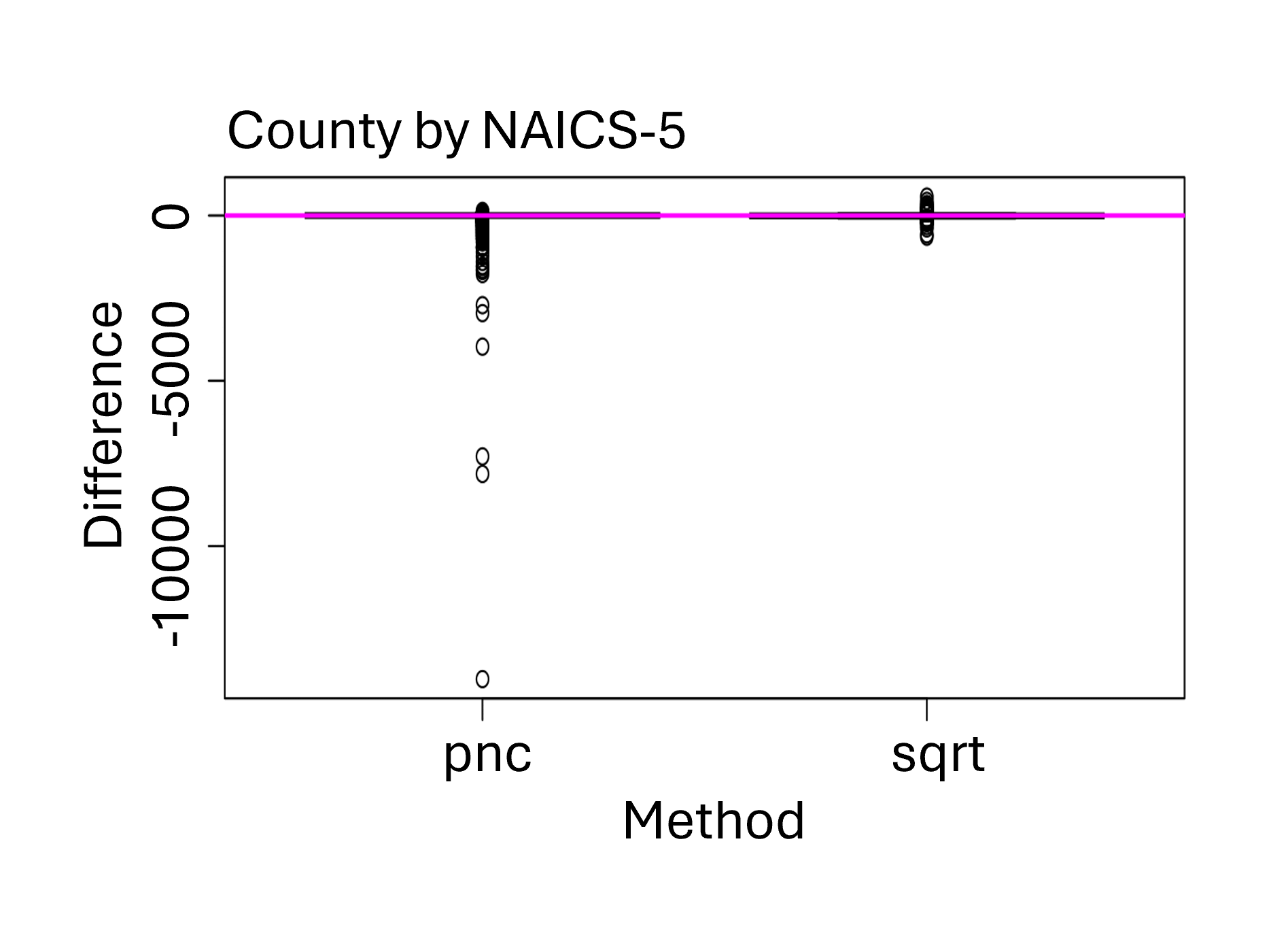}
\caption{Comparison of the differences of the output from the two disclosure limitation methods and the original values for the state at different levels of aggregation. Going from left to right we can 
see the distribution of the difference for the totals of cells by super-sector (aggregation level 53), 
 5-digit NAICS (aggregation level 57), super-sector by county (aggregation level 73), and 5-digit NAICS by county (aggregation level 57). }
\label{fig:appendix_extra_boxplots}
\end{figure}

In Section \ref{sec:experiments_realQCEW}, the quartiles and mean difference between the sanitized and original aggregate cell values for six aggregation levels are reported (Table \ref{tab:qcew_comp_diff}). For a more complete picture of the distribution, the corresponding boxplots for four of these aggregation levels are include in Figure \ref{fig:appendix_extra_boxplots}.

\begin{figure}[!htp] 
\includegraphics[width=0.24\linewidth,clip=true,trim=1.1cm 1.6cm 0.8cm 1.3cm]{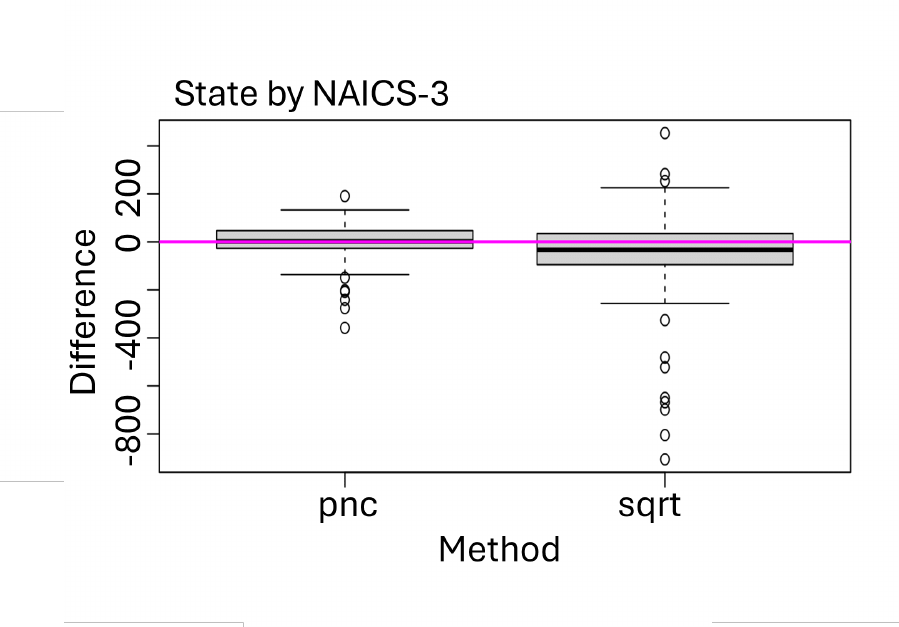}
\includegraphics[width=0.24\linewidth,clip=true,trim=1.0cm 1.6cm 1.1cm 1cm]{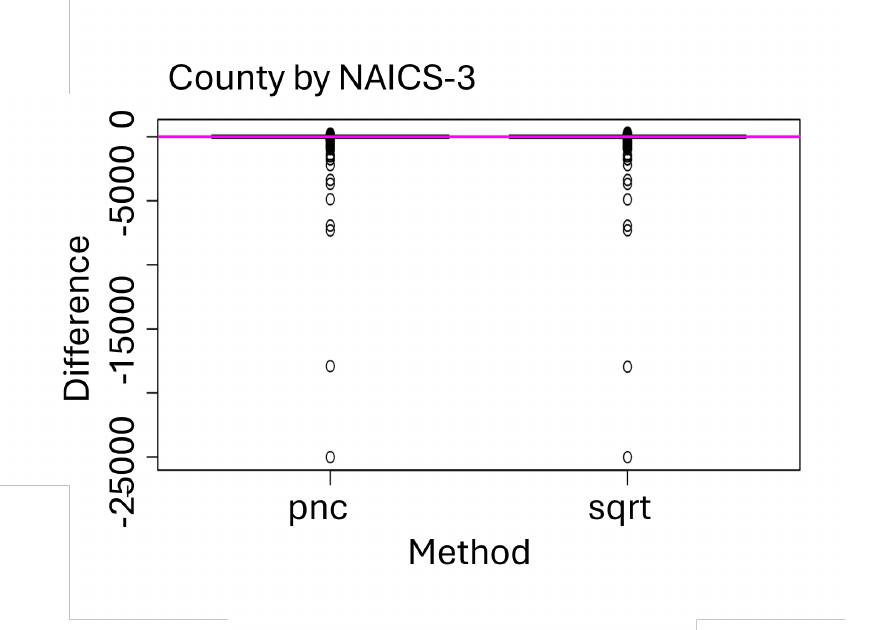}

\caption{(Employment) Box-plot of signed difference for group-by 3-digit NAICS prefix (left) vs. group-by county and 3-digit NAICS prefix (right). Box-plots show median, 1st and 3rd quartiles, and the outliers. Both plots show downward bias. Median and quartiles for the right plot are all near 0.
}
\label{fig:boxplots}
\end{figure}

We can also compare what happens when groupings get more fine-grained.
Figure~\ref{fig:boxplots} shows boxplots (median, 1st and 3rd quartile, and outliers) of the signed differences for the group by queries ``3-digit NAICS prefix'' (left) and ``3-digit NAICS prefix within each county'' (right). Thus, each group in the right-most plot (all establishments with the same 3-digit NAICS prefix) gets subdivided in the left-most plot (a group consists of all establishments in the same county having the same 3-digit NAICS prefix). Figure \ref{fig:boxplots} shows high utility (the quartiles and median are near 0), except for the outliers caused by confidentiality concerns as explained above. By comparing to Figure \ref{fig:agg55}, we see that outliers are caused when the true employment total within a group is small (and hence would raise confidentiality problems if the answer in such a group was too accurate, especially if the group only had 1 establishment). For this reason, plots of confidentiality-preserving vs. true value (e.g., Figure~\ref{fig:agg55}) appear to be more informative than the distribution of raw differences across groups (e.g., Figure~\ref{fig:boxplots}) as they present context that is not present in Figure~\ref{fig:boxplots}.

\section{Additional results from synthetic evaluation data}\label{app:tabsynth}

\begin{table}[!htb]
\small
\addtolength{\tabcolsep}{-0.05em}
    \begin{tabular}{|>{\centering\arraybackslash}p{0.09\linewidth} >{\raggedright\arraybackslash}p{0.29\linewidth}|>{\raggedleft\arraybackslash}p{0.049\linewidth} >{\raggedright\arraybackslash}p{0.059\linewidth} >{\raggedleft\arraybackslash}p{0.062\linewidth} >{\raggedleft\arraybackslash}p{0.062\linewidth}  >{\raggedleft\arraybackslash}p{0.105\linewidth}|}
    \toprule
     $q$ & Query & $\queryprivbudget{q}$ & $\queryprivbudget{q}^2/\mu^2$ & $\queryprivbudgetemp{q}$ & $\queryprivbudgetw{q}$ & $\queryprivbudgetw{q}^2/\queryprivbudget{q}^2$\\
    \midrule
        $I$ & Identity & 1.22 & .281 & 0.70 & 0.15 & .015 \\
       $\Sigma$& State Total  & 0.36 & .024 &0.20 & 0.10 &.077\\
       $N5$ & State NAICS-5 & 1.05 & .207 & 0.60 & 0.15 & .020\\
      $C$ &  County Total & 1.05 & .207 & 0.60 & 0.15 & .020\\
      $C\text{x}N5$ &  County by NAICS-5 & 1.22 & .281 & 0.70 & 0.15 &.015\\
     \bottomrule
    \end{tabular}
    \caption{The baseline privacy loss budget allocation to queries. The overall privacy budget $\mu\approx 2.31$ is allocated to five queries. The query level budget is denoted $\queryprivbudget{q}$ for query $q$. Within each query $q$, the budget is allocated to each month's employment (denoted $\queryprivbudgetemp{q}$) and quarterly wage (denoted $\queryprivbudgetw{q}$). Columns with headings $\queryprivbudgetemp{q}$ and $\queryprivbudgetw{q}$ are exact values; everything else is computed from them using the composition rule (square root of the sum of squares) and rounded.}\label{tab:baseline-parameters}
    \addtolength{\tabcolsep}{0.05em}
\end{table}

 \let\oldclearpage\clearpage

\let\clearpage\relax

For our experiments, we focus on synthetic data for New Jersey and look at error metrics for quarterly wage and month 3 employment level query values. The creation of this synthetic dataset is  described in Section \ref{app:synthetic}. We use two additional utility metrics in the appendix: absolute difference and relative absolute difference. We define absolute difference as $|x-\widetilde{x}|$ where $x$ is a confidential cell value of a query and $\widetilde{x}$ is the corresponding sanitized cell value. The relative absolute difference is the absolute difference scaled by the original value (i.e. $|x-\widetilde{x}|/(x+1)$). When we summarize the utility of cells across a query we use the mean for absolute difference and the median for the relative absolute difference. Median metrics are more robust to outliers than mean metrics so the combination of the two can be helpful to capture error behavior. However, the state total (51) and state by domain (52) queries only have one and two cells respectively, so these summary statistics should be identical in these cases. Additionally, we may be interested in how the distribution of the cell values within a query and the number of establishments aggregated into a cell will influence the utility of the queries. In Table \ref{tab:agg_nj_estnum} and Table \ref{tab:agg_estnum_summaryri_all}, for each query we show the number of cells and some summary statistics for the number of establishments in each cell. The 3-,4-, and 5- digit NAICS codes at both the state and county level have a wide range of establishments per cell with a minimum of 1 and the maximum over 1,000. Most queries result in the standard deviation of establishments per cell that is larger than the mean number of establishments. All this suggests highly variable cells in terms of the number of establishments. 

\begin{table}
\centering
\small

\begin{tabular}{lrrrrr}
\toprule
\multicolumn{6}{c}{\textbf{Synthetic New Jersey State-level Queries}}\\ \midrule
 & Min & Median & Max & Mean & Std.Dev.\\
\midrule
\multicolumn{6}{l}{\textbf{State by Domain (52) has 2 cells}}\\
\hspace{1em} $n$ & 28476 & 103884 & 179292 & 103884 & 106643\\
\hspace{1em} $emp$ & 389796 & 1513961 & 2638126 & 1513961 & 1589809\\
\hspace{1em} $w$ & 56870451 & 65671540 & 74472630 & 65671540 & 12446620\\
\multicolumn{6}{l}{\textbf{State by Supersector (53) has 10 cells}}\\
\hspace{1em} $n$ & 250 & 22542 & 44429 & 20777 & 14141\\
\hspace{1em} $emp$ & 3457 & 236286 & 648521 & 302792 & 245261\\
\hspace{1em} $w$ & 1164792 & 5928548 & 51028965 & 13134308 & 16676271\\
\multicolumn{6}{l}{\textbf{State NAICS Sector (54) has 22 cells}}\\
\hspace{1em} $n$ & 67 & 5070 & 29073 & 9444 & 9538\\
\hspace{1em} $emp$ & 1261 & 97925 & 561919 & 137633 & 129971\\
\hspace{1em} $w$ & 34977 & 2829067 & 24074883 & 5970140 & 7101329\\
\multicolumn{6}{l}{\textbf{State NAICS 3-digit (55) has 74 cells}}\\
\hspace{1em} $n$ & 1 & 641 & 29073 & 2808 & 5253\\
\hspace{1em} $emp$ & 9 & 18359 & 298154 & 40918 & 61407\\
\hspace{1em} $w$ & 97 & 596856 & 19955782 & 1774906 & 3123361\\
\multicolumn{6}{l}{\textbf{State NAICS 4-digit (56) has 258 cells}}\\
\hspace{1em} $n$ & 1 & 144 & 16950 & 805 & 1716\\
\hspace{1em} $emp$ & 1 & 3679 & 200556 & 11736 & 21559\\
\hspace{1em} $w$ & 39 & 161771 & 9017144 & 509082 & 1250695\\
\multicolumn{6}{l}{\textbf{State NAICS 5-digit (57) has 571 cells}}\\
\hspace{1em} $n$ & 1 & 79 & 16950 & 364 & 1075\\
\hspace{1em} $emp$ & 1 & 1530 & 200556 & 5303 & 14205\\
\hspace{1em} $w$ & 39 & 40925 & 8276039 & 230023 & 747805\\
\multicolumn{6}{l}{\textbf{State NAICS 6-digit (58) has 853 cells}}\\
\hspace{1em} $n$ & 1 & 33 & 8172 & 244 & 692\\
\hspace{1em} $emp$ & 0 & 819 & 140800 & 3550 & 10231\\
\hspace{1em} $w$ & 0 & 21351 & 8276039 & 153978 & 591643\\
\bottomrule
\end{tabular}\hspace{3em}
\begin{tabular}{lrrrrr}
\toprule

\multicolumn{6}{c}{\textbf{Synthetic New Jersey County Queries}}\\ \midrule
\midrule
 & Min & Median & Max & Mean & Std.Dev.\\
\midrule
\multicolumn{6}{l}{\textbf{County Total (71) has 21 cells}}\\
\hspace{1em} $n$ & 999 & 9385 & 28689 & 9894 & 6968\\
\hspace{1em} $emp$ & 15581 & 157835 & 370634 & 144187 & 103475\\
\hspace{1em} $w$ & 938253 & 5954245 & 16382122 & 6254432 & 5017341\\
\multicolumn{6}{l}{\textbf{County by Supersector (73) has 208 cells}}\\
\hspace{1em} $n$ & 1 & 676 & 6170 & 999 & 1069\\
\hspace{1em} $emp$ & 5 & 6844 & 99154 & 14557 & 18051\\
\hspace{1em} $w$ & 279 & 131295 & 9177485 & 631457 & 1344169\\
\multicolumn{6}{l}{\textbf{County NAICS Sector (74) has 454 cells}}\\
\hspace{1em} $n$ & 1 & 177 & 4146 & 458 & 638\\
\hspace{1em} $emp$ & 1 & 3280 & 77820 & 6669 & 9323\\
\hspace{1em} $w$ & 56 & 77536 & 8096919 & 289302 & 777652\\
\multicolumn{6}{l}{\textbf{County NAICS 3-digit (75) has 1324 cells}}\\
\hspace{1em} $n$ & 1 & 33 & 4146 & 157 & 348\\
\hspace{1em} $emp$ & 0 & 694 & 44129 & 2287 & 4458\\
\hspace{1em} $w$ & 5 & 13925 & 8096919 & 99202 & 440438\\
\multicolumn{6}{l}{\textbf{County NAICS 4-digit (76) has 3992 cells}}\\
\hspace{1em} $n$ & 1 & 12 & 2110 & 52 & 121\\
\hspace{1em} $emp$ & 0 & 209 & 28438 & 758 & 1704\\
\hspace{1em} $w$ & 0 & 3920 & 5935630 & 32902 & 201878\\
\multicolumn{6}{l}{\textbf{County NAICS 5-digit (77) has 8026 cells}}\\
\hspace{1em} $n$ & 1 & 7 & 2110 & 26 & 79\\
\hspace{1em} $emp$ & 0 & 84 & 25265 & 377 & 1159\\
\hspace{1em} $w$ & 0 & 1180 & 5935630 & 16365 & 137917\\
\multicolumn{6}{l}{\textbf{County NAICS 6-digit (78) has 10067 cells}}\\
\hspace{1em} $n$ & 1 & 5 & 1289 & 21 & 55\\
\hspace{1em} $emp$ & 0 & 63 & 25265 & 301 & 933\\
\hspace{1em} $w$ & 0 & 834 & 5929093 & 13047 & 120849\\
\bottomrule
\end{tabular}
\caption{Summary statistics for number of establishments in a cell ($n$), month 3 employment count ($emp$), and quarterly wages ($w$) for each aggregate level code for the synthetic New Jersey. Overall, synthetic New Jersey has 207,768 establishments.}
\label{tab:agg_nj_estnum}
\end{table}

\begin{table}[!htb]
\centering
\small

\begin{tabular}{lrrrrr}
\toprule
\multicolumn{6}{c}{\textbf{Synthetic Rhode Island State-level Queries}}\\ \midrule
 & Min & Median & Max & Mean & Std.Dev.\\
\midrule
\multicolumn{6}{l}{\textbf{State by Domain (52) has 2 cells}}\\
\hspace{1em}$n$ & 4439 & 12710 & 20981 & 12710 & 11697\\
\hspace{1em}$emp$ & 57050 & 188016 & 318981 & 188016 & 185213\\
\hspace{1em}$w$ & 5985933 & 5989127 & 5992321 & 5989127 & 4517\\
\multicolumn{6}{l}{\textbf{State by Supersector (53) has 10 cells}}\\
\hspace{1em}$n$ & 47 & 2994 & 4903 & 2542 & 1570\\
\hspace{1em}$emp$ & 297 & 34772 & 100286 & 37603 & 30426\\
\hspace{1em}$w$ & 64248 & 548588 & 5309640 & 1197825 & 1606737\\
\multicolumn{6}{l}{\textbf{State NAICS Sector (54) has 22 cells}}\\
\hspace{1em}$n$ & 14 & 687 & 3272 & 1155 & 1165\\
\hspace{1em}$emp$ & 111 & 14654 & 79298 & 17092 & 17778\\
\hspace{1em}$w$ & 35043 & 240442 & 3079880 & 544466 & 769112\\
\multicolumn{6}{l}{\textbf{State NAICS 3-digit (55) has 68 cells}}\\
\hspace{1em}$n$ & 4 & 109 & 2986 & 374 & 634\\
\hspace{1em}$emp$ & 8 & 2604 & 42028 & 5530 & 7873\\
\hspace{1em}$w$ & 104 & 73695 & 1250070 & 176151 & 247533\\
\multicolumn{6}{l}{\textbf{State NAICS 4-digit (56) has 224 cells}}\\
\hspace{1em}$n$ & 1 & 32 & 2409 & 113 & 227\\
\hspace{1em}$emp$ & 1 & 674 & 36147 & 1679 & 3292\\
\hspace{1em}$w$ & 27 & 19038 & 670281 & 53474 & 104387\\
\multicolumn{6}{l}{\textbf{State NAICS 5-digit (57) has 484 cells}}\\
\hspace{1em}$n$ & 1 & 16 & 2409 & 53 & 144\\
\hspace{1em}$emp$ & 1 & 200 & 36147 & 777 & 2278\\
\hspace{1em}$w$ & 1 & 4243 & 670281 & 24748 & 71079\\
\multicolumn{6}{l}{\textbf{State NAICS 6-digit (58) has 663 cells}}\\
\hspace{1em}$n$ & 1 & 10 & 1144 & 38 & 94\\
\hspace{1em}$emp$ & 1 & 147 & 20274 & 567 & 1617\\
\hspace{1em}$w$ & 1 & 2604 & 670281 & 18067 & 58239\\
\bottomrule
\end{tabular}\hspace{3em}
\begin{tabular}{lrrrrr}
\toprule
\multicolumn{6}{c}{\textbf{Synthetic Rhode Island County Queries}}\\ \midrule
Variable & Min & Median & Max & Mean & Std.Dev.\\
\midrule
\multicolumn{6}{l}{\textbf{County Total (71) has 5 cells}}\\
\hspace{1em}$n$ & 1102 & 3354 & 14414 & 5084 & 5340\\
\hspace{1em}$emp$ & 10955 & 38191 & 237653 & 75206 & 92665\\
\hspace{1em}$w$ & 699819 & 1959523 & 4255187 & 2395651 & 1470599\\
\multicolumn{6}{l}{\textbf{County by Supersector (73) has 50 cells}}\\
\hspace{1em}$n$ & 1 & 274 & 2764 & 508 & 634\\
\hspace{1em}$emp$ & 1 & 2936 & 68784 & 7521 & 12431\\
\hspace{1em}$w$ & 1232 & 63424 & 2346055 & 239565 & 432476\\
\multicolumn{6}{l}{\textbf{County NAICS Sector (74) has 107 cells}}\\
\hspace{1em}$n$ & 1 & 104 & 1957 & 238 & 384\\
\hspace{1em}$emp$ & 1 & 1462 & 54369 & 3514 & 6694\\
\hspace{1em}$w$ & 87 & 25122 & 1314614 & 111946 & 241131\\
\multicolumn{6}{l}{\textbf{County NAICS 3-digit (75) has 303 cells}}\\
\hspace{1em}$n$ & 1 & 17 & 1698 & 84 & 197\\
\hspace{1em}$emp$ & 0 & 328 & 22958 & 1241 & 2834\\
\hspace{1em}$w$ & 0 & 6722 & 1175186 & 39532 & 109617\\
\multicolumn{6}{l}{\textbf{County NAICS 4-digit (76) has 854 cells}}\\
\hspace{1em}$n$ & 1 & 8 & 1323 & 30 & 73\\
\hspace{1em}$emp$ & 0 & 104 & 19551 & 440 & 1204\\
\hspace{1em}$w$ & 0 & 2232 & 510039 & 14026 & 48852\\
\multicolumn{6}{l}{\textbf{County NAICS 5-digit (77) has 1620 cells}}\\
\hspace{1em}$n$ & 1 & 4 & 1323 & 16 & 49\\
\hspace{1em}$emp$ & 0 & 42 & 19551 & 232 & 861\\
\hspace{1em}$w$ & 0 & 702 & 510039 & 7394 & 34446\\
\multicolumn{6}{l}{\textbf{County NAICS 6-digit (78) has 1996 cells}}\\
\hspace{1em}$n$ & 1 & 4 & 610 & 13 & 34\\
\hspace{1em}$emp$ & 0 & 35 & 12845 & 188 & 674\\
\hspace{1em}$w$ & 0 & 498 & 510039 & 6001 & 30892\\
\bottomrule
\end{tabular}
\caption{Summary statistics for number of establishments ($n$), employment count ($emp$), and quarterly wages ($w$) for each aggregate level code for the synthetic Rhode Island. Overall, synthetic Rhodes Island has 25,420 establishments.}
\label{tab:agg_estnum_summaryri_all}
\end{table}

\begin{figure}
    \centering
    \includegraphics[width=0.7\linewidth]{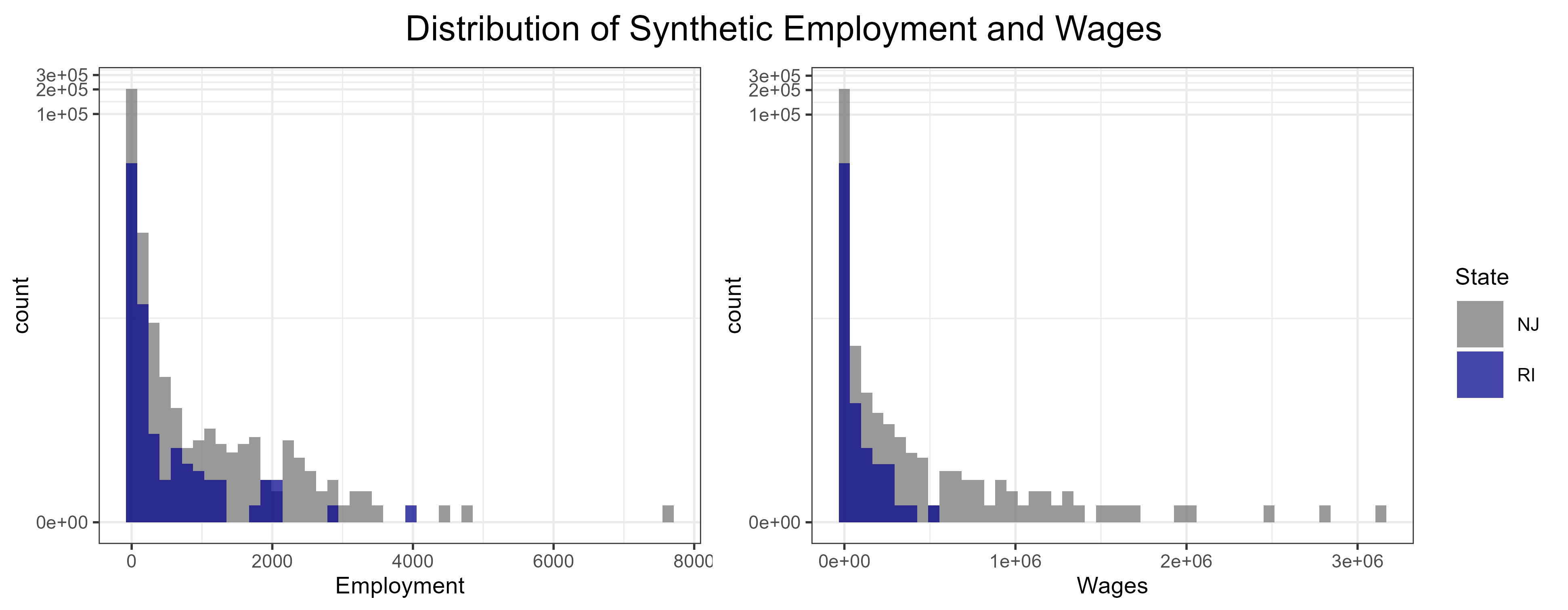}
    \caption{The distributions of wages and employment of the establishments in the Synthetic New Jersey and Rhode Island Datasets features a heavy tail. The skew of the data is so intense that the scale of the y-axis have to be transformed so the counts of the higher values bins are visible.}
    \label{fig:nj_distribution}
\end{figure}

\begin{figure}[!htp]
    \centering
    \includegraphics[width=0.3\linewidth]{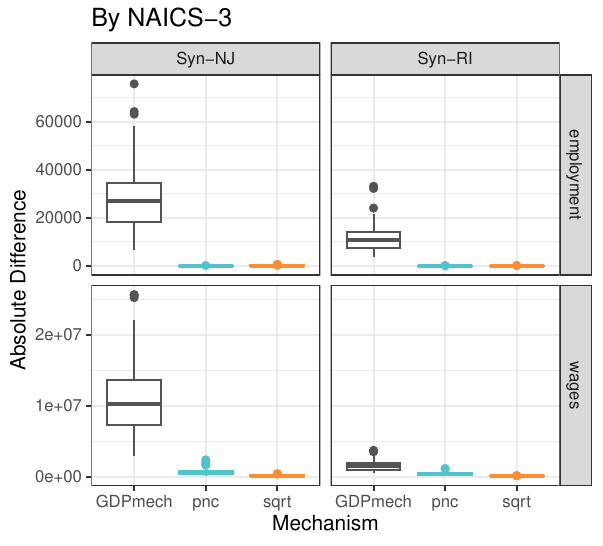}
    \includegraphics[width=0.3\linewidth]{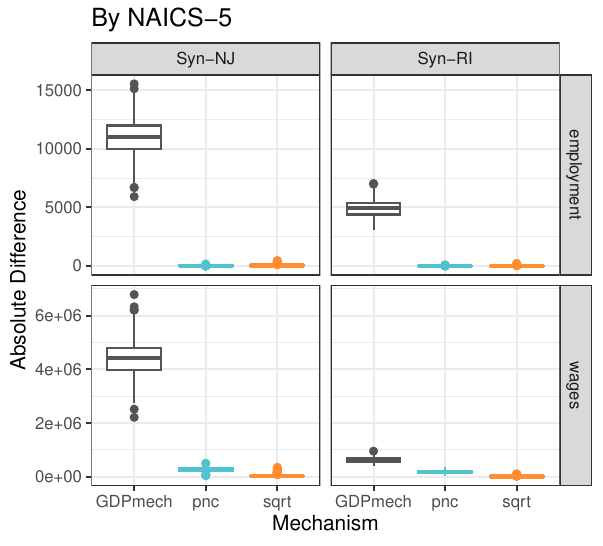}
    \caption{Absolute error for each query group averaged across 34 replicates including the GDP mechanism.}
    \label{fig:appendix_query_wGDP}
\end{figure}

\begin{figure}[!htp]
    \centering
    \includegraphics[width=0.3\linewidth]{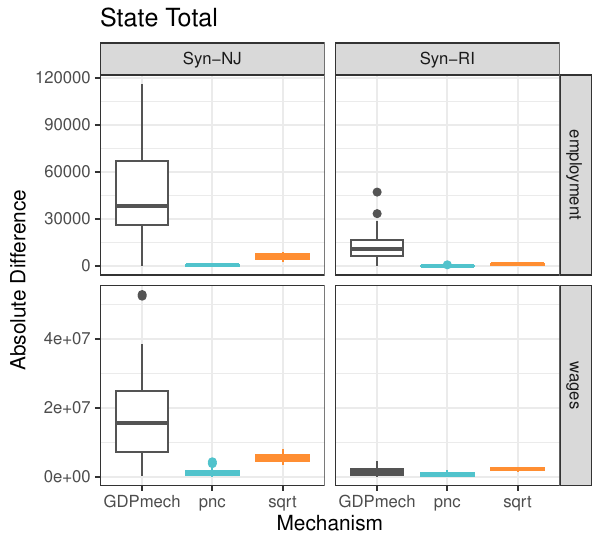}
       \caption{Absolute error across 34 replicates including the GDP mechanism.}
    \label{fig:appendix_state_wGDP}
\end{figure}

In Section \ref{sec:experiment_synth}, the GDP mechanism is omitted from several plots (Center and right plots of Figure \ref{fig:eqempwage} and Figure \ref{fig:eqempwage2}) because the scale of its error is so large it obscures the differences between the \pncmech and \sqrtmech. The plots including the GDP mechanism is shown in Figures \ref{fig:appendix_query_wGDP} and Figures \ref{fig:appendix_state_wGDP}.

For all additional experiments, we use a baseline \pncmech implementation with $\confparam=0.01$ and \sqrtmech implementation for comparison. This baseline matches the parameters used in Section \ref{sec:comps}. The baseline overall privacy budget is $\mu=2.31$ and the privacy distance for monthly employment is $\distparam_{emp}=0.5$ and for quarterly wages is $\distparam_w=50$. The baseline implementations use the  queries and privacy parameters described in Table \ref{tab:baseline-parameters}

\begin{figure}[!htp]
    \centering
    Syn-NJ Employment:
    \includegraphics[width=0.98\linewidth,clip=true,trim=0.57cm 0.5cm 0.6cm 0.7cm]{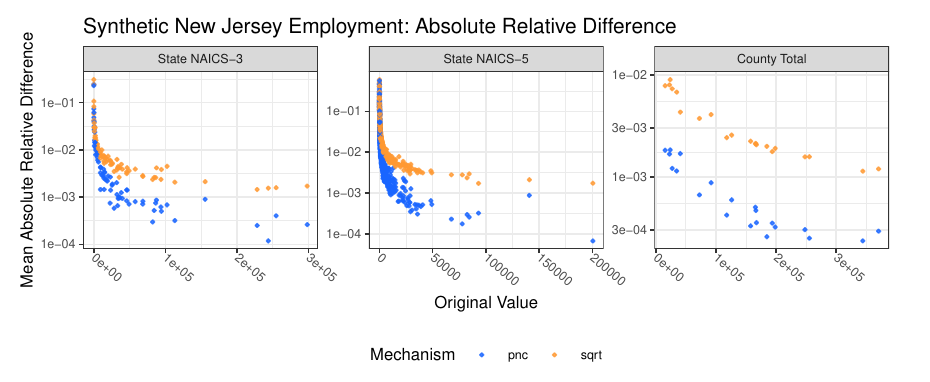}
    
    \vspace{-1.1em}
    Syn-NJ Wages:
    \includegraphics[width=0.98\linewidth,clip=true,trim=0.57cm 0.3cm 0.3cm 0.7cm]{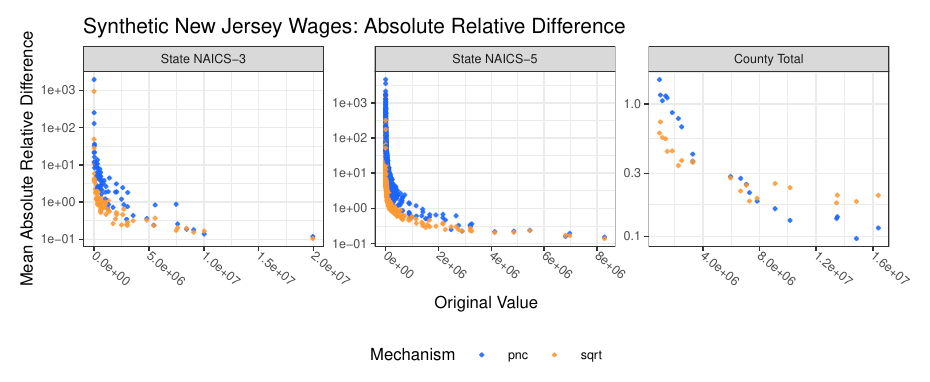}
    \caption{Comparison of \pncmech in blue vs. \sqrtmech  in orange on Synthetic NJ for employment and wages, using the baseline configuration from Table \ref{tab:baseline-parameters} ($\mu\approx 2.31$). The first column is the query that groups by 3-digit NAICS prefix (not a measurement query), while the other two columns evaluate accuracy on 2 measurement queries. $x$-axis: ground truth answer for a group, $y$-axis: corresponding relative error for that group. %
    }
    \label{fig:baserel}
\end{figure}

When we vary privacy allocation with a fixed overall budget, we scale the privacy share of a query or confidential value type (i.e. wages or employment). This share is described as the squared ratio of the privacy cost to its next highest cost level. In other words, for query $k$ the wage privacy share is the squared privacy cost to wage ($\mu_{Q,w}^2$) over the query-level privacy budget ($\mu_Q^2$). For the baseline parameters this value is in the last column of Table \ref{tab:baseline-parameters}. When we look at privacy allocation to queries, we look at the squared query budget $\mu_Q^2$ over the squared total budget $\mu^2$. For the baseline parameters, this is fourth column labeled $\mu^2_Q/\mu^2$ in Table \ref{tab:baseline-parameters}.

 \begin{figure}[!htp]
     \centering
     \includegraphics[width=0.57\linewidth]{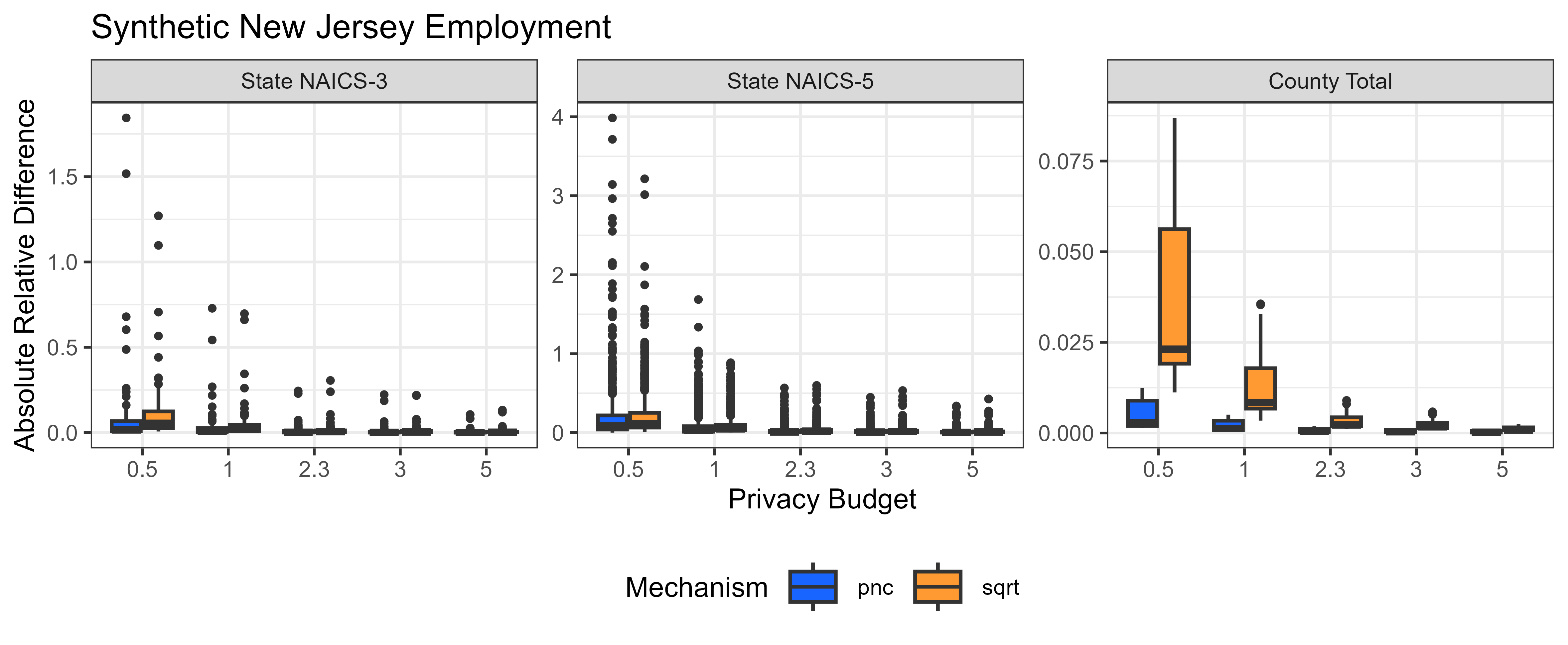}\\
 \includegraphics[width=0.57\linewidth]{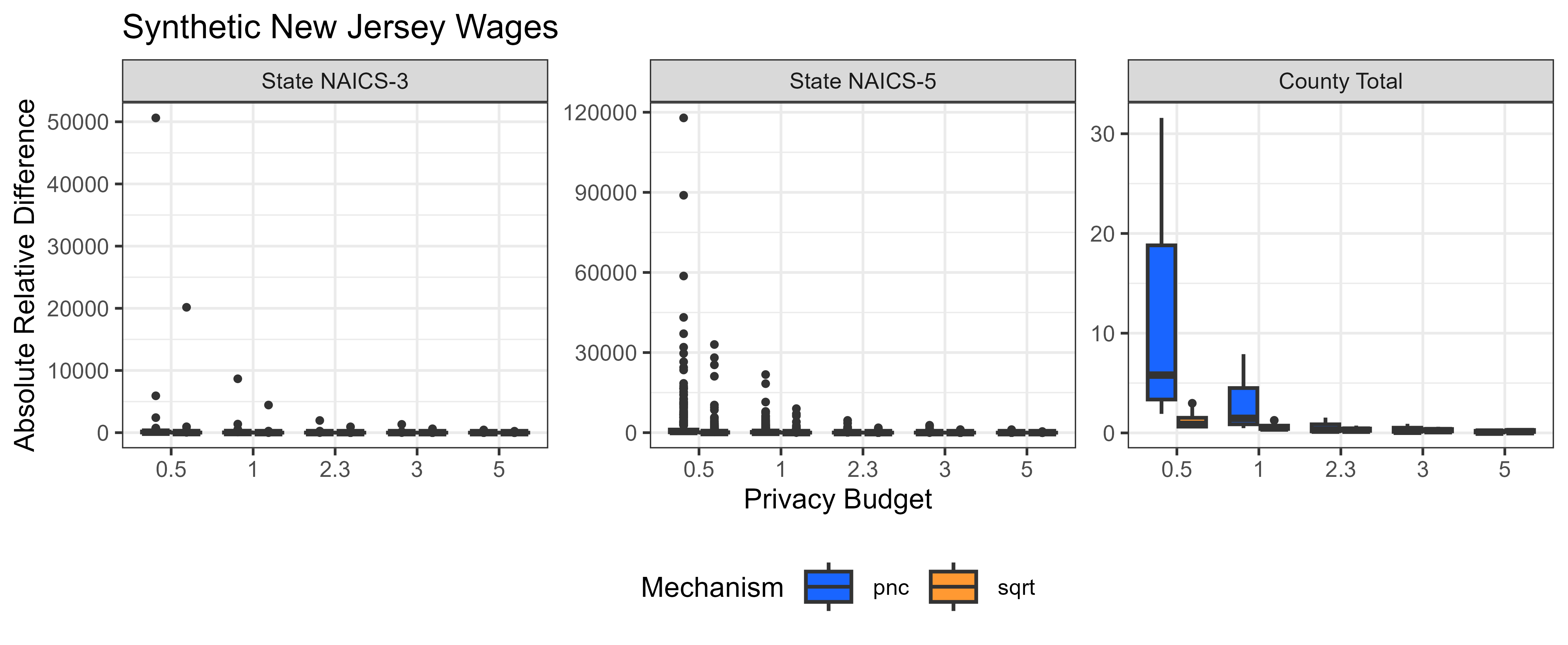}
     \caption{Employment and Wages relative errors, as overall privacy budget varies, using the base configuration from Table \ref{tab:baseline-parameters}.}
     \label{fig:budgets_boxplots_outliers}
 \end{figure}

\let\clearpage\oldclearpage

\clearpage

\end{document}